\documentclass[3p]{elsarticle}
\usepackage{graphicx}
\usepackage{amssymb}
\usepackage{subfiles}
\usepackage{subcaption}
\usepackage{amsmath}

\usepackage{floatrow}
\usepackage[colorlinks]{hyperref}

\usepackage{booktabs}
\usepackage{array}
\usepackage{arydshln}
\newcommand{\PreserveBackslash}[1]{\let\temp=\\#1\let\\=\temp}
\newcolumntype{C}[1]{>{\PreserveBackslash\centering}p{#1}}
\newcolumntype{M}[1]{>{\centering\arraybackslash}m{#1}}
\newcolumntype{L}[1]{>{\arraybackslash}m{#1}}
\newcolumntype{N}{@{}m{0pt}@{}}

\newcommand{\beq}{\begin{equation}}
\newcommand{\eeq}{\end{equation}}
\newcommand{\bea}{\begin{eqnarray}}
\newcommand{\eea}{\end{eqnarray}}
\newcommand{\ev}[1]{\left\langle #1 \right\rangle}
\renewcommand{\d}{{\rm d}}
\renewcommand{\i}{{\rm i}}
\newcommand{\e}{{\rm e}}

\newcommand{\equref}[1]{Eq.~(\ref{Eq:#1})}
\newcommand{\figref}[1]{Fig.~\ref{fig:#1}}
\newcommand{\tabref}[1]{Tab.~\ref{tab:#1}}
\newcommand{\secref}[1]{Sec.~\ref{sect:#1}}

\def\CO{{\mathcal O}}
\def\CZ{{\mathcal Z}}
\def\CD{{\mathcal D}}

\journal{\ }

\begin{document}

\begin{frontmatter}

\title{Complex Langevin and other approaches to \\ the sign problem in quantum many-body physics}

\author[unc]{C. E.~Berger\corref{cor1}}
\ead{caseyb@bu.edu}

\author[tud,gsi]{L.~Rammelm\"uller\corref{cor1}}
\ead{lrammelmueller@theorie.ikp.physik.tu-darmstadt.de}

\cortext[cor1]{These authors contributed equally to this work.}

\author[unc]{A. C.~Loheac}
\ead{loheac@live.unc.edu}

\author[tud]{F.~Ehmann}
\ead{fehmann@theorie.ikp.physik.tu-darmstadt.de}

\author[tud,emmi,fair]{J.~Braun}
\ead{jens.braun@physik.tu-darmstadt.de}

\author[unc]{J. E.~Drut}
\ead{drut@email.unc.edu}

\address[unc]{Department of Physics and Astronomy, University of North Carolina, Chapel Hill, North Carolina 27599, USA}
\address[tud]{Institut f\"ur Kernphysik (Theoriezentrum), Technische Universit\"at Darmstadt, D-64289 Darmstadt, Germany}
\address[gsi]{GSI Helmholtzzentrum f\"ur Schwerionenforschung GmbH, Planckstra\ss e 1, D-64291 Darmstadt, Germany}
\address[fair]{FAIR, Facility for Antiproton and Ion Research in Europe GmbH, Planckstra\ss e 1, D-64291   Darmstadt, Germany}
\address[emmi]{ExtreMe Matter Institute EMMI, GSI, Planckstra{\ss}e 1, D-64291 Darmstadt, Germany}

\begin{abstract}
We review the theory and applications of complex stochastic quantization to the quantum many-body problem. Along the way, we present a brief overview of a number of ideas that either ameliorate or in some cases altogether solve the sign problem, including the classic reweighting method, alternative Hubbard-Stratonovich transformations, dual variables (for bosons and fermions), Majorana fermions, density-of-states methods, imaginary asymmetry approaches, and Lefschetz thimbles. We discuss some aspects of the mathematical underpinnings of conventional stochastic quantization, provide a few pedagogical examples, and summarize open challenges and practical solutions for the complex case. Finally, we review the recent applications of complex Langevin to quantum field theory in relativistic and nonrelativistic quantum matter, with an emphasis on the nonrelativistic case.
\end{abstract}


\end{frontmatter}

\tableofcontents

\newpage

\section{\label{sect:intro}Introduction}

\subsection{The challenge of many-body quantum mechanics: memory and statistics}

In the early days of quantum mechanics it was quickly discovered that the Schr\"odinger equation could be solved analytically for hydrogen and hydrogen-like atoms in a straightforward manner~\cite{BetheSalpeterBook}. However, each new particle added to the problem came at a dauntingly steep price, leaving the vast majority of the periodic table of the elements unattainable due to the complexity of the equations and the accompanying high cost of computation. Indeed, the presence of more than two interacting particles yields equations that are analytically intractable. The quantum few-body problem thus appeared to be very difficult, and the chances of solving the quantum {\it many-body} problem seemed dire. At the heart of the problem is the fact that, while the influence of the massive atomic nucleus on the much lighter electrons can be approximated (as a static external field \`a la Born-Oppenheimer), addressing the Coulomb interaction among the electrons is far more challenging. Dirac famously remarked in 1929 that, while the underlying physical laws were then completely known, ``the difficulty is only that the exact application of these laws leads to equations much too complicated to be soluble''\cite{DiracQuote}. This difficulty could hardly be overemphasized then, and remains a challenge to this day. In facing that challenge, a wide variety of algorithms was -- and continues to be -- developed by specialists around the world to fit the paradigms of their specific area of physics or chemistry.

The most common first-principles approaches to the quantum many-body problem can be roughly divided into two sets: memory intensive and statistics intensive. The former include methods such as exact diagonalization (see e.g.~\cite{BARRETT2013131,Johnson:2018hrx}) and coupled cluster (see e.g.~\cite{RevModPhys.79.291, Hagen:2013nca}), while the latter include a set of stochastic techniques generally known as quantum Monte Carlo (QMC) methods. Within that QMC set, this review focuses on a large class of approaches for which the many-body problem is expressed in the language of second-quantization or quantum field theory, such that expectation values of operators are written as a path integral over continuous fields living on a spacetime lattice. That formulation is in fact very general -- it is natural in relativistic quantum field theory as well as nuclear and condensed matter physics, either in the form of low-energy effective field theories (see e.g.~\cite{Burgess:2007pt,RevModPhys.81.1773,Machleidt:2011zz}), or as a reformulation of traditional Hamiltonians like the Hubbard model (see e.g.~\cite{Shankar:1996vk,doi:10.1142.6826}). Regardless of the application, the computational cost of path-integral QMC methods scales at face value (see below) polynomially with particle number and basis size (i.e. the size of the spacetime lattice), which makes them exceptionally well-suited for the many-body problem.

An essential component of QMC techniques is that they rely on a stochastic process governed by the Metropolis accept-reject algorithm~\cite{Metropolis}, which itself requires a well-defined probability measure to guarantee convergence to the correct result. Simply put, the algorithm requires that the
partition function $\mathcal Z$ be written as a sum of positive weights $W(C)$ (which play the role of the probability mentioned above) over some set of configurations $C$:
\beq
\mathcal Z = \sum_C W(C).
\eeq
The Metropolis algorithm, by construction, provides samples of the configurations $C$ distributed according to $W(C)$. Under many circumstances, however, a serious issue arises for this kind of algorithm, which has hindered computation in a wide range of situations: the infamous {\it sign problem}.
In those cases, $W(C)$ does not have a well-defined sign or even becomes complex (as explained in further detail below). Unfortunately, by far most systems of interest suffer from such a problem: high-$T_c$ superconductors (due to strong repulsive interaction away from half filling, see e.g.~\cite{PhysRevB.41.9301}), nuclear structure (strong repulsive core, finite spin-isospin polarization, see e.g.~\cite{KOONIN19971, Alhassid:2016ojg}), and quantum chromodynamics (finite quark density see e.g.~\cite{Gattringer:2016kco, Bongiovanni:2016ess, Aarts:2015tyj}), to name only a few.

Over the last few decades, many ideas have been proposed to overcome the sign problem in quantum many-body physics and field theory. This review covers some of them briefly and focuses on the so-called complex Langevin (CL) approach, as applied to the calculation of equilibrium properties of quantum many-body systems in relativistic and nonrelativistic physics, with an emphasis on the latter. The next section sketches out the basic path-integral formalism involved in Metropolis-based and stochastic quantization approaches, with the goal of showing where and how the sign problem arises.

\subsection{Path integrals and the sign problem}

The central quantity of the field theoretical approach to the quantum many-body problem is the partition function,
which in the grand canonical ensemble is given by
\beq
\label{Eq:Z}
\mathcal Z = \textrm{Tr}\left[ {\rm e}^{-\beta (\hat H - \mu \hat N)} \right] ,
\eeq
where $\hat H$ is the Hamiltonian of the system, $\hat N$ the particle number operator, $\beta$ the inverse temperature, $\mu$ the chemical potential, and the trace is over all multiparticle states (i.e. Fock space). Note that $\mathcal Z$ is shown here for a single particle species, but is straightforwardly generalized to multiple chemical potentials, etc. As written, $\mathcal Z$ contains the thermodynamic information of the system: by differentiating with respect to $\beta$ and $\mu$ one obtains expectation values of the Hamiltonian and the particle number operators. More detailed information, such as momentum distributions and other correlation functions, can be obtained by adding sources to the Hamiltonian as is common in quantum field theory. The direct evaluation of \equref{Z} is impossible for interacting systems, as it requires \textit{a priori} knowledge of the full energy spectrum.

The path-integral approach to the many-body problem provides an alternative route. Either using an operator-based approach or coherent states, one arrives at an expression for $\mathcal Z$ which is written generically as
\beq
\label{Eq:PartitionZ}
\mathcal Z = \int \mathcal D \phi\ {\rm e}^{-S[\phi]},
\eeq
where $\phi$ is a field living in $(d+1)$-dimensional spacetime and represents any degrees of freedom in the system.
One thus replaces the problem of evaluating \equref{Z} with that of calculating the above path integral.
In practice, boundary conditions in the spatial directions can be chosen in a variety of ways, but those in the time
direction, which is compact and runs in the range $\tau \in [0,\beta)$, are set by the quantum statistics of the problem:
bosonic fields will obey periodic boundary conditions and fermionic fields anti-periodic.

From this point on, our discussions will focus on nonrelativistic systems unless otherwise specified.
In purely bosonic theories the action $S[\phi]$ will typically take a simple local form such as
\beq
\label{Eq:BosonAction}
S[\phi] = \int \d\tau \d^dx \left\{ \phi^* \left (\partial_\tau + {\mathcal H}\right) \phi + V[\phi] \right\},
\eeq
where $\mathcal H$ represents the noninteracting Hamiltonian (including external trapping potentials) and $V[\phi]$ represents the interactions
(i.e. terms cubic and beyond in $\phi$).

For real bosonic variables $\phi$, the action $S[\phi]$ is also real and therefore ${\rm e}^{-S[\phi]}$ can be used as a probability measure in a stochastic process.
For complex $\phi$, however, the fact that $\partial_\tau$ is an antisymmetric operator results in a complex $S[\phi]$, which is the source of
a sign problem in this formulation (see below) and has a counterpart in relativistic bosons at finite chemical potential (see Secs.~\ref{sect:DualVariables} and~\ref{sect:RQFT}).
The problem of the antisymmetry of $\partial_\tau$ can be circumvented for an even number of species with attractive interactions,
which however render bosons unstable (but not fermions, see below).

\begin{figure}[h]
  \centering
  \includegraphics[width=\columnwidth]{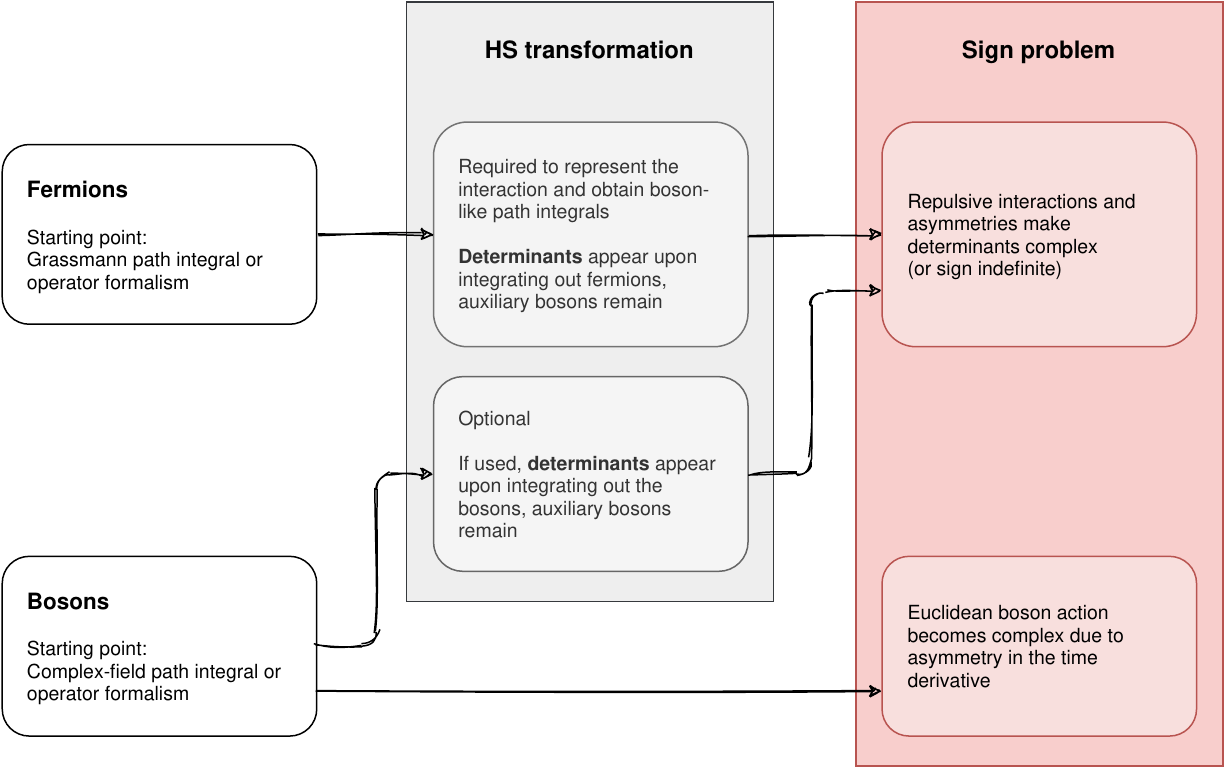}
  \caption{\label{fig:PathIntegralsFermionsBosons} Pathways of the appearance of the sign problem in path-integral approaches to non-relativistic many-body systems.}
\end{figure}

To make contact with the fermionic case discussed below (see also \figref{PathIntegralsFermionsBosons}), it is instructive to rewrite the interaction using a Hubbard-Stratonovich (HS) transformation, although it is not strictly needed in the bosonic case. Consider for instance the case of a complex field $\phi$. Schematically, one introduces
an auxiliary field $\sigma$ such that
\beq
{\rm e}^{-S_{\rm int}[\phi]} =  \int \mathcal D \sigma\;  {\rm e}^{-S_{\rm aux}[\phi,\sigma] - S_{0}[\sigma]},
\eeq
where
\beq
S_{\rm int}[\phi] = \int \d\tau \d^dx\ V[\phi] \, ,
\eeq
$S_{\rm aux}[\phi,\sigma]$ is a quadratic functional of both $\phi$ and $\sigma$, and $S_{0}[\sigma]$ is a pure-$\sigma$ term; both $S_{\rm aux}$ and $S_{0}$ depend on the specific
choice of HS transformation. Since the action is now quadratic in $\phi$, the corresponding path integral can be carried out, which results in a $\sigma$-dependent determinant, i.e. the full partition function can now be expressed as
\beq
\label{Eq:PartitionZ}
\mathcal Z = \int \mathcal D \sigma \;  {\rm e}^{-S_{\rm HS}[\sigma]},
\eeq
where
\beq
\label{Eq:SHSBosonDeterminant}
{\rm e}^{-S_{\rm HS}[\sigma]} = \frac{{\rm e}^{-S_{0}[\sigma]}}{\det M[\sigma]}.
\eeq

Usually it is possible to factor the determinant into a determinant for each particle species (flavors), such that if $N_f$ flavors are present then
\beq
\label{Eq:BosonDeterminant}
\det M[\sigma] = \det M_1[\sigma] \det M_2[\sigma]\, \cdots\, \det M_{N_f}[\sigma].
\eeq
Naturally, calculations with bosons are not carried out using the action of \equref{SHSBosonDeterminant}.
The formulation based on $S[\phi]$ of \equref{BosonAction} is considerably easier to work with as
there are no determinants involved. However, the appearance of the boson determinant in \equref{SHSBosonDeterminant}
shows that, if $\det M$ is real and an even number of species is present, then the sign problem can be avoided if $S_{0}[\sigma]$ is real.
Unfortunately, that situation is not relevant for bosons as it corresponds to attractive interactions, which make a many-boson system unstable.
On the other hand, the determinant-based representation of \equref{SHSBosonDeterminant} has a direct counterpart in the fermion case,
which we discuss next.

In theories with fermions, the action will {\it require} the much more complicated (non-linear, non-local) form based on
determinants because the fermionic analogue of \equref{BosonAction} is written in terms of anticommuting objects, i.e. Grassmann numbers,
which are not amenable to numerical computation.
We therefore assume that fermionic degrees of freedom (i.e. said Grassmann variables) have been integrated out. Taking such a step {\it requires} a HS
transformation of some kind to decouple the interaction, i.e. one introduces auxiliary fields to obtain a quadratic action in the fermion fields, which
are then integrated and result in a fermion determinant. Assuming such steps have already been taken, we have the schematic form,
\beq
\label{Eq:FermionAction}
{\rm e}^{-S[\phi]} = \det M[\phi] {\rm e}^{-S_g[\phi]},
\eeq
where $M$ encodes the dynamics of the fermions (quarks, electrons, atoms) in the external field $\phi$, and $S_g[\phi]$
is the ``pure HS" part of the action (often called ``pure gauge" part in QED and QCD); for the
latter, the form of $S_g$ will depend on the kind of HS transformation utilized (see \secref{AlternativeHS}).
Parameters like the fermion mass and chemical potential appear in $M$. In particular, in many cases it is possible to choose a HS transformation that decouples $N_f$ species
(i.e. flavors) of fermions such that, as in the bosonic case described above,
\beq
\label{Eq:FermionDeterminant}
\det M[\phi] = \det M_1[\phi] \det M_2[\phi]\, \cdots\,  \det M_{N_f}[\phi].
\eeq

The above path integral formulation, being a rewriting of $\mathcal Z$, inherits the usual mechanisms to access expectation values of operators,
namely differentiating $\mathcal Z$ with respect to a chosen parameter. For instance, the average particle number is given by
\beq
\langle \hat N \rangle = \frac{\partial \ln \mathcal Z}{\partial (\beta \mu)} = \frac{1}{\mathcal Z} \textrm{Tr}\left[ \hat N {\rm e}^{-\beta (\hat H - \mu \hat N)} \right],
\eeq
such that in the bosonic case of \equref{BosonAction},
\beq
\label{Eq:Nboson}
\langle \hat N \rangle = \int \mathcal D \phi\;  P[\phi] \left[-\frac{\partial S[\phi]}{\partial (\beta \mu)} \right ],
\eeq
while in the fermionic case of \equref{FermionAction},
\beq
\label{Eq:Nfermion}
\langle \hat N \rangle = \int \mathcal D \phi\;  P[\phi]\, \mathrm{Tr}\left [M^{-1} \frac{\partial M}{\partial (\beta \mu)} \right],
\eeq
where in either case
\beq
P[\phi] = \frac{{\rm e}^{-S[\phi]}}{\mathcal Z}.
\eeq
In evaluating Eqs.~(\ref{Eq:Nboson}) or~(\ref{Eq:Nfermion}), the natural course of action is to sample field configurations
$\phi$ according to the probability $P[\phi]$ and evaluate the quantities of
interest that appear between square brackets.
It is for that reason that the identification of $P[\phi]$ as a probability measure is a central aspect of conventional,
Metropolis-based approaches to the evaluation of expectation values in quantum systems with many degrees of freedom.
More specifically, in those cases where the sign (or phase) of $P[\phi]$ does not depend on $\phi$, one samples $\phi$ according to
$P[\phi]$ using the Metropolis algorithm (combined with a suitable field updating procedure, e.g. Wolff, worm, or
hybrid Monte Carlo algorithms) to obtain a set of $\mathcal N_\phi$ decorrelated samples $\{ \phi \}$, which in turn
are used to estimate expectation values as
\beq
\langle \mathcal O \rangle = \int \mathcal D \phi \; P[\phi] \mathcal O[\phi]
\simeq \frac{1}{\mathcal N_\phi}\sum_{\{ \phi \}} \mathcal O [\phi],
\eeq
for a given operator $\mathcal O$.

As mentioned above, by far for most systems of interest in physics face a sign problem, as the
sign (or more generally complex phase) of $P[\phi]$ varies with $\phi$. Then, $P[\phi]$ simply cannot be interpreted
as a probability and the Metropolis algorithm is not applicable.

For nonrelativistic fermionic systems, the sign problem typically happens at finite polarization (i.e. chemical potential asymmetry)
or when interactions contain a repulsive component. The problem therefore affects essentially all of condensed matter,
nuclear physics, and quantum chemistry. There are notable exceptions, such as a large class of
systems in one spatial dimension and the Hubbard model at half filling, for which the sign problem can be eliminated
completely. For relativistic fermions, such as quarks at finite chemical potential, the sign problem has obstructed the
investigation of the phase diagram of QCD.

The case of bosons is markedly different from that of fermions. Here, the nonrelativistic case presents a sign problem
even in the absence of interactions or chemical potentials: it is the asymmetry of the single time derivative, see \equref{BosonAction},
that creates the problem, as we will explain in further detail in \secref{formalism}.
This is to be contrasted with the relativistic case, which develops a sign problem when a chemical potential is turned on
(see however \secref{DualVariables}).

The remainder of this review is organized as follows. \secref{genformal} reviews a broad (but by no means complete)
set of approaches to the sign problem. \secref{formalism}
introduces the formal aspects of stochastic quantization and the complex Langevin method in more detail, including pedagogical examples as well as
a brief discussion of the challenges and shortcomings in the mathematical underpinnings. Secs.~\ref{sect:RQFT} and \ref{sect:NRQFT} review the recent and
emerging applications of CL in relativistic
and nonrelativistic physics, with an emphasis on the latter. Finally, \secref{outlook} concludes the review with a summary and outlook.


\section{\label{sect:genformal} Approaches to the sign problem: from reweighting to complex Langevin}

There have been multiple approaches suggested to solve or ameliorate the sign problem.  Some of these methods aim at solving the problem directly,
typically by rewriting the partition function in new and clever ways that remove the sign problem entirely. Other approaches involve rewriting the original problem
so it can be solved stochastically but with controlled sign fluctuations. Below we present a selection of those methods in a logical sequence that starts with
the simplest idea, namely reweighting, and concludes with complex plane methods.
Along the way, we present an elementary discussion of each method and comment on their advantages and shortcomings, which often result in valuable insights on
the nature of the sign problem.

\begin{figure}[t]
  \centering
  \includegraphics[width=0.95\columnwidth]{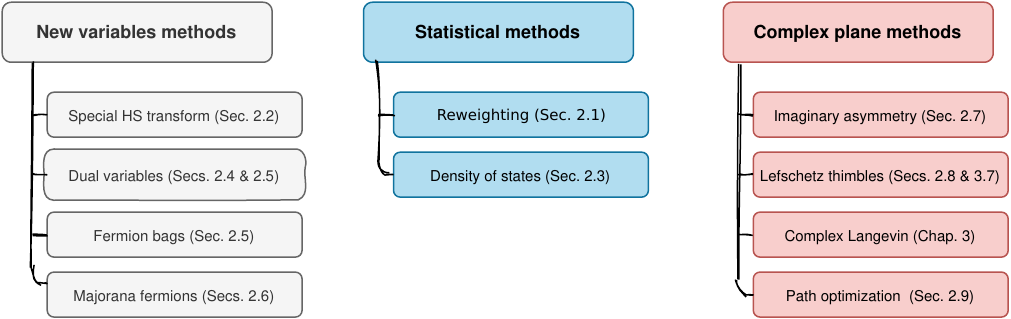}
  \caption{\label{fig:SPApproaches} A proposed map of the many approaches to the sign problem.}
\end{figure}

In \figref{SPApproaches} we propose a visual organization of the various approaches to the sign problem. Although we do not follow the proposed
taxonomy in this review in a linear fashion, we do find it helpful to organize the information in this manner. On the left column of that figure we list ``new variables'' methods as those that attempt to tackle the sign problem by switching from the conventional path-integral formulation to a new set of
variables. We begin by reviewing a classic work that looks for sign problem free HS transformations in \secref{AlternativeHS}.
That work shows that a choice of HS transformation may not solve the problem but may help in addressing it
(and certainly choosing the wrong one can be a recipe for trouble). Dual-variable and Majorana-fermion representations
succeed in completely solving the sign problem in many cases, as shown in Secs.~\ref{sect:DualVariables}
and~\ref{sect:Majorana}.

The middle column of \figref{SPApproaches} lists the set of what we call ``statistical" approaches, which attempt to tackle
the sign problem in a head-on manner. The simplest of those methods is by far the oldest and most commonly applied one across
all areas of physics: reweighting, which we describe in \secref{reweighting}. More recent statistical approaches,
commonly referred to as ``density-of-states'' methods, proceed by probing the shape of the probability distribution at fixed action
phase, and then integrating over that phase at the end; we describe those ideas in \secref{DensityOfStates}

Finally, the right column lists ``complex plane" methods, which also come in different flavors. \secref{ImaginaryAsymmetry} reviews the use of imaginary asymmetries in the parameters of a given theory (e.g. chemical potential, mass imbalance) to carry out calculations without a sign problem, necessarily followed by some kind of analytic continuation to return to the real physical values. Complex Langevin methods, the focus of this review, start in \secref{CL}, while in \secref{Thimbles} and \secref{pom} we mention ideas based on contour deformations (Lefschetz thimbles and the path optimization method, respectively). All of those methods rely on complexifying the integration variables; based on that idea, there exist constructive approaches (mentioned in \secref{CL}) that aim to define a real action in such a complex space.

\subsection{Reweighting~\label{sect:reweighting}}

The simplest (and likely oldest~\cite{PhysRevLett.61.2635}) idea to overcome the sign problem is that of {\it reweighting}, which amounts to sampling $\phi$
using the magnitude of $P[\phi]$ as a probability measure. In such an approach, one rewrites the expectation
value of $\mathcal O$ as
\begin{equation}
\label{Eq:ReweightedO}
\langle  \mathcal O \rangle = \frac{1}{\mathcal{Z}} \int \mathcal D \phi \; |P[\phi]| {\rm e}^{i\theta[\phi]} \mathcal O[\phi] =
\frac{\int \mathcal D \phi \; |P[\phi]| {\rm e}^{i\theta[\phi]} \mathcal O[\phi]}{\int \mathcal D \phi \; |P[\phi]|}
\frac{\int \mathcal D \phi \; |P[\phi]|}{\int \mathcal D \phi \; |P[\phi]| {\rm e}^{i\theta[\phi]}} =
\frac{\langle \langle \mathcal O [\phi] {\rm e}^{i\theta[\phi]}\rangle\rangle}{\langle\langle {\rm e}^{i\theta[\phi]}\rangle\rangle},
\end{equation}
where ${\rm e}^{i\theta[\phi]}$ is the phase of $P[\phi]$, and the double angle bracket denotes
an expectation value taken with respect to $|P[\phi]|$.

Reweighting thus provides a way forward for systems that have a sign problem: simply compute the numerator and
denominator of \equref{ReweightedO}) and then take their ratio. In practice, however,
both the numerator and the denominator of \equref{ReweightedO}) vanish {\it exponentially}
as the physical extent of the spacetime lattice is increased. The phase average is
\begin{equation}
\langle\langle {\rm e}^{i\theta[\phi]}\rangle\rangle =
\frac{\int \mathcal D \phi \; |P[\phi]| {\rm e}^{i\theta[\phi]}}{\int \mathcal D \phi \; |P[\phi]|} = \frac{\mathcal Z}{{\mathcal Z}_{pq}} = {\rm e}^{-\beta(\Omega - \Omega_{pq})},
\end{equation}
where ${\mathcal Z}_{pq}$ is the partition function of the ``phase-quenched" theory\footnote{In the phase-quenched theory, the probability distribution is replaced by its absolute value, such that it is a positive-definite measure.}, and $\Omega_{pq}$ and $\Omega$ are the corresponding grand thermodynamic potentials of the phase-quenched and original theory. Both $\mathcal Z$ and ${{\mathcal Z}_{pq}}$ are real quantities, but since ${\mathcal Z}_{pq}$ is a sum over nonnegative real numbers, while ${\mathcal Z}$ accounts for the phase, we necessarily have ${\mathcal Z} \leq {{\mathcal Z}_{pq}}$.
More importantly, thermodynamic potentials are extensive quantities in the spatial volume $V$ of the system, such that we may write the above in terms of intensive potentials $\omega$ and $\omega_{pq}$ as
\begin{equation}
\label{Eq:PhaseAvg}
\langle\langle {\rm e}^{i\theta[\phi]}\rangle\rangle = {\rm e}^{-\beta V (\omega - \omega_{pq})},
\end{equation}
which exposes the exponential nature of the sign problem in the thermodynamic ($V \to \infty$) and ground-state ($\beta \to \infty$) limits.
This can be seen more clearly by examining the statistical uncertainty $\Delta$, which in a Monte Carlo calculation with ${\mathcal N}_\mathrm{s}$ samples decreases as
${\mathcal N}_\mathrm{s}^{-1/2}$.In the case of re-weighting, the relative statistical uncertainty on the average phase is overpowered by the exponential
behavior coming from \equref{PhaseAvg}:
\begin{equation}
\frac{\Delta}{\langle\langle {\rm e}^{i\theta[\phi]}\rangle\rangle} \sim \frac{{\rm e}^{\beta V (\omega - \omega_{pq})}}{\sqrt{\mathcal N_s}}.
\end{equation}
This last equation shows the difficulty in approaching the sign problem with a simple technique such as re-weighting: an exponentially
large number of samples is needed in order to determine the average phase with any reasonable accuracy as the volume of spacetime
is increased. Viewed through the lens of this simple idea, the sign problem may be regarded as the reappearance of an exponential type
of computational wall, which affects non-stochastic methods (see Introduction) in the guise of memory requirements and statistical methods
in the form of a signal-to-noise problem.

\subsection{Alternative Hubbard-Stratonovich transformations\label{sect:AlternativeHS}}

The partition function $\mathcal Z$ of \equref{Z} is naturally a sum of positive quantities ${\rm e}^{-\beta(E - \mu N)}$.
The path-integral representation of $\mathcal Z$, while exact, introduces a large number of degrees of freedom to represent the
same quantity. It therefore seems natural to expect that such a formulation would {\it require} massive cancellations
(i.e. the sign problem) to yield correct physical answers. On the other hand, there are many ways to choose a HS representation,
which may in turn yield different kinds of cancellations (i.e. more or less dramatic, by some measure).
 Even in the absence of a sign problem, different kinds of continuous or discrete HS transformations display varying behavior
(see e.g.~\cite{PhysRevB.28.4059, PhysRevC.78.024001}).
It therefore makes sense to ask whether efficient representations exist, i.e. HS transformations which can substantially reduce the
difference $\omega - \omega_{pq}$ in \equref{PhaseAvg} or even eliminate it completely.

As an example, consider the fermionic Hubbard model given by
\begin{equation}
  \label{Eq:HubbardH}
  \hat H = -t \sum_{s=\uparrow,\downarrow}\sum_{\langle {\bf i} {\bf j}\rangle} \hat c^\dagger_{s,\bf i} \hat c^{}_{s,\bf j} + U \sum_{{\bf i}} (\hat n_{\uparrow, {\bf i}} - 1/2)(\hat n_{\downarrow, {\bf i}} - 1/2)\, ,
\end{equation}
where $t$ is the nearest-neighbor hopping, $U > 0$ is the repulsive coupling, $\hat c^{(\dagger)}_{s,\bf i}$ is the annihilation (creation) operator for particles of spin $s=\uparrow,\downarrow$ at location ${\bf i}$, and $\hat n_{s,{\bf i}}$ is the corresponding density operator.

The work of Ref.~\cite{PhysRevB.48.589} showed that a general HS transformation for the above model resulting in positive weights
(i.e. a real action $S[\phi]$) does not exist. While attractive interactions -- such as in the negative-$U$ Hubbard model -- feature no sign problem,
repulsive interactions and in general any finite polarization (i.e. non-zero chemical potential asymmetry) do yield sign oscillations.
More specifically, the problem arises because the determinant in \equref{FermionAction} becomes a product of two determinants which are
real or can be made real by choosing a proper HS transformation, but which will generally have different signs.
Reference~\cite{PhysRevB.48.589} showed that it is possible to isolate the origin of the signs in such a way that the determinants are real and
identical, i.e. one ends up with a square of a determinant, and the signs are not eliminated but can be predicted. This remarkable property is
illustrated below.

We begin by implementing a Trotter-Suzuki factorization of the Boltzmann weight with imaginary time step $\tau$, such as
\begin{equation}
{\rm e}^{-\tau \hat H} \simeq {\rm e}^{-\tau \hat T}  {\rm e}^{-\tau \hat V},
\end{equation}
where $\hat T$ contains the hopping terms and $\hat V$ the on-site interaction, as they appear in \equref{HubbardH}. It is to address the latter that an
HS transformation is used. The two most common HS representations, used in calculations of the repulsive Hubbard model, proceed by
writing (omitting the spatial indices)
\begin{equation}
\label{Eq:HSDiscrete}
{\rm e}^{-\tau U(\hat n_{\uparrow} - 1/2)(\hat n_{\downarrow} - 1/2)} = \frac{{\rm e}^{-\tau U/4}}{2} \sum_{\phi = \pm 1} {\rm e}^{-\lambda_d \phi (\hat n_{\uparrow} - \hat n_{\downarrow})},
\end{equation}
\begin{equation}
\label{Eq:HSContinuous}
{\rm e}^{-\tau U(\hat n_{\uparrow} - 1/2)(\hat n_{\downarrow} - 1/2)} = \frac{{\rm e}^{-\tau U/4}}{\sqrt{2\pi}} \int_{-\infty}^{\infty} \d\phi \; {\rm e}^{-\phi^2/2 -\lambda_c \phi (\hat n_{\uparrow} - \hat n_{\downarrow})},
\end{equation}
where $\phi$ is the auxiliary field, $\lambda_d$ is set by $\cosh(\lambda_d) = {\rm e}^{\tau U /2}$, and $\lambda_c = \sqrt{\tau U}$.
Both of these ``density-channel'' transformations successfully decouple the two spin species $\uparrow$ and $\downarrow$, and the
resulting determinants are real, but they are generally different from each other, such that
\begin{equation}
P[\phi] = \det M_\uparrow[\phi] \det M_\downarrow[\phi],
\end{equation}
will generally vary in sign with $\phi$. Here we omit the pure-$\phi$ part for the continuous case, which is real and positive anyway.
It should be pointed out that there are more general ways than the above factorized form that result in a sign problem free situation;
for an exploration of more general conditions based on time-reversal invariance, see Refs.~\cite{PhysRevB.71.155115, doi:10.1146/annurev-conmatphys-033117-054307}
and further discussion in \secref{Majorana}.

A more general transformation that aims to preserve the up-down symmetry of the Hubbard Hamiltonian, and therefore provide the square of a
real determinant in $P[\phi]$, can be written as
\begin{equation}
{\rm e}^{-\tau U(\hat n_{\uparrow}\hat n_{\downarrow} - \hat n_{\uparrow}/2 - \hat n_{\downarrow}/2)} =
\int_{-\infty}^{\infty} \d \phi \; p[\phi] \; {\rm e}^{\phi(\hat n_{\uparrow} + \hat n_{\downarrow})},
\end{equation}
where we want $p[\phi]$ to be real and positive and, evaluating both sides at the eigenvalues of the density operators (i.e. setting $\hat n_{s} \to 0,1$), we see that
\bea
\int_{-\infty}^{\infty} \d \phi \; p[\phi] &=& 1, \\
\int_{-\infty}^{\infty} \d \phi \; p[\phi] {\rm e}^\phi &=& {\rm e}^{\tau U /2}, \\
\int_{-\infty}^{\infty} \d \phi \; p[\phi] ({\rm e}^\phi)^2 &=& 1.
\eea
Unfortunately, the last two equations can only be satisfied simultaneously if ${\rm e}^{\tau U /2} \leq 1$, i.e. if $U \leq 0$, which is not the case we are interested in here as there is then no sign problem.

The above shows that, at least within the rather general form proposed, it is not possible to generate the square of a determinant {\it and} avoid the
sign problem at the same time for the repulsive Hubbard model.
On the other hand, if $p[\phi]$ is allowed to vary in sign, then there is no constraint on $U$ and we obtain the square of a real determinant. In that
case, the sign problem comes not from the fermion determinant but from $p[\phi]$, which means that it is completely predictable as soon as $\phi$
is known, without computing determinants. Such predictability of the sign or phase of the determinant has not been exploited in the literature beyond
the work of Ref.~\cite{PhysRevB.48.589}, but it could be of interest in the context of the density-of-states methods discussed in the next section.
In those methods, the knowledge of the precise form of the imaginary part of the action as a functional of the field is essential and has been used
with some success to characterize a class of relativistic field theories.

Generally speaking, the fact that there exists a family of HS transformations representing the same partition function,
especially if they are non-trivially related to each other, provides in effect a variety of calculations that can be used as checks against each other.
As explored in Ref.~\cite{PhysRevB.42.2282},
the density-channel decompositions mentioned above can be replaced by their ``pairing channel'' counterparts
(of which a new family exists, with discrete and continuous members, as above), which display different sign properties.
Other kinds of useful HS transformations have been discussed in Refs.~\cite{PhysRevB.56.15001,doi:10.1143/JPSJ.66.1872}

Crucially, the availability of different HS transformations with different sign behavior shows that the sign problem is not an intrinsic property of
a given Hamiltonian, but rather depends on the decoupling scheme. Therefore, the search for a link between the physics of a given system and the
sign problem should be taken with caution, as such a link may be entirely an artifact of the formulation of the problem. An interesting example
in that regard is the elimination of the sign problem by way of a fermionic reformulation of a bosonic problem in the case of
a frustrated Kondo model coupled to fermions~\cite{PhysRevLett.120.107201}, followed by a HS transformation on the resulting fermionic interaction
(see also~\cite{PhysRevB.63.155114, PhysRevLett.83.796}).

It is worth pointing out, however, that there is a link between the sign problem and phase transitions. Indeed, with the path integral formulation
at hand, one can reasonably argue that the sign problem can be expected to be severe close to a critical point. One way to
visualize that concept is in terms of the Lee-Yang zeros of the partition function $\mathcal Z$, written as a path integral (i.e. \equref{PartitionZ}.
When sampled over the relevant configurations of $\phi$, the integrand ${\rm e}^{-S[\phi]}$ must reflect the existence of an accumulation point of roots of $\mathcal Z$
when approaching the phase transition. By itself, that property would not pose a problem. However, the natural scale of the integrand is
the exponential of an extensive quantity; therefore, ${\rm e}^{-S[\phi]}$ must oscillate dramatically in order to generate the large collection of zeros
(in the thermodynamic limit), and the corresponding high sensitivity to the parameter values, around the transition point. For cases that do not have
a sign problem, the integrand must necessarily tend to zero when approaching a phase transition (again, in the thermodynamic limit), which is
often reflected in the appearance of zero modes in fermion matrices.

\subsection{Density-of-states methods~\label{sect:DensityOfStates}}

The density-of-states (DoS) approaches are a class of methods that attempt to tackle the sign problem in a head-on
manner, as opposed to rewriting the partition function in terms of new variables or straightforward reweighting (although it may be argued that DoS
methods are actually a kind of reweighting). The original idea of sampling the density of states as an alternative to Metropolis-based
methods is due to Wang and Landau~\cite{PhysRevLett.86.2050} and has been applied to a wide variety of systems including gauge
theories~\cite{PhysRevD.85.056010, PhysRevLett.109.111601}, but its generalization to systems with a sign problem was
explored later on in Refs.~\cite{Fodor:2007vv, PhysRevD.90.094502, 1742-6596-631-1-012063, PhysRevD.88.071502} (see also Refs.~\cite{PhysRevLett.61.2054, Schmidt:2006us}). The result of those explorations is now known
in the literature as the logarithmic linear regression (LLR) algorithm or the functional fit approach (FFA), both of which are very closely related
but differ on specific details. We will restrict ourselves here to those approaches (which have also been reviewed recently in
Ref.~\cite{Gattringer:2016kco}) but it is worth pointing out that DoS methods have also been applied to finite density QCD
in different forms which involve histograms of the phase of the fermion determinant (see e.g.~\cite{PhysRevD.77.014508, PhysRevD.78.074507, 10.1093/ptep/pts005}).

The idea common to all DoS approaches is that, in the presence of a sign problem where the action can be decomposed into real and imaginary parts
$S = S_R + iS_I$, the partition function can be written as
\begin{equation}
\label{Eq:ZZ}
\CZ =   \int \CD \phi \, {\rm e}^{-S[\phi]} = \int \d s \, \rho(s) \, {\rm e}^{-i s} = 2 \int \d s \, \rho(s)\, \cos(s),
\end{equation}
where we used the fact that the partition function is real and took the real part in the last step, and
\begin{equation}
\rho (s) = \int \CD \phi \, \delta (S_I[\phi] - s) {\rm e}^{-S_R[\phi]}.
\end{equation}
The determination of $\rho (s)$ is then carried out by combining two ingredients: first, propose a functional form that can account for its variation over vast
orders of magnitudes; second, carry out restricted calculations at constant or approximately constant imaginary action $S_I[\phi]$ in order to determine the
coefficients in the proposed functional form for $\rho (s)$. A key aspect of the method is that, using these elements, it can deliver exponential accuracy in the calculation of $\rho(s)$.

In the approach of Refs.~\cite{PhysRevD.90.094502, GATTRINGER2015545}, the parametrization of $\rho(s)$ is done in a piecewise-linear fashion:
\begin{equation}
\rho(s) = A_n {\rm e}^{-k_n s},
\end{equation}
for $s \in I_n$, $I_n = [s_n, s_{n+1}]$, where the partitioning and sizes of the intervals $I_n$, i.e. the set of numbers $\{ s_j \}$, can be chosen at will to
reflect the desired precision in describing $\rho(s)$. By requiring continuity of $\rho(s)$ and a normalization condition $\rho(0) = 1$, the constants $A_n$
can be determined as a function of $k_n$:
\begin{equation}
A_n = \exp\left(-\sum_{j=0}^{n-1} \Delta_j (k_j - k_n)\right),
\end{equation}
where $\Delta_j = s_{j+1} - s_{j}$ is the size of the $j$-th interval.
In order to determine the constants $k_n$, the FFA uses restricted expectation values defined by
\begin{equation}
\label{Eq:Xaverage}
\langle \langle X\rangle \rangle_n(\lambda) = \frac{\partial \ln \CZ_n(\lambda)}{\partial \lambda},
\end{equation}
where the restricted partition function is
\begin{equation}
\CZ_n(\lambda) = \int \CD \phi\ {\rm e}^{-S_R[\phi] + \lambda S_I[\phi]}\theta_n(S_I[\phi]),
\end{equation}
with $\theta_n(x) = 1$ for $x \in I_n$ and $0$ otherwise.

\begin{figure}[t]
  \centering
  \includegraphics[width=0.53 \columnwidth]{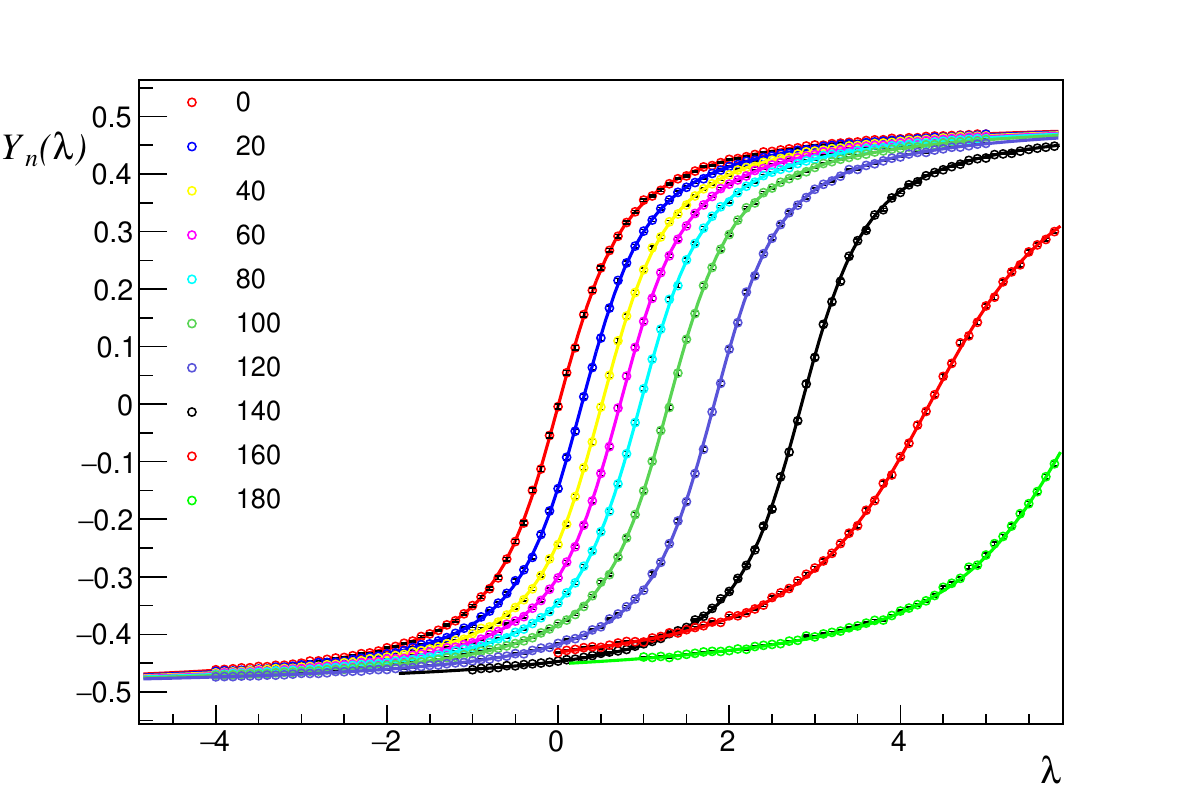}
  \includegraphics[width=0.46 \columnwidth]{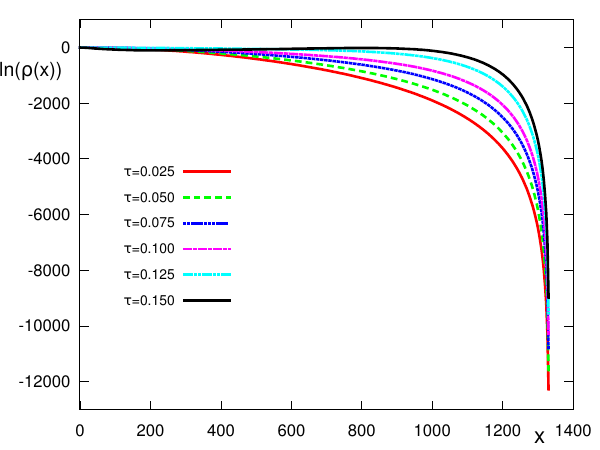}
  \caption{\label{fig:FFAPlots} (Left)  $Y_n(\lambda)$ as a function of $\lambda$ for several discretizations $n$.
  (Right) Density of states for several values of the nearest-neighbor coupling $\tau$. Both plots correspond to
  the SU(3) spin model of Ref.~\cite{GIULIANI2016627}.
  }
\end{figure}

With the above piecewise-linear parametrization of $\rho(s)$, the restricted partition function and expectation values can be
computed in closed form. It turns out that
\begin{equation}
Y_n(\lambda) \equiv \frac{\langle \langle X\rangle \rangle_n(\lambda) - D_{n-1}}{\Delta_n} - \frac{1}{2} = h((\lambda - k_n) \Delta_n),
\end{equation}
where $D_{n-1} = \sum_{j=0}^{n-1} \Delta_j$ and
\begin{equation}
h(x) = \frac{1}{1 - {\rm e}^{-x}} - \frac{1}{x} - \frac{1}{2}.
\end{equation}
These equations encode a crucial aspect of the method: once the intervals $I_n$ are chosen (such that the $\Delta_n$ are fixed constants),
each of the functions $Y_n(\lambda)$ is entirely determined by the single parameter $k_n$ and must follow the shape dictated by $h(x)$.
Thus, the function $Y_n(\lambda)$ is a kind of response function in which the source parameter $\lambda$ is coupled to the imaginary
part of the action $S_I$ to constrain the value of $k_n$ for each $n$. If the one-parameter fit to $h(x)$ is unsatisfactory, that signals
a poor choice of the discretization $\{ s_j \}$, such that a more refined mesh is likely needed.
An example of the typical shape of $Y_n(\lambda)$ is shown in the left panel of \figref{FFAPlots} for several values of $n$
for the SU(3) spin model of Ref.~\cite{GIULIANI2016627}.

Once the $k_n$ are known, one may reconstruct $\rho(s)$ and calculate the full partition function as a Fourier transformation via \equref{ZZ}).
This last step will be sensitive to the large oscillations due to the $\cos(s)$ factor. As an example, the density of states obtained in Ref.~\cite{GIULIANI2016627}
for the SU(3) spin model is shown in the right panel of \figref{FFAPlots}, for several values of the nearest-neighbor coupling $\tau$. The
variations of $\rho(s)$ over many orders of magnitude are evident from that figure.

While we have focused here on the FFA, the LLR method~\cite{PhysRevLett.109.111601} accomplishes exponential error suppression by
calculating the slopes $k_n$ of the distribution using a fixed-point iteration method. The latter is based on the work of Robbins and Monro~\cite{robbins1951},
which guarantees that the values obtained in subsequent iterations are Gaussian-distributed around the exact answer.
The above-mentioned exponential error suppression amounts to a constant relative error in the determination of the density of states over the full domain
of the phase, which is crucial in order to carry out the Fourier integral in \equref{ZZ}).

The FFA and the LLR methods have been used to analyze several models on the lattice at finite
density such as the $Z_3$ spin model~\cite{GATTRINGER2015545}, the SU(3) gauge theory with static color sources~\cite{GIULIANI2016627, GIULIANI2017166},
and two-color QCD with heavy quarks~\cite{PhysRevD.88.071502}. One of the most interesting advantages of DoS methods is that they are extremely
parallelizable on modern computers. These methods require a large set of calculations, e.g. several independent calculations as a function of $\lambda$
and $n$ to determine $k_n$, but each of those is an independent run and can therefore be done in a perfectly scalable fashion. As long as the number of $\lambda$
and $n$ points does not grow exponentially with the spacetime volume (there is not indication thus far in the literature that that is the case, due largely to the
smoothness and lack of sharp features in $\ln \rho(s)$), the computational cost will scale better than that of reweighting methods.

On the other hand, DoS approaches present a paradigm that is quite different from conventional MC methods when it comes to calculating different observables.
For observables $\mathcal O$ that do not depend explicitly (and only) on the action $S[\phi]$, it will not be enough to know $\rho(s)$. In such a case,
a source term $j \mathcal O$ would have to be included in the action and a family of densities $\rho(s,j)$ would need to be calculated at least for small $j$.
Once $\rho(s,j)$ is thus obtained, numerical differentiation of $\ln \mathcal Z[j]$ yields the desired $\langle O\rangle$ in the limit $j \to 0$. Such an approach may seem
slightly cumbersome or costly, but it amounts to multiple applications of the same idea which moreover retains the full parallelizability property mentioned above.

\subsection{Dual variables for bosons~\label{sect:DualVariables}}

Dualization is another approach involving rewriting the partition function $\mathcal Z$ so as to eliminate or ameliorate the sign problem.
The dual variables are a new set of variables, typically discrete, that may yield a representation of $\mathcal Z$ entirely in terms of positive quantities.
While the concept of duality is in itself an old one, the use of dual variables in quantum Monte Carlo calculations first appeared in the 1980's: for instance, Ref.~\cite{ROSSI1984105} showed that that the strong-coupling limit of QCD could be represented as a system of dimers.
Later on, Ref.~\cite{PhysRevB.61.10725} used dual variables to analyze the bosonic Hubbard model. The concept was later applied
to fermions as well, which we review in the next section. For pure bosonic theories at finite density, it was shown in Ref.~\cite{PhysRevD.75.065012} that dual
variables are not just an alternative representation: they successfully solve the sign problem for both relativistic as well as non-relativistic systems. Moreover,
the dual variables can be efficiently sampled using the worm algorithm~\cite{PhysRevLett.87.160601, PhysRevE.66.046701}. Since the early 2010's, a few groups have pursued the
study of several quantum field theories (from simple models to effective theories of QCD) at finite temperature and density (see e.g.~\cite{PhysRevD.81.125007, PhysRevLett.106.222001, GATTRINGER2011242, Fromm:2011qi, MERCADO20121920, MERCADO2012737, Gattringer:2012df, Gattringer:2012ap}).

Following the notation and steps of Ref.~\cite{Gattringer:2016kco}, we show how to introduce dual variables first in relativistic bosons and then in the non-relativistic case.
The lattice action for a relativistic complex-valuedued field $\phi_x^{}$ is
\begin{equation}
S[\phi] = \sum_{x} \left (
\eta |\phi_x^{}|^2
+
\lambda |\phi_x^{}|^4
-
\sum_{\nu = 1}^4
\left[
{\rm e}^{\mu \delta _{\nu,4}}
\phi_x^{*}\phi_{x+\hat \nu}^{}
+
{\rm e}^{-\mu \delta _{\nu,4}}
\phi_x^{*}\phi_{x-\hat \nu}^{}
\right]
\right),
\end{equation}
where $x$ is a spacetime lattice point and $\hat \nu$ denotes a unit vector in the $\nu$-th direction ($\nu = 4$ being the imaginary-time direction).
At finite $\mu$, the quantity in square brackets ceases to be real, which exposes the sign problem. Exponentiating the action, as one would normally
do to compute the partition function, we note that
\begin{equation}
{\rm e}^{-S} = \prod_x {\rm e}^{-\eta |\phi_x^{}|^2 -\lambda |\phi_x^{}|^4}
\prod_{x,\nu} \exp ({\rm e}^{\mu \delta _{\nu,4}} \phi_x^{*}\phi_{x+\hat \nu}^{}) \exp ({\rm e}^{-\mu \delta _{\nu,4}} \phi_x^{}\phi_{x+\hat \nu}^{*}).
\end{equation}
At each spacetime-Lorentz point $x,\nu$, the offending term becomes a factor that can be rewritten by expanding each of the exponentials
in a Taylor series as
\begin{equation}
\prod_{x,\nu} \exp ({\rm e}^{\mu \delta _{\nu,4}} \phi_x^{*}\phi_{x+\hat \nu}^{}) \exp ({\rm e}^{-\mu \delta _{\nu,4}} \phi_x^{}\phi_{x+\hat \nu}^{*})
= \sum_{\{n, \bar n \}} \mathcal N_{n, \bar n} \prod_x {\rm e}^{\mu (n_{x,4} - \bar n_{x,4})}
\phi_x^{*\sum_\nu (n_{x,\nu} + \bar n_{x-\hat \nu,\nu})}
\phi_x^{\sum_\nu (\bar n_{x,\nu} + n_{x-\hat \nu,\nu})},
\end{equation}
where $\mathcal N_{n, \bar n} = \prod_{x,\nu} {1}/({n_{x,\nu}! \bar n_{x,\nu}!})$ and the sum $\sum_{\{n, \bar n \}}$ denotes a sum over all
configurations of the Taylor indices $n_{x,\nu} \geq 0$ and $\bar n_{x,\nu} \geq 0$.
Using the above and the polar form $\phi_x = r_x {\rm e}^{i\theta_x}$, we obtain for the partition function
\begin{equation}
\mathcal Z = \sum_{\{n, \bar n \}} \mathcal N_{n, \bar n}
\prod_x {\rm e}^{\mu (n_{x,4} - \bar n_{x,4})} R[n,\bar n,x] T[n,\bar n,x],
\end{equation}
where
\begin{equation}
R[n,\bar n,x] = \int_0^{\infty} \d r_x r_x^{1 + \sum_\nu \left(n_{x,\nu} + n_{x-\hat \nu,\nu} + \bar n_{x,\nu} + \bar n_{x-\hat \nu,\nu}\right)}
{\rm e}^{-\eta r_x^2 -\lambda r_x^4},
\end{equation}
which is a non-negative, local function of the index configuration $\{n, \bar n\}$ (note also that the exponent of $r_x$ is strictly positive), and
\begin{equation}
T[n,\bar n,x] = \int_{-\pi}^{\pi} \frac{\d \theta_x}{2\pi} {\rm e}^{-i\theta_x \sum_\nu \left(n_{x,\nu} - \bar n_{x,\nu} - n_{x-\hat \nu,\nu}  + \bar n_{x-\hat \nu,\nu}\right)},
\end{equation}
which results in Kronecker delta functions for each $x$ imposing constraints on the configurations.
In principle, the job is done at this point: we have shown that there is a discrete-field representation of the partition function as a sum
over positive quantities. It is useful, however, to take a few more steps towards simplifying the calculation, specifically towards
implementing the constraints imposed by the function $T[n,\bar n,x] $. To that end, one parameterizes the sum
and difference of $n$ and $\bar n$ via two new `dual' variables $k, \ell$ defined via
\begin{equation}
n_{x,\nu} - \bar n_{x,\nu} \equiv k_{x,\nu} \ \ \ \ \ \ \ \textrm{and} \ \ \ \ \ \ \ \  n_{x,\nu} + \bar n_{x,\nu} \equiv |k_{x,\nu}| + 2\ell_{x,\nu},
\end{equation}
which take values over all integers and all non-negative integers, respectively. We finally obtain
\begin{equation}
\mathcal Z = \sum_{k,\ell} \mathcal N_{|k| + \ell, \ell} \prod_x W(s_x) {\rm e}^{\mu \sum_x k_{x,4}} \prod_x \delta(\nabla_v k_{x,\nu}),
\end{equation}
where
\begin{equation}
W(n) = \int_0^{\infty} \d r r^{n+1}{\rm e}^{-\eta r^2 -\lambda r^4},
\end{equation}
\begin{equation}
s_x = \sum_\nu \left[ |k_{x,\nu}| + |k_{x-\bar\nu,\nu}| + 2(\ell_{x,\nu} + \ell_{x-\nu,\nu}) \right], \ \ \ \ \ \ \textrm{and} \ \ \ \ \ \
\nabla_v k_{x,\nu} \equiv \sum_\nu \left[k_{x,\nu} - k_{x-\hat \nu,\nu} \right].
\end{equation}
The constraint $\nabla_v k_{x,\nu} = 0$ enforces that the field $k_{x,\nu}$ be solenoidal, such that the flux of $k_{x,\nu}$ is conserved.

Pure gauge theories can also be written in terms of dual variables and, as above, results in a new representation for the partition function in
terms of purely positive terms. This approach yielded the first real and positive dualization of {\it abelian} gauge theories with a so-called $\theta$
term~\cite{PhysRevD.92.114508}, which is a term in the action coupled to a topological charge (see Refs.~\cite{Gattringer:2018dlw, Sulejmanpasic:2019ytl} for updates on that work). Because that term is necessarily complex, it results in a sign
problem in conventional path integral representations. The current challenge for this approach is the extension to {\it non-abelian} gauge theories and the
inclusion of fermions (see however next section).

The case of non-relativistic bosons can also be addressed with dual variables with some modifications with respect to the
relativistic case. We show some of steps of that derivation here as a pedagogical example; they closely follow the relativistic case.
As we will show in more detail in \secref{NRQFT},
the problem appears in non-relativistic bosons not because of the chemical potential but because of the asymmetry in the time derivative: there are only
particles and no antiparticles. (The physical source of the problem is thus the same as in the relativistic case: the breaking of time-reversal invariance.)
Starting with the lattice action for the complex-valued field $\phi_x^{}$ in $3+1$ dimensions (although this example can be easily generalized
to $d+1$ dimensions), we have
\begin{equation}
S = \sum_{x} \left (
\lambda |\phi_x^{}|^4
+
\phi_x^{*}\phi_{x}^{}
-
{\rm e}^{\tau \mu}
\phi_x^{*}\phi_{x+\hat 4}^{}
-\frac{1}{2}
\sum_{k = 1}^3
\left [
\phi_x^{*}\phi_{x+\hat k}^{}
+
\phi_x^{*}\phi_{x-\hat k}^{}
-
2 \phi_x^{*}\phi_{x}^{}
\right]
\right),
\end{equation}
Then,
\begin{equation}
{\rm e}^{-S} = \prod_x {\rm e}^{-4 |\phi_x^{}|^2 -\lambda |\phi_x^{}|^4}
\prod_{x} \exp ({\rm e}^{\tau \mu} \phi_x^{*}\phi_{x+\hat 4}^{})
\prod_{x,k} \exp (\phi_x^{*}\phi_{x+\hat k}^{}/2) \exp (\phi_x^{*}\phi_{x-\hat k}^{}/2),
\end{equation}
where we now expand the exponentials of the derivative terms in a power series, such that
\bea
&& \!\!\!\!\!\!\!\!\!\!\!\!\!\!\!\!\!\!\!\!\!\!\!\!
\prod_{x} \exp ({\rm e}^{\tau \mu} \phi_x^{*}\phi_{x+\hat 4}^{})
\prod_{x,k} \exp (\phi_x^{*}\phi_{x+\hat k}^{}/2) \exp (\phi_x^{*}\phi_{x-\hat k}^{}/2) \nonumber \\
&=&
\sum_{\{n,m,\bar m\}}\;
\mathcal N_{n,m,\bar m}
{\rm e}^{\tau \mu n_x}
{\phi^*_x}^{n_x + \sum_k (m_{x,k} + \bar m_{x,k})}
{\phi^{}_x}^{n_{x-\hat 4} + \sum_k (m_{x-\hat k,k} + \bar m_{x+\hat k,k})},
\eea
where $\mathcal N_{n, m, \bar m} = \prod_{x} {1}/{n_{x}}! \prod_{x,k} {1}/(2^{m_{x,k}+ \bar m_{x,k}} {m_{x,k}! \bar m_{x,k}!})$,
and $n_x$ is a site variable, whereas $m_{x,k}$ and $\bar m_{x,k}$ are link variables in the spatial directions.
As in the previous example, one may now write the fields in terms of their polar representation $\phi_x = r_x {\rm e}^{i\theta_x}$ to obtain constraints for the
integer fields ${n,m,\bar m}$ and eventually arrive at a sum of positive definite terms for $\mathcal Z$. Explicitly,
\begin{equation}
\mathcal Z = \sum_{n,m,\bar m}\;
\mathcal N_{n,m,\bar m}
\prod_{x} {\rm e}^{\tau \mu n_x}
R[n,m,\bar m,x]
T[n,m,\bar m,x],
\end{equation}
where
\begin{equation}
R[n,m,\bar m,x] = \int_0^{\infty} \d r_x r_x^{1 + n_x + n_{x-\hat 4} + \sum_k \left( m_{x,k} + \bar m_{x,k} + m_{x-\hat k,k} + \bar m_{x+\hat k,k} \right)}
{\rm e}^{-4 r_x^2 -\lambda r_x^4},
\end{equation}
which is a non-negative, local function of the index configuration $\{n, m, \bar m\}$ (also, as before, the exponent of $r_x$ is strictly positive), and
\begin{equation}
T[n,m,\bar m,x] = \int_{-\pi}^{\pi} \frac{\d \theta_x}{2\pi} {\rm e}^{-i\theta_x \left[n_x - n_{x-\hat 4} - \sum_k \left( m_{x,k} + \bar m_{x,k} - m_{x-\hat k,k} - \bar m_{x+\hat k,k}\right)\right]},
\end{equation}
which results in Kronecker delta functions that impose constraints on the configurations.

It is then clear that the dual-variable formulation avoids the sign problem for non-relativistic bosons, as first noted in Ref.~\cite{PhysRevD.75.065012}.
It is worth noting, however, that other cases such as coupling to angular momentum, are not obviously solvable with this technique.

While this method completely solves the sign problem in the cases shown above (and some others, e.g. bosons with non-abelian spin-orbit coupling), the calculation
of specific observables acquires a new degree of complexity due to the dramatic change of variables. This is merely an algebraic
inconvenience but, in practice, the change from the original fields $\phi$ to the discrete fields $n,m,\bar{m}$ implies that any operator expression
in the $\phi$ language needs to be re-derived (e.g. by inserting sources in the original action or using the parameters of the theory).

In this section we have focused on exact alternative representations of scalar bosons. However, such representations have also been found for gauge
fields~\cite{Vairinhos:2014uxa, Vairinhos:2015ewa} which are valid also away from the strong coupling limit.

\subsection{Dual variables for fermions and fermion bags \label{sect:fermion_bags}}

One of the first uses of dual variables for fermions in Monte Carlo calculations was unrelated to the sign problem: it was found in
Ref.~\cite{ROSSI1984105} that QCD in the strong-coupling limit could be represented as a system of dimers, which
inspired multiple studies~\cite{WOLFF2009491, WOLFF2009549, WOLFF2010254, WOLFF2010520, PhysRevLett.104.112005,Unger:2011it}.
However, dual variables by themselves do not necessarily avoid the fermion sign problem. As first described in Ref.~\cite{PhysRevLett.83.3116},
in what is now known as the `meron cluster' approach, one must sum analytically over configurations in a given cluster (where the type of configuration
cluster must be cleverly identified) and then stochastically over clusters. When the clusters are properly chosen, they contain configurations that may vary
in sign but such that the overall contribution of a cluster is of constant sign across clusters.

The fermion bag approach of Refs.~\cite{PhysRevD.82.025007, Chandrasekharan:2013rpa} (extended to continuous time in Ref.~\cite{PhysRevD.96.114502}; see also
Refs.~\cite{Chandrasekharan2011, PhysRevLett.108.140404, PhysRevD.86.021701, PhysRevB.89.111101}) extends the meron cluster approach to a larger class of theories. Rather than following those derivations (based on Grassmann numbers), here we connect with them from a different perspective.
We begin by re-writing the partition function as
\begin{equation}
\mathcal Z = \textrm{Tr}\left[ {\rm e}^{-\beta (\hat H - \mu \hat N)} \right] = \textrm{Tr}\left[ \left( {\rm e}^{-\tau (\hat H - \mu \hat N)} \right)^{N_\tau} \right],
\end{equation}
where $\beta = \tau N_\tau$. Using a Suzuki-Trotter decomposition, we may approximate
\begin{equation}
\label{Eq:TransfMatrix}
{\rm e}^{-\tau (\hat H - \mu \hat N)} \simeq {\rm e}^{-\tau (\hat T - \mu \hat N)} {\rm e}^{-\tau \hat V},
\end{equation}
where $\hat T$ is the kinetic energy and $\hat V$ the interaction. As an example, we specialize to the case where $\hat V = -U \sum_x \hat n_\uparrow(x) \hat n_\downarrow(x)$, and then we have
\begin{equation}
{\rm e}^{-\tau \hat V} = \prod_x {\rm e}^{\tau U \hat n_\uparrow(x) \hat n_\downarrow(x)} = \prod_x \left( 1 + B\hat n_\uparrow(x) \hat n_\downarrow(x)\right)
=\sum_{m_{x} = 0,1} B^{\sum_{x} m_{x}} \prod_{x} \left ( \hat n_\uparrow(x) \hat n_\downarrow(x)\right)^{m_{x}}
,
\end{equation}
where  $B = {\rm e}^{\tau U} -1$. We may use the above at each point in imaginary time by inserting this
expression into \equref{TransfMatrix}) to obtain
\begin{equation}
{\rm e}^{-\beta (\hat H - \mu \hat N)}
= \prod_{t=1}^{N_\tau} \hat {\mathcal T}_t \prod_{x}\left( 1 + B\hat n_\uparrow(x) \hat n_\downarrow(x)\right)_t
=  \sum_{m_{x,t} = 0,1} B^{\sum_{x,t} m_{x,t}} \prod_{t=1}^{N_\tau} \left [ \hat {\mathcal T}_t \prod_{x} \left ( \hat n_\uparrow(x) \hat n_\downarrow(x)\right)^{m_{x,t}}_t \right ],
\end{equation}
where $\hat {\mathcal T}_t \equiv {\rm e}^{-\tau (\hat T - \mu \hat N)}$ and we note that the product over the time slices actually factorizes across flavors as
\begin{equation}
\prod_{t=1}^{N_\tau}
\left [ \hat {\mathcal T}^\uparrow_t \prod_{x} \hat n_\uparrow(x)^{m_{x,t}}_t \right ]
\prod_{t=1}^{N_\tau}
\left [ \hat {\mathcal T}^\downarrow_t \prod_{x} \hat n_\downarrow(x)^{m_{x,t}}_t \right ].
\end{equation}
Below we will show the following trace-determinant identity
\begin{equation}
\label{Eq:TraceDet}
\textrm{Tr} \prod_{t=1}^{N_\tau}
\left [ \hat {\mathcal T}^\uparrow_t \prod_{x} \hat n_\uparrow(x)^{m_{x,t}}_t \right ] = \det W_\uparrow[\{ m\}],
\end{equation}
where $W_s[\{ m\}]$ is the free fermion matrix for spin $s$ in which the rows and columns ${x_0,t_0}$ for which $m_{x_0,t_0} = 1$ are dropped
(see below for details on the form of $W_s[\{ m\}]$).
Using the above in the definition of $\mathcal Z$ yields
\begin{equation}
\mathcal Z = \sum_{m_{x,t} = 0,1} B^{\sum_{x,t} m_{x,t}} \det W_\uparrow[\{ m\}]\det W_\downarrow[\{ m\}],
\end{equation}
where we finally have completely re-written the full partition function as a sum over configurations of the monomer field $m_{x,t}$.

\begin{figure}[t]
\floatbox[{\capbeside\thisfloatsetup{capbesideposition={right,top},capbesidewidth=0.4\textwidth}}]{figure}[\FBwidth]
{\caption{\label{fig:FermionBags} Fermion bag interpretation of the dual-variable approach to many-fermion systems. The blue dots represent the points on the
  spacetime lattice for which the discrete field acquires values $m_{x,t} = 1$. The regions enclosed by continuous black lines represent the fermion bags, within
  which $m_{x,t} = 0$ and where fermions move freely. This picture is not general but meant only as an illustration. The true shape of the allowed bag configurations will depend on the specific form of the Hamiltonian; in particular, the form of the kinetic energy operator is crucial in determining whether such a bag decomposition is possible.}}
{\includegraphics[width=0.55\textwidth]{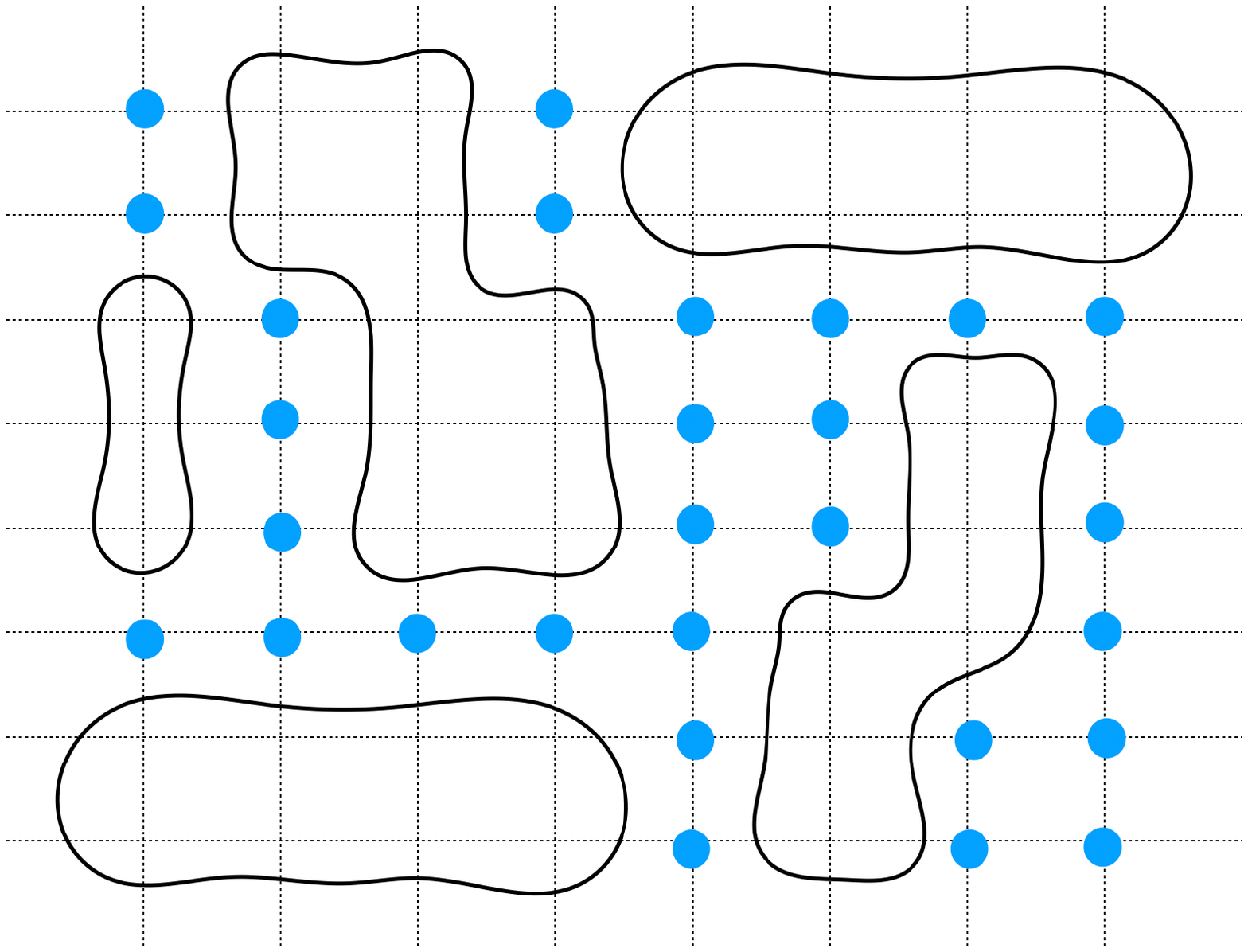}}
\end{figure}

For unpolarized non-relativistic systems, $W_s[\{ m\}]$ is real and takes on the same value for $s = \uparrow,\downarrow$, such that there is no sign problem,
as long as $B \geq 0$, i.e. for attractive interactions. Thus far, the same conditions apply for the auxiliary field formulation of the problem: repulsive interactions
($B < 0$) or polarization ($W_\uparrow[\{ m\}] \neq W_\downarrow[\{ m\}]$) would lead to a sign problem. This formulation, however, lends itself to an interpretation of the sum
in terms of clusters known as fermion bags, which are disjoint regions ${\mathcal B}_i$ of the discrete field $m_{x,t}$ within which $m_{x,t} = 0$ (see
\figref{FermionBags}). As the corresponding interaction vertices (i.e. insertions of $U$) are thus absent in $W_s[\{ m\}]$, fermions are free to move about inside the
bag. Using that property, Ref.~\cite{PhysRevD.82.025007} argued that, while the contributions to a given fermion bag configuration may vary in sign, the overall contribution
of each bag to the full partition function is actually positive. Thus, if one is able to add up the terms within each bag, then it is possible to use Metropolis-based importance
sampling to sum over all possible fermion bag configurations.

The right-hand side of \equref{TraceDet}) can also be calculated using Wick's theorem, if interpreted as the expectation value of a time-dependent operator
in a noninteracting system. In that weak-coupling interpretation, it is possible to show that
\begin{equation}
\label{Eq:TraceDet2}
\textrm{Tr} \prod_{t=1}^{N_\tau}
\left [ \hat {\mathcal T}^\uparrow_t \prod_{x} \hat n_\uparrow(x)^{m_{x,t}}_t \right ] = \det M_\uparrow  \det G_\uparrow[\{ m\}],
\end{equation}
where $M_\uparrow$ is the noninteracting spacetime fermion matrix and $G_\uparrow[\{ m\}]$ is a propagator matrix whose size depends on the
monomer configuration $m_{x,t}$ and which contains noninteracting propagators connecting the monomer sites where $m_{x,t} = 1$.
As explained in Ref.~\cite{Chandrasekharan:2013rpa}, the equality between Eqs.~(\ref{Eq:TraceDet}) and (\ref{Eq:TraceDet2}) represents a
duality relation between strong coupling and weak coupling; in the former case, a large number of monomers appear and \equref{TraceDet}) is easier
to calculate than \equref{TraceDet2}), which becomes easier at weak coupling.

In combination with the hopping expansion (which amounts to expanding the exponential of the
kinetic energy rather than the potential energy), one arrives at other useful sign problem-free representations of fermionic partition functions. One such example is well-known and is the
case of non-relativistic fermions in 1D with two-body interactions~\cite{PhysRevB.26.5033}. Another more recently discovered case is that of non-relativistic fermions in 1D
with four-body interactions, shown and used in Refs.~\cite{PhysRevA.85.063624, PhysRevLett.109.250403, PhysRevA.87.063617}, where also a fermion-bag type idea is
used to sum over configuration clusters. Interestingly, baryons at strong coupling can also be described by bags where three quarks propagate coherently as a single free
fermion (i.e. a baryon) inside bags, while the complementary domain displays quark- and di-quark-type excitations~\cite{PhysRevD.97.074506}.
Other relevant examples can be found in Refs.~\cite{WOLFF2009549, PhysRevD.80.071503, PhysRevD.97.054501}.

\newpage
\paragraph{Proof of trace-determinant identity}
For completeness, we outline the proof of the trace-determinant identity \equref{TraceDet}) used above.
We have not seen this way to approach the proof anywhere else, and since we find it particularly clear, we include it here.
First we quote the auxiliary identity
\bea
\label{Eq:AuxIDTraceDet}
\textrm{Tr}
\left[
\prod_{k=1}^{N}
{\rm e}^{\hat B_k}{\rm e}^{\hat A_k}
\right]
= \det
\left(
\begin{array}{c c c c c c}
1 & 0 &0 &0 & \cdots & {\rm e}^{B_N} \\
-{\rm e}^{A_1} & 1 &0 & 0 & \cdots & 0 \\
0 & -{\rm e}^{B_1} & 1 & \ddots & 0 & 0 \\
\vdots & \dots & \ddots & \ddots & \vdots \\
0 & 0 & -{\rm e}^{A_{N-1}} & 1  & 0\\
0 & 0 & 0 & -{\rm e}^{B_{N-1}} & 1  & 0\\
0 & 0 & 0 & \cdots & -{\rm e}^{A_{N}} & 1
\end{array}
\right ),
\eea
which is a reformulation of a well-known identity in the operator formulation of quantum Monte Carlo
(see e.g.~\cite{Drut:2012md}).
Here, the left-hand side trace is over Fock space, the entries of the Fermi matrix on the right-hand side are themselves matrices (i.e. the above is shown in
block form), and $\hat A_k = \sum_{i,j} [A_k]_{ij} \hat c^\dagger_i \hat c_j$ and $\hat B_k = \sum_{i,j} [B_k]_{ij} \hat c^\dagger_i \hat c_j$ are
(generally non-commuting) one-body operators.
For our purposes, $k$ represents a particular time slice, $\hat B_k = j_{x,k} \hat n_{x,k}$, such that $[{\rm e}^{B_k}]_{x,x'} = \delta_{x,x'} {\rm e}^{j_{x,k}}$, and
$\hat A_k$ encodes the kinetic energy, i.e. it is actually a $k$-independent operator. Our focus is on the $B_k$ factors.

The crucial step in proving \equref{TraceDet}) is in differentiating both sides of \equref{AuxIDTraceDet}) with respect to $j_{x,k}$ in as
many points $\{x,k\}$ as needed (namely the points where $m_{x,k} = 1$; recall each point appears only once in the matrix) to match the desired insertions of
$\hat n(x)$ at the desired time slices $k$.
To carry out the differentiations on the right-hand side, we first use the Laplace cofactor expansion of the determinant and
set the corresponding $j_{x,k}$ to zero once the derivative is taken. Once all the desired derivatives are applied, what remains is the corresponding cofactor
determinant. The latter is the determinant of the matrix on the right-hand side of \equref{AuxIDTraceDet}) where the rows and columns
of a given ${\rm e}^{B_k}$ containing the differentiated points $\{x,k\}$ are simply dropped, and the sources $j$ in any remaining terms are set to zero; that
prescription defines the square matrix $W_s[\{m\}]$, whose number of rows (and columns) is reduced from the original matrix by the number of non-zero monomers.
Note that there will, typically, be more than one spatial point $x$ affected within a given temporal block ${\rm e}^{B_k}$; similarly,
any temporal block not affected by the derivatives will turn into an identity matrix once the sources are set to zero.
Note also that any overall signs are unimportant because they cancel against the corresponding expression for the other fermion species.


\subsection{Majorana fermions~\label{sect:Majorana}}

In Ref.~\cite{PhysRevB.91.241117} a fermion representation was introduced that does not display a sign problem for a broad class of
systems. Those developments were precipitated in part by the work of Huffman and Chadrasekharan of Ref.~\cite{PhysRevB.89.111101}
and were further investigated by several authors in different ways (see in particular Refs.~\cite{PhysRevLett.115.250601, PhysRevLett.116.250601}).
The main result of that line of research, which we will explain in this section, is that there is a new class of systems that do not have a sign
problem, and that that class goes beyond the well-known time-reversal-symmetric situation.

To understand the main principle behind this new class of systems, note that for a typical non-relativistic, single-species Fermi system, the discretization of
the time direction into $N_\tau$ slices followed by a Hubbard-Stratonovich transformation yield a partition function of the form
\begin{equation}
\mathcal Z = \int \mathcal D \sigma \det M[\sigma],
\end{equation}
where
\bea
M[\sigma]
=
\left(
\begin{array}{c c c c c}
1 & 0 & \cdots & 0  & {\rm e}^{A_{N_\tau}} \\
-{\rm e}^{A_1} & 1 & 0 &\dots & 0 \\
0 & -{\rm e}^{A_2} & 1 & \vdots &  0 \\
\vdots & \dots & \vdots & \ddots & \vdots  \\
0 & 0 & 0 & -{\rm e}^{A_{N_\tau-1}} & 1
\end{array}
\right ),
\eea
which satisfies $\det M[\sigma] = \det\left( 1 + U[\sigma]\right)$, where
\begin{equation}
U[\sigma] = {\rm e}^{A_1}{\rm e}^{A_2}\dots {\rm e}^{A_{N_\tau}},
\end{equation}
and the ${A_k}$ factors encode the kinetic or potential energy contributions (the latter in the form given by the choice
of Hubbard-Stratonovich transformation). It is then clear that the question of which systems display a sign problem amounts to asking
under what conditions expressions of the form $\det\left( 1 + U[\sigma]\right)$ have a constant sign.
That question was ``crowd sourced'' by L. Wang on the MathOverflow website and then analyzed in detail by multiple authors, leading to the work
of Ref.~\cite{PhysRevLett.115.250601} (see also Ref.~\cite{doi:10.1146/annurev-conmatphys-033117-054307}), whose discussion we parallel next.

Under specific conditions on the $A_k$ matrices, reviewed below, the product $U[\sigma]$ lies in the split-orthogonal group $O(n,n)$,
which is defined as the set of real matrices $R$ such that $R^T \eta R = \eta$, where the metric is
\begin{equation}
\eta = \text{diag}(1,1,\dots,1, -1, -1,\dots,-1),
\end{equation}
where $1$'s and $-1$'s appear $n$ times each. It is easy to see that $|\det R | = 1$ and it is also possible to show that, writing
$R$ in terms of $n\times n$ blocks
\begin{equation}
R =
\left(
\begin{array}{cc}
R_{11} & R_{12}\\
R_{21} & R_{22}\\
\end{array}
\right),
\end{equation}
that $|\det R_{11} | \geq 1$ and $|\det R_{22} | \geq 1$, which defines four different sectors in $O(n,n)$ labeled by the signs of these
two determinants, typically denoted $O^{\pm\pm}(n,n)$. Of those, only $O^{++}(n,n)$ contains the identity and forms a subgroup.
Crucially, it can be shown that, if $U[\sigma] \in O^{++}(n,n)$ then $\det\left( 1 + U[\sigma]\right) \geq 0$, and if
$U[\sigma] \in O^{--}(n,n)$ then $\det\left( 1 + U[\sigma]\right) \leq 0$. If $U[\sigma]$ is in not in those sectors, then the determinant vanishes.

Finally, we see that, if the generating matrices $A_k$ are in the algebra of $O(n,n)$, i.e. if $\eta A_k \eta = -A_k^T$, then their exponentials will be group elements
and the product of such exponentials will be in $O(n,n)$, in particular it will be in $O^{++}(n,n)$. Furthermore, parametrizing one such algebra generator as
\begin{equation}
\label{Eq:ADecomp}
A =
\left(
\begin{array}{cc}
C & B\\
X & D\\
\end{array}
\right),
\end{equation}
the condition of it being in the algebra of $O(n,n)$ implies $X = B^T$, $C^T = -C$, and $D^T = -D$, such that
the diagonal of $A$ must be entirely zero. Below we will consider bipartite systems and $C$ and $D$ will contain
matrix elements corresponding to same-lattice indices, whereas $B$ and $B^T$ will connect different sublattices.


As an example of a model whose auxiliary-field representation satisfies the above constraints, Ref.~\cite{PhysRevLett.115.250601} considers
the spinless $t$-$V$ model on a bipartite lattice:
\begin{equation}
  \label{Eq:HubbardHspinless}
  \hat H = \sum_{{\bf i}, {\bf j}} \hat c^\dagger_{\bf i} K_{{\bf i}{\bf j}} \hat c^{}_{\bf j}
  + \sum_{\langle {\bf i},{\bf j}\rangle} \left[V \left(\hat n_{{\bf i}}\hat n_{{\bf j}} - \frac{\hat n_{{\bf i}} + \hat n_{{\bf j}}}{2} \right) - \Gamma \right],
\end{equation}
where $\hat c^\dagger_{\bf i}$ and $\hat c^{}_{\bf i}$ are the creation and annihilation operators and $\hat n_{{\bf i}}$ is the number density operator
at site $\bf i$. The angle brackets denote a sum over nearest neighbors, which are assumed to belong to different sublattices. The two key properties of this
model are: a) the kinetic matrix $K$ only connects terms across the two sublattices and is zero on the diagonal; b) the interaction term can be decoupled using
a Hubbard-Stratonovich transformation which again only connect different sublattices (and to that end the constant $\Gamma$ plays a crucial role).
Specifically, Ref.~\cite{PhysRevLett.115.250601} proposed the following auxiliary-field representation:
\begin{equation}
\label{Eq:HStransformNN}
\left[V \left(\hat n_{{\bf i}}\hat n_{{\bf j}} - \frac{\hat n_{{\bf i}} + \hat n_{{\bf j}}}{2} \right) - \Gamma \right] =
- \frac{\Gamma}{2} \sum_{\sigma = \pm 1} \exp\left[\sigma \lambda  (\hat c^\dagger_{\bf i} \hat c^{}_{\bf j}  + \hat c^\dagger_{\bf j} \hat c^{}_{\bf i} )\right],
\end{equation}
where $\lambda = \text{acosh}(1 + V/(2 \Gamma))$, which is a real number for $V, \Gamma \geq 0$, i.e. repulsive interactions. This is easily checked by using
the fact that, when $\bf i \neq \bf j$, all positive even powers of the combination $(\hat c^\dagger_{\bf i} \hat c^{}_{\bf j}  + \hat c^\dagger_{\bf j} \hat c^{}_{\bf i} )$
take the same operator value, namely
\begin{equation}
(\hat c^\dagger_{\bf i} \hat c^{}_{\bf j}  + \hat c^\dagger_{\bf j} \hat c^{}_{\bf i} )^2 = \hat n_{{\bf i}} + \hat n_{{\bf j}} -2\hat n_{{\bf i}}\hat n_{{\bf j}},
\end{equation}
and all odd powers of $(\hat c^\dagger_{\bf i} \hat c^{}_{\bf j}  + \hat c^\dagger_{\bf j} \hat c^{}_{\bf i} )$ also take on one operator value,
which then vanishes upon summing over $\sigma = \pm 1$.

As usual, there are two different kinds of matrices $A_k$: one for the kinetic energy operator $A^{(K)}$, and one for the potential energy $A^{(V)}$ (which includes the Hubbard-Stratonovich field). Since $K$ is a real symmetric matrix and only connects different sublattices, $A^{(K)}$ features
$C=D=0$ and $B=B^T$ [c.f. Eq.~(\ref{Eq:ADecomp})]. On the other hand, the potential energy factor resulting from Eq.~(\ref{Eq:HStransformNN}) is also real and symmetric and is designed to connect different sublattices only, such that also in this case $C=D=0$ and $B=B^T$. Thus, with the above choice
one is within the purview of the theorem of Ref.~\cite{PhysRevLett.115.250601} and there is no sign problem for the spinless $t$-$V$ model on a bipartite lattice.
As anticipated in a previous section, the choice of Hubbard-Stratonovich transformation (of which there exist an infinite number for any given system) can determine
the appearance or not of a sign problem, and that is the case here.


It is possible and useful to recast the above discussion in terms of Majorana variables. Reference~\cite{PhysRevB.91.241117} showed that, by writing the fermion operators in the original Hamiltonian as
\begin{equation}
\hat c_{\bf j}^{} = \frac{1}{2}(\hat \gamma_{1,{\bf j}} + i \hat \gamma_{2,{\bf j}} ), \\
\end{equation}
where $\hat \gamma_{i,{\bf j}}$ are Majorana fermions, it is possible to avoid the sign problem in certain classes of spinless fermion models on bipartite lattices.
%
Moreover, using this type representation, it is possible to generalize the conclusions
obtained using the split-orthogonal group regarding the types of systems that do not display a sign problem. That generalization
was carried out in Ref.~\cite{PhysRevLett.116.250601}. There, it was shown that a system displays no sign problem if it admits a
Majorana decomposition in which the usual kinetic and potential energy factors (after the Hubbard-Stratonovich transformation) take the bilinear form
\begin{equation}
\hat H_{bl} = {\hat \gamma}^T V \hat \gamma,
\end{equation}
where the vector $\hat \gamma$ contains the operators $\hat \gamma_{i,{\bf j}}$ above (in some order), and
\begin{equation}
\label{Eq:VDecomp}
V =
\left(
\begin{array}{cc}
C & iB\\
-iB^T & C^*\\
\end{array}
\right),
\end{equation}
where $C = -C^T$ is complex antisymmetric and $B$ is Hermitian positive (or negative) semidefinite.
Based on the above result, Ref.~\cite{PhysRevLett.116.250601} showed that not only models like the above $t-V$ case can be made sign problem free,
but also cases in which coupling to a pairing channel is present can be shown to have no sign problem using Majorana fermions. A precursor to that result appeared early on in Ref.~\cite{PhysRevLett.92.257002}.

In a second theorem, Ref.~\cite{PhysRevLett.116.250601} also showed that there is another class of systems which, though partially overlapping with the
above, represents a new class of sign problem-free systems as a result of Majorana-Kramers positivity. For the latter to hold, operators $S$ and $P$
must exist such that $V$ satisfies
\bea
S^T V S &=& V^* \\
P V P^{-1} &=& V
\eea
where $S$ is a real antisymmetric matrix satisfying $S^2 = -I$ and $S^T = -S$, $P$ is a symmetric or antisymmetric Hermitian matrix satisfying $P^2 = I$ and $PS = -SP$. The first of the above equations ensures that $V$ is time-reversal symmetric, which by itself is not a sufficient condition
to avoid the sign problem. The second equation enforces a Kramers degeneracy that ensures that there is no sign problem.

A characterization of the classes of sign problems that can be addressed with this technique can be found in
Refs.~\cite{PhysRevLett.116.250601, PhysRevLett.117.267002} (see also~\cite{doi:10.1146/annurev-conmatphys-033117-054307} for a recent review). Notably, exponentiating the generator $A$
of Eq.~(\ref{Eq:VDecomp}) yields elements of $O(n,n)$, which is a real group, while the exponentials of $V$ yield elements of $O(m, \mathbb{C})$. Reference~\cite{Wei:2017wns} presented a more general approach to such a group-theoretic characterization (so far the most general, to the best of our
knowledge), based on Lie semigroups.
Various applications were explored in Refs.~\cite{Li_2015, PhysRevB.92.235129, s41467-017-00167-6, PhysRevB.96.155112, PhysRevB.96.035129}.

Rather than a new method or an algorithm, one may regard the Majorana representation as a way to discover systems that do
not have a sign problem (and for which it is not otherwise obvious that this is the case in conventional fermion formulations). Once that
property is established, conventional algorithms can be used to carry out the calculation.

\begin{figure}[t]
  \centering
  \includegraphics[width=0.49 \columnwidth]{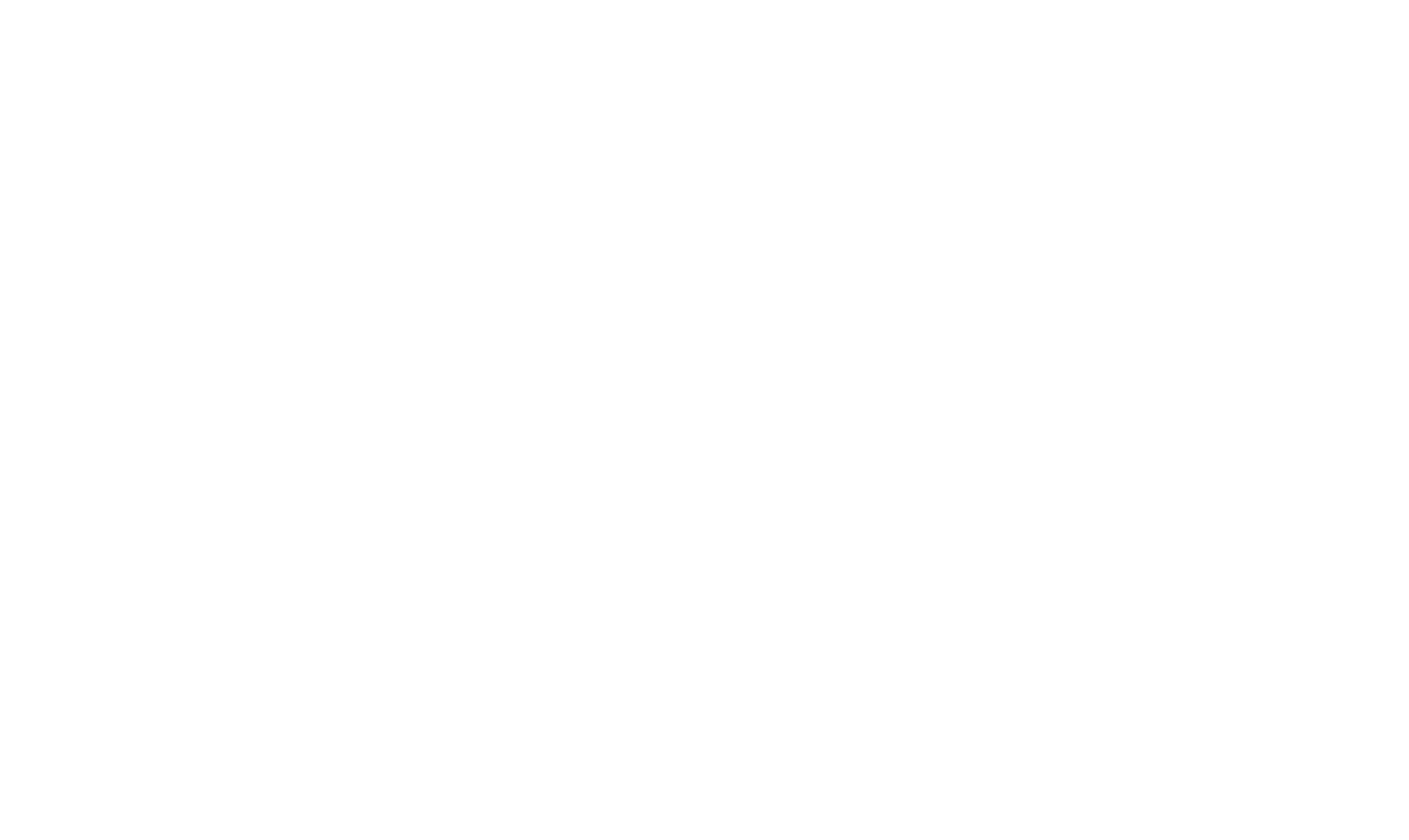}
  \includegraphics[width=0.49 \columnwidth]{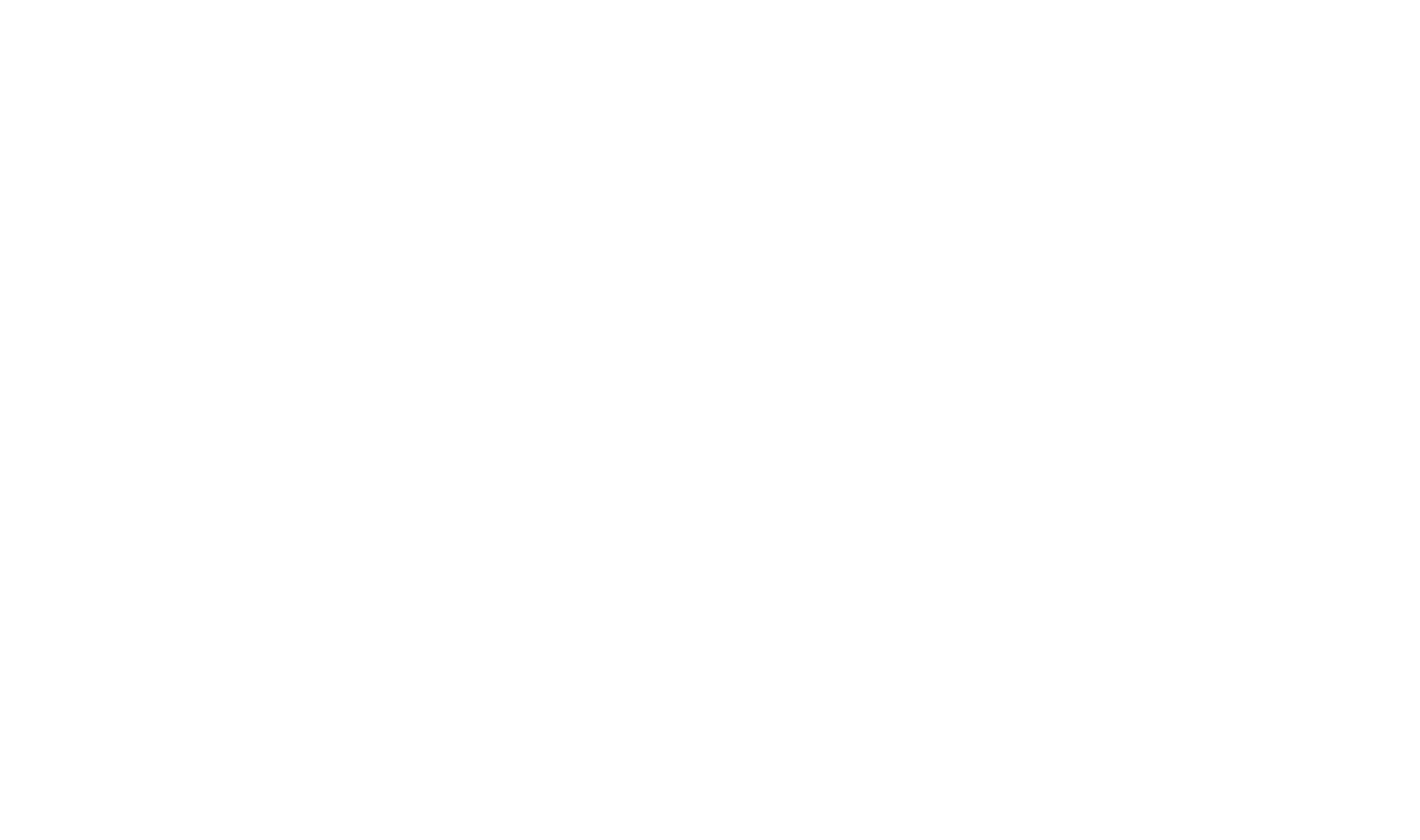}
  \caption{\label{fig:ComplexPlaneMethodsCartoons} (Left) Approaches based on imaginary chemical potential asymmetry,
  as parametrized by $i \delta \mu_R$, where $\delta \mu_R$ is real, avoid the sign problem (see text) but the results must then
  be analytically continued to the physical axis where the asymmetry is real. (Right) Approaches based on deformed contour or Lefschetz thimbles
  aim to minimize the phase oscillations by numerically modifying the domain of integration of the path integral, allowing the field to acquire an imaginary part.}
\end{figure}

\subsection{Imaginary asymmetry~\label{sect:ImaginaryAsymmetry}}
In non-relativistic physics, fermions at finite polarization or mass imbalance present a sign problem because although the determinant in \equref{FermionDeterminant} factorizes
and the factors are real, they will not typically be equal for all values of the HS field. A trick to overcome that problem can be borrowed from condensed matter theory~\cite{PhysRevB.41.811} which is to make the asymmetry imaginary.
For instance, if there are two spin flavors with corresponding chemical potentials $\mu_{\uparrow,\downarrow} = \mu \pm \delta \mu$, then taking $\delta \mu \to i \delta \mu_R$, where $\delta \mu_R$ is real, such that $\delta \mu$ is purely imaginary, makes the fermion determinants complex conjugates of one another and their product is real and positive, i.e.,
\begin{equation}
  \det M_\uparrow \det M_\downarrow \to |\det M_\uparrow |^2.
\end{equation}
This trick allows efficient sampling of the modified path integral via conventional Monte Carlo approaches. Below, we shall refer to imaginary asymmetry methods in general by the abbreviation ``iHMC''.

Naturally, a caveat of this approach is that one must return to the real asymmetry axis (see left panel of \figref{ComplexPlaneMethodsCartoons}). This usually involves a fit of an ansatz to the numerical data which then needs to be analytically continued. At this point, some degree of uncontrolled approximation enters the analysis as the ansatz for the fit is not unique.
Nevertheless, the fact that investigations on the imaginary asymmetry axis are done in an entirely nonperturbative and controlled way is certainly an attractive feature.

In relativistic physics, the equivalent of the above is the introduction of a finite chemical potential that breaks the particle-antiparticle symmetry, thus inducing a finite difference between the densities of particles and antiparticles. The associated breaking of the  charge-conjugation symmetry creates a sign problem which can be cured by rendering the chemical potential entirely imaginary, not only the difference of the chemical potentials as in non-relativistic field theories.
This idea was originally put forward in Ref.~\cite{Alford:1998sd} in the late 1990s and was very successfully employed in lattice QCD shortly thereafter~\cite{, DEFORCRAND2003170, PhysRevD.67.014505, PhysRevD.70.074509}. Since then, this approach has proven to be very valuable for studies of thermodynamics as well as the phase structure of QCD, see, e.g., Refs.~\cite{PhysRevLett.105.152001, Philipsen:2012nu, PhysRevD.85.094512, Bonati:2014kpa, Philipsen:2016hkv, Gunther:2016vcp, Borsanyi:2020fev} for more recent results.
For a discussion of the analytic continuation and suitable functional parametrizations of the data in the case of QCD, we refer the reader to Refs.~\cite{Lombardo:2006yc,DElia:2007bkz,Gunther:2016vcp} and to Ref.~\cite{Karbstein:2006er} for a detailed analysis of this issue with the aid of an exactly solvable field-theoretical model.

\begin{figure}[t]
  \centering
  \includegraphics[width=0.45 \columnwidth]{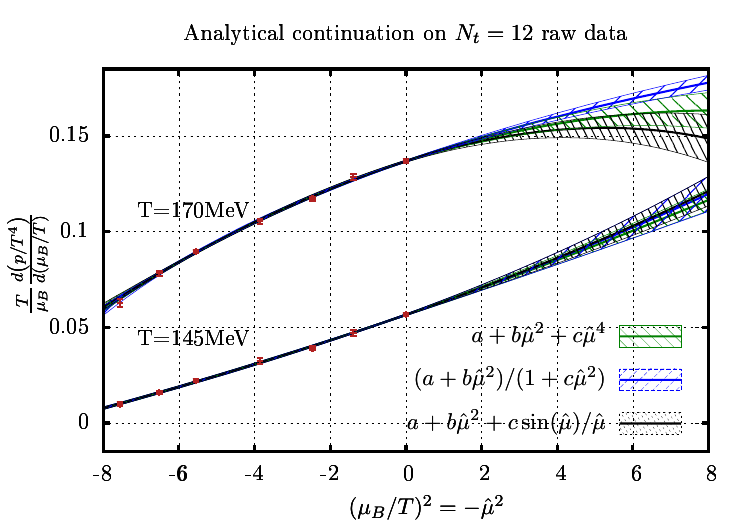}
  \includegraphics[width=0.50 \columnwidth]{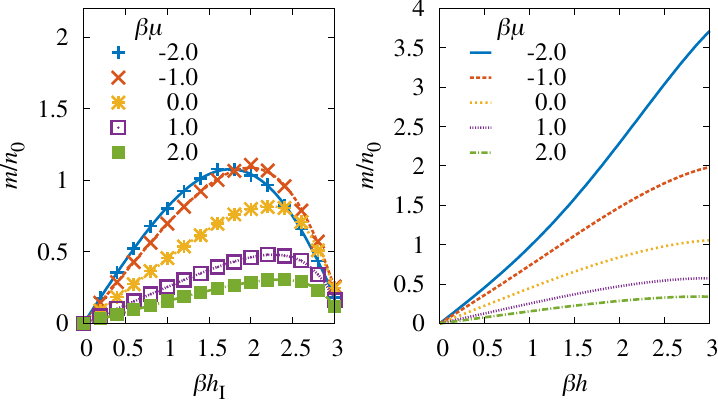}
  \caption{\label{fig:ComplexPlaneMethodsResults} (Left) Dimensionless baryon density~$(1/T^3)({\rm d}p/{\rm d}\mu_{\rm B})$
  rescaled by a factor~$T/\mu_{\rm B}$ for (2+1)-flavor QCD, see Ref.~\cite{Gunther:2016vcp}. The figure shows the analytical continuation of the
  lattice data for imaginary chemical potential (negative horizontal axis) to real chemical potential corresponding to $(\mu_B/T)^2>0$ based on the use of different functional forms to
  fit the data. (Middle and Right) Magnetization of a non-relativistic 1D gas of two-component fermions interacting via contact interaction
  before (middle panel) and after (right panel) analytic continuation from imaginary chemical
  potential difference $h_I={\rm i} h$ with $h=\mu_{\uparrow}-\mu_{\downarrow}$, see Ref.~\cite{PhysRevA.92.063609} for details. Note that
  the magnetization is the difference in the densities of the two fermion components.}
\end{figure}

In~\figref{ComplexPlaneMethodsResults} (left panel), for illustration purposes, we show results for the dimensionless baryon density~$(1/T^3){(\rm d}p/{\rm d}\mu_{\rm B})$, rescaled by a factor of~$T/\mu_{\rm B}$, for (2+1)-flavor QCD, see Ref.~\cite{Gunther:2016vcp}.
Here,~$\mu_B$ is the baryon chemical potential. To obtain the results for~$(\mu_B/T)^2>0$ (physical case) in~\figref{ComplexPlaneMethodsResults} (left panel), various functional forms have first been fitted to the points on the negative side of the horizontal axis and then analytically continued to the positive side. The invariance of QCD under a sign flip of the baryon chemical potential has been used in constructing the fit functions. Whereas the predictions for~$(\mu_B/T)^2>0$ are impressively independent of the used functional form at small chemical potential, the uncertainty grows large when the chemical potential is increased, illustrating the limitations of this approach. In any case, we note that it is not possible to study the zero-temperature limit of QCD at finite baryon chemical potential with this approach. Indeed, because of the Roberge-Weiss symmetry~\cite{Roberge:1986mm}, only the regime~$|\mu_{\rm B}|/T <\pi$ is accessible.

Inspired by these studies, the same principle has been followed in non-relativistic physics in recent years. More specifically, it has been applied to chemical potential and mass
asymmetries~\cite{Roscher:2013aqa, PhysRevLett.110.130404, PhysRevA.92.063609, PhysRevLett.114.050404, PRD96094506}.
An example is shown in the center and right panel of \figref{ComplexPlaneMethodsResults} where the magnetization $m$ (in units of the noninteracting density $n_0$) of one-dimensional non-relativistic spin-$1/2$ fermions at several values of the chemical potential $\mu$ (given in units of the inverse temperature $\beta=1/T$) is depicted. In this case, calculations avoiding the sign problem are possible at imaginary chemical potential asymmetry $h_I$ (center panel). To obtain the results for real asymmetry~$h$ (right panel), an ansatz for the magnetization in terms of a function is first fitted to the data for imaginary asymmetry (center panel) and then analytically continued to real asymmetry~$h$.

Here, $h=\mu_{\uparrow}-\mu_{\downarrow}$ is the difference in the chemical potentials of the spin-up and spin-down component. Similarly to the case of finite baryon chemical potential in QCD, it is not possible to compute the magnetization as a function of chemical potential asymmetry at zero temperature with this approach. In fact, it is only possible to reach values in the regime~$|h|/T < \pi$ because of the $2\pi$-periodicity of non-relativistic fermions at finite temperature~\cite{PhysRevLett.110.130404}. As already stated above, the basic idea of ``taking a detour in the complex plane of the parameter space" is not limited to chemical potentials. In fact, the energy equation of state of 1D non-relativistic two-component fermions coming with different masses has been successfully studied in Ref.~\cite{PRD96094506} using an imaginary mass difference. Other interesting applications, such as imaginary angular velocity coupled to angular momentum, remain unexplored to date but could in principle also be studied in this way.

It should be pointed out that, for small asymmetries, one can avoid the problem of analytic continuation completely by performing a Taylor expansion of the path integral around vanishing asymmetry, where calculations can be carried out without a sign problem. This approach has been successfully employed in many finite-temperature lattice QCD studies at finite baryon chemical potential to compute the equation of state and extract the phase structure, at least at small baryon chemical potential. See Refs.~\cite{Allton:2003vx,Ejiri:2005uv,deForcrand:2007rq,Endrodi:2011gv} for ground-breaking studies of 2- and (2+1)-flavor QCD with this approach and Refs.~\cite{Bazavov:2017dus,Sharma:2017jwb} for recent state-of-the-art results. For example, we can exploit the relation between the baryon density~$n_{\rm B}$ and the QCD partition function~$\mathcal Z$:
\begin{equation}
  \frac{n_{\rm B}}{T^3} = \frac{1}{VT^2} \frac{{\partial}\ln {\mathcal Z}}{{\partial}\mu_{\rm B}}\,,
\end{equation}
where~$V$ is the spatial volume and the temperature is introduced to render the expression dimensionless. At finite baryon chemical potential~$\mu_{\rm B}$, the partition function is invariant under~$\mu_{\rm B}\to-\mu_{\rm B}$ and therefore the density can be written as a series in odd powers of~$({\mu_B}/T)^2$. The coefficients of this series can then be computed rigorously with stochastic methods by realizing that they are directly related to derivatives of~$\ln {\mathcal Z}$ with respect to~$\mu_{\rm B}$ evaluated at~$\mu_{\rm B}=0$. Moreover, the pressure equation of state can eventually be obtained by integrating the density with respect to~$\mu_{\rm B}$ since~$p = (T/V)\ln{\mathcal Z}$.

Essentially, this amounts to the computation of static response functions. This approach can indeed be very efficient provided that these functions can be calculated in a statistically controlled manner, see also Refs.~\cite{deForcrand:2007rq,Brandt:2018omg} for a discussion of the reliability of this approach. Presently, the effort has been pushed to sixth order in the baryon chemical potential~\cite{Bazavov:2017dus}. Of course, a similar approach can also been applied to non-relativistic theories to study equations of state as well as the phase structure and appears to be a worthwhile endeavor, also to cross-check results obtained by using imaginary asymmetries.

\subsection{Lefschetz thimbles~\label{sect:Thimbles}}
In the presence of a phase problem, i.e. when the action $S[\phi]$ has real and imaginary parts: $S[\phi] = S_R[\phi] + i S_I[\phi]$, it may be possible to deform the integration contour away from the real line $\phi \in (-\infty,\infty)$ and into the complex $\phi$ plane (see right panel of \figref{ComplexPlaneMethodsCartoons}) such that $S_I[\phi]$ is constant, or approximately so. If our problem concerns a simple one-site model, such that the partition function is a one-dimensional integral
\begin{equation}
  \mathcal Z = \int_{-\infty}^{\infty} \d\phi \; \e^{-S[\phi] }
\end{equation}
then observables take the form
\begin{equation}
  \langle \mathcal O \rangle = \frac{1}{\mathcal Z}\int_{-\infty}^{\infty} \d\phi \; \e^{-S[\phi] }\CO[\phi].
\end{equation}
The goal of the Lefschetz thimbles method is to achieve a deformation of the integration region into a new region $\mathcal C$ such that we may alculate the above as
\begin{equation}
  \label{Eq:ZnoImS}
  \mathcal Z_0 = \int_{\mathcal C} \d\phi \; \e^{-S_{\rm R}[\phi] }
\end{equation}
\begin{equation}
  \label{Eq:OnoImS}
  \langle \mathcal O \rangle =  \frac{1}{\mathcal Z_0}\int_{\mathcal C} \d\phi \; \e^{-S_{\rm R}[\phi] }\CO[\phi],
\end{equation}
where $\phi$ is now regarded as a complex variable and we have used the fact that $S_{\rm I}[\phi]$ is constant along $\mathcal C$, i.e. we have cancelled it in both Eqs.~(\ref{Eq:ZnoImS}) and~(\ref{Eq:OnoImS}).

Such a contour deformation changes neither the theory nor the observables, as the integrand in the path integral of any theory of interest will be an analytic function of $\phi$. If such a contour can be determined either \textit{a priori} or dynamically during a calculation, the sign problem could potentially be solved or at least tamed. As a Monte Carlo method, the idea can be traced back to the work of Ref.~\cite{ROM1997382} where the so-called shifted contour auxiliary-field Monte Carlo method was put forward for electronic systems (see also Ref.~\cite{Witten:2010cx, Witten:2010zr}).

To extend the case of the simple one-dimensional integral discussed above to general QFTs, one complexifies the field variable in accordance with complex Langevin (which will be explored in the remainder of this review). In such a complexified configuration space, the Lefschetz thimbles approach aims to find the stationary points for which $\delta S_{\rm I}[\phi]/\delta \phi = 0$, as those points feature reduced phase oscillations for $\exp\left (-\i S_{\rm I}[\phi]\right )$. Such an approach is of course the generalization of the saddle-point (or critical point) method of evaluating complex integrals and the higher-dimensional version is often referred to as Picard-Lefschetz theory. The corresponding deformed, high-dimensional integration contours of steepest descent are called Lefschetz thimbles.

In practice, the locations of such (stable) points of steepest descent are found by evolving the field along a fictitious time $t$, which is similar in spirit to the fictitious Langevin time (although unrelated to the imaginary time $\tau$). This propagation proceeds according to the holomorphic gradient flow equation
\begin{equation}
  \frac{\d\phi}{\d t} = -\left(\frac{\delta S[\phi]}{\delta\phi}\right)^*
\end{equation}
%
or, more explicitly,
\bea
  \label{Eq:HGF}
  \frac{\d \phi_{\rm R}}{\d t} &=& - \textrm{Re}\left[\frac{\delta S[\phi]}{\delta \phi} \right],\\
  \frac{\d \phi_{\rm I}}{\d t} &=& + \textrm{Im}\left[\frac{\delta S[\phi]}{\delta \phi} \right],
\eea
which, as we shall see below, is remarkably similar to the CL equations. In fact, the above expression corresponds, up to a sign in the imaginary part, to the drift term in the Langevin equations~\equref{complex_langevin_equations}, given by $d \phi / d t = -{\delta S[\phi]}/{\delta \phi}$. Ref.~\cite{Aarts2013c} presents a particularly lucid side-by-side discussion of CL versus Lefschetz thimbles approaches for a simple quartic integral. There, it is shown that there are similarities in that the location of the distributions in the complex plane obtained from CL and thimbles follow each other closely. However, there are also important differences with regard to the weight distribution and the role of the residual phase across the thimble (see below), suggesting that, despite the structural similarity of the above equations, the methods behave quite differently in practice.

The crucial advantages of deforming the path integral to capture, effectively, a set of mean-field configurations and the corresponding fluctuations, are that $S_I[\phi]$ is locally constant and that the real part, which determines the weight $\exp \left( - S_R[\phi]\right)$, is maximally localized around the saddle point. In other words: Lefschetz thimbles are the best locations to carry out stochastic evaluations of path integrals. While the constant-$S_I[\phi]$ property is crucial, the method does not rely on finding the precise location of the critical points. Rather, it is based on finding a useful deformation which may or may not be close to the stationary phase contours attached to the critical points (wherever those may be), but where the variations in $\textrm{Im} S[\phi]$ are small~\cite{Alexandru:2015sua}. Those ideas were crucial for the application of Refs.~\cite{PhysRevLett.117.081602, PhysRevD.95.114501} where they were used to calculate the properties of a low-dimensional field theory in real time.
Furthermore, Ref.~\cite{PhysRevD.96.034513} found that, even if the holomorphic gradient flow of \equref{HGF}) is used to push the deformation very close to the thimbles (which is naively the ideal situation), then the high barriers separating the thimbles would make the sampling very challenging in practice.

One of the difficulties of deforming a contour in a functional integral is the calculation of the associated Jacobian factor $J(t)$ due to the curvature of the thimble, which shows up explicitly upon parametrizing the complex-plane integral of Eq.~(\ref{Eq:ZnoImS}) by a parameter $t \in (-\infty,\infty)$, namely
\begin{equation}
  \mathcal Z_0 = \int_{-\infty}^{\infty} dt \; J(t)\; e^{-S_R[\phi(t)] }.
\end{equation}
While the above is a computational issue that is difficult but tractable~\cite{PhysRevD.93.094514}, a more serious challenge lies in accounting for all the possible thimbles, which turns the above into a sum with unknown weights, i.e. in that case
\begin{equation}
  \label{Eq:ZImS}
  \mathcal Z = \sum_{k} m_k \int_{\mathcal C_k} d\phi \; e^{-S_R[\phi] }
\end{equation}
and
\begin{equation}
  \label{Eq:OImS}
  \langle \mathcal O \rangle =  \frac{1}{\mathcal Z}\sum_{k} m_k \int_{\mathcal C_k} d\phi \; e^{-S_R[\phi] }\mathcal O[\phi],
\end{equation}
where, crucially, the phases in the numerator and denominator of Eqs.~(\ref{Eq:ZImS}) and~(\ref{Eq:OImS}) do not cancel. This issue is often referred to as a `global' sign problem, as opposed to the `residual' sign problem coming from the remaining curvature (i.e, variations in the imaginary part of the Jacobian across the thimble). In that regard, it is useful to note that (for a fixed set of input parameters), only a subset of thimbles contribute to the partition function. The global sign problem depends on the subset and the relative weights of each thimble, both of which are difficult to determine in practice (see e.g. Ref.~\cite{Bluecher:2018sgj}). However, the holomorphic flow bypasses that complication. It is then possible to track the global and the residual sign problems on the deformed surface by measuring the average phase.

Recently, contour deformation and Lefschetz thimbles have re-emerged as a method in quantum field theory and its applications have extended to condensed matter physics as well. References~\cite{PhysRevD.86.074506, PhysRevD.88.051501, PhysRevD.88.051502} were the first ones to propose that the sign problem in finite density QCD could be overcome using Lefschetz thimbles. Refs.~\cite{Fujii:2013sra, PhysRevD.89.114505} proposed ways to overcome the residual sign problem across a thimble (see below). Reference~\cite{Kanazawa2015} studied the structure of thimbles in fermionic theories.
Ref.~\cite{PhysRevD.91.101701} used thimbles to avoid the sign problem in a mean-field analysis of the QCD partition function. while Ref.~\cite{Tanizaki:2016cou} interpreted the Silver Blaze problem of finite-density QCD in a one-site fermion model. Reference~\cite{PhysRevD.95.014502} used a generalization of the Lefschetz thimbles method to study the finite-density Thirring model in two spacetime dimensions, and Ref.~\cite{PhysRevD.98.034506} generalized that study to QED in two spacetime dimensions.

Other interesting connections have also been drawn. For instance, Ref.~\cite{Fukushima:2015qza} showed that the steepest descent trajectories (i.e. the deformed integration contours mentioned above) can be interpreted as ground-state wave-functions of a supersymmetric Hamilton dynamics. Connections to the complex Langevin method and its convergence shortcomings (see also below) were pointed out in Ref.~\cite{Hayata:2015lzj} and Ref.~\cite{Nishimura:2017eiu}.

On the condensed matter side, Refs.~\cite{Ulybyshev:2017hbs, Ulybyshev2019fte, Ulybyshev2019hfm} studied the Lefschetz thimbles representation of the Hubbard model and studied it in a hexagonal lattice away from half filling. Ref.~\cite{Fukuma2019wbv} applied a variant of the Lefschetz thimbles method (so-called ``tempered'' Lefschetz thimbles method, developed in Ref.~\cite{Fukuma:2017fjq}) to the Hubbard model away from half filling.

In light of the excellent recent review article on Lefschetz thimbles and its applications~\cite{Alexandru:2020wrj}, we limit ourselves to the above discussion.


\subsection{Path optimization method~\label{sect:pom}}
The path optimization method (POM) refers to a relatively novel complex plane approach, whose aim is to shift the integration away from the real line in order to minimize the sign fluctuations along the new integration path in the complex plane~\cite{Ohnishi2018,PhysRevD96.111501,MoriLattice2019}. Although the sign problem will not be completely absent on the new integration contour, the hope is to ameliorate it enough such that the signal-to-noise ratio is manageable.
%
%
%

Similar to the above discussed method of Lefschetz thimbles, the path integral can be expressed in terms of a shifted path $\phi(t)$ parametrized by the real parameter $t$ and under consideration of the Jacobian $J[\phi(t)]$:
\begin{equation}
  \label{Eq:POMPathIntegral}
  \mathcal{Z} = \int \mathcal{D}\phi\ \e^{-S[\phi]} = \int_\mathcal{C} \mathcal{D}t\ J[\phi(t)]\e^{-S[\phi(t)]}
\end{equation}
Of course, in order for the Cauchy theorem to hold, the integration path should not enclose singular points of the Boltzmann factor. The method relies on a trial function which parameterizes the integration path in the complex plane by some parameters, collectively denoted as $\alpha$. Following Ref.~\cite{Ohnishi2018}, we may for instance expand the path as a sum over a complete set of polynomials
\begin{equation}
  \phi_\alpha(t) = \phi_\alpha^{\rm R}(t) + \i \phi_\alpha^{\rm I}(t) = t + \sum_{n} \left(\alpha_{n}^{\rm R}+ \i \alpha_{n}^{\rm I}\right)H_{n}(t),
\end{equation}
although this is only one particular choice. The important features to consider (besides some technical aspects, see, e.g.,~\cite{Alexandru:2020wrj}) is that the chosen parameterization allows contours with a more suitable sign-structure and that the Jacobian may be evaluated efficiently. To find the path with the mildest sign problem, the parameters $\alpha_{n}^{\rm R}$ and $\alpha_{n}^{\rm I}$ will be optimized by minimizing some cost function. In Ref.~\cite{Ohnishi2018} the cost function
\begin{equation}
  F[\phi_\alpha(t)] = \frac{1}{2} \int \d t\ |\e^{\i \theta_\alpha(t)} - \e^{\i \theta_{0}}|^{2}\ |J(\phi_\alpha(t)) \e^{-S[\phi_\alpha(t)]}| 
\end{equation}
was applied, but other cost functions can be used. Here $\theta_\alpha(t)$ is the complex phase of our parameterized integrand $J[\phi_\alpha(t)] \e^{-S[\phi_\alpha(t)]}$, as defined in~\equref{POMPathIntegral}, and $\theta_{0}$ is the complex phase of the original integrand.
The cost function is then used to tune the parameters $\alpha_{n}^{R}$ and $\alpha_{n}^{I}$, such that $F[\phi_\alpha(t)]$ is minimized.
This results in an enhanced phase factor, thus reducing or ideally eliminating the oscillations discussed in~\secref{reweighting}, which are the source of the sign problem.

For the optimization, several algorithms may be used, which mostly rely on the computation of the gradient in $\alpha$-space $\nabla_\alpha \ev{\e^{\theta_\alpha(t)}}$. A convenient (and important for practical implementations) feature is that it is typically enough to have a rough knowledge of the gradient such that efficient optimization is ensured. Naturally there is significant overlap with methods in Machine Learning (ML). A simple steepest descent method can be used, but other methods for minimizing the cost function have been explored in ML and neural network (NN) applications~\cite{Ohnishi2019POMandNN,MoriPOMandNN}.

While this method is very new, early work using shows it is able to overcome a severe sign problem in a one-dimensional toy model where CL fails~\cite{Ohnishi2018}, one dimensional Bose gases with chemical potential~\cite{Bursa2018}, as well as in a U(1) gauge theory and complex scalar field theory~\cite{PhysRevD102.014514}. Work towards applications to full $3+1$ dimensional QCD has included $0+1$ dimensional QCD at finite density~\cite{MoriPOMinQCD} and effective models for QCD~\cite{PhysRevD99.014033,PhysRevD99.114005} and the Thirring model~\cite{PhysRevD.97.094510,PhysRevLett.121.191602}.


\section{\label{sect:formalism}The Langevin method for real and complex variables}
Stochastic quantization as a method for treating Euclidean field theories has been around since Parisi and Wu first proposed the connection between the Euclidean field theories and statistical systems coupled to a heat bath~\cite{ParisiWu}. It is now well-established as a successful tool for treating quantum many-body systems with a real Euclidean action~\cite{PhysicsReportsStochasticQuantization}. This section examines the method for systems without a sign problem (referred to from here on as ``real Langevin") and its extension to systems with complex actions as a possible circumvention of the sign problem (``complex Langevin" or CL, the focus of this review). We present a pedagogical example to illustrate the real and complex Langevin methods using a simple toy model, and discuss some of the challenges that arise in using the complex Langevin method along with proposed solutions to those problems.

\subsection{Complex Langevin: origins and modern re-emergence~\label{sect:CL}}
\begin{figure}
  \includegraphics{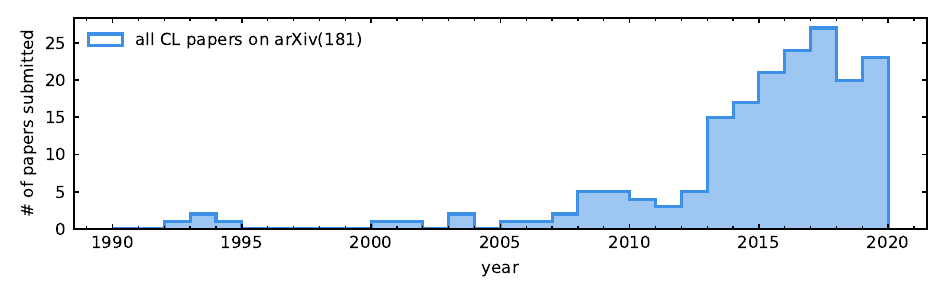}
  \caption{\label{fig:cl_arxiv_history} Histogram of the number of preprints on arXiv per year since its introduction in 1993 (the search query was ``complex Langevin''). This overview neglects progress made in the 1980s; however, it highlights the recent surge in inerest in this promising method.}
\end{figure}
Shortly after the introduction of the concept of real stochastic quantization, it was realized that the approach could be extended to the case of complex actions.
Loosely speaking, using a Langevin equation rather than an importance sampling approach eliminates the restriction to real and positive semidefinite measures. This is due to the ability of the probability measure used in a Langevin method to be complexified -- at least formally. In 1983 such a strategy was discussed independently by Klauder~\cite{Klauder1983a,Klauder1983b} and Parisi~\cite{Parisi1983} as an alternative to existing Monte Carlo methods and marked the first investigations of how the complex Langevin equation can be used to address the complex phase problem.

The elegant form of the approach as well as the potential to circumvent the sign problem drew considerable interest in the years after the initial proposals. Following the first successful numerical application for the quantum Hall effect by Klauder \cite{Klauder1984}, the method was employed in many studies, albeit with mixed success. Unfortunately, the convergence of the method cannot be guaranteed \textit{a priori} and even if convergence is achieved, spurious solutions with biased expectation values might be found~\cite{Ambjorn1985,Ambjorn1986}. This is connected to subtle mathematical issues that arise in the case of complex weight and the structure of the associated complex Fokker-Planck equation. As a consequence, the initial flurry of interest stalled and progress on these matters slowed down over the years, despite early attempts to understand these shortcomings~\cite{Klauder1985,Gausterer1993,Gausterer1998,GaustererNPA1998}.
Nevertheless, progress was made in several directions and the applicability of CL was investigated in a set of simplified models such as the chiral Schwinger model~\cite{Gausterer1986,Kieu1995} as well as toy problems for relativistic~\cite{Haymaker1988} and non-relativistic fermionic theories~\cite{PRC2001026303}. Interestingly, the method has also spread to the realm of physical chemistry and was used in simulations of polymeric fluids \cite{Ganesan_2001,Duechs2003,Fredrickson2002} as well as reaction simulations in the context of physical chemistry \cite{HOCHBERG200654,DELOUBRIERE2002135,Jonghoon2008,Popov2007,Xingkun2014,Duechs2014} as a way to include beyond mean-field corrections.


As recently as the mid 2000s to early 2010s, the CL approach re-emerged as a method of interest in relativistic physics, particularly in the study of lattice QCD, when it was realized that some of the initially encountered problems are treatable with an improved integration strategy. In this new era, early applications to relativistic physics examined non-equilibrium QFT~\cite{PhysRevLett95202003,PhysRevD75045007}, which can provide insights into high-energy physics, particularly heavy ion collisions.
Moreover, in 2008, Aarts and Stamatescu demonstrated that CL could be applied to models of finite density QCD that exhibit a sign problem~\cite{JHEP200809018}. Shortly after that, Aarts demonstrated that CL could be used to circumvent the sign problem in the relativistic Bose gas with finite chemical potential~\cite{Aarts2009, AartsPRL102131601, JHEP200905052}. This began a resurgence of interest in this method in the field of finite density Lattice QCD (LQCD), in which nonperturbative calculations of strongly interacting matter with finite baryon chemical potential are inhibited by the sign problem. This renewed interest led to work in the next few years on optimization of the method to prevent runaways and improve stability, using stochastic reweighting, gauge fixing, and adaptive step size algorithms~\cite{BERGES2008306,AARTS2010154,JHEP20100820}, 
see also~\figref{cl_arxiv_history} for a history of the ``CL activity on the arXiv".

The successes of the method, and advances made in treating instabilities and singularities in the fermion determinant, have generated interest in applying CL to non-relativistic systems, particularly many-fermion systems, in which sign problems arise frequently. Work with the CL method in the context of non-relativistic systems is just beginning, but is already showing great promise. Discussion of these recent applications to relativistic and non-relativistic physical systems can be found in Secs.~\ref{sect:RQFT} and~\ref{sect:NRQFT}.

Despite the formal challenges of CL, it should be pointed out that, at least in principle, the existence of a well-defined probability measure on a complexified field space
is guaranteed under certain conditions (often referred to as Weingarten's theorem~\cite{PhysRevLett.89.240201}). However, the challenge lies in constructing such a measure, which has been investigated by Salcedo and others ~\cite{Salcedo:1996sa, Salcedo_2007, Wosiek:2015iwl, Wosiek:2015bqg, PhysRevD.94.074503, Seiler_2017, Salcedo_2018, Wosiek:2018jht}. While this is a very attractive area of research, we will not pursue it further in this review.

\subsection{Stochastic quantization: path integrals and the Langevin process}
The main ingredient of the CL method is the concept of {\it stochastic quantization}. The idea was introduced in the seminal 1981 paper of Parisi and Wu~\cite{ParisiWu} and a few years later summarized nicely in the famous review article by Damgaard and H\"{u}ffel~\cite{PhysicsReportsStochasticQuantization} \footnote{See also Ref.~\cite{jona-lasinio1985} for a more formal introduction to stochastic quantization}.
Notably, stochastic quantization has played an important role both theoretically as well as computationally; in particular, it has been used extensively in field theory (see e.g. Ref.~\cite{PhysRevD.32.2736}) and condensed matter
(see e.g. Ref.~\cite{PhysRevB.99.035114}), and is the precursor of the
hybrid Monte Carlo algorithm~\cite{Duane:1987de, PhysRevD.35.2531}, which has been the workhorse of LQCD for decades. In the following we present a brief introduction to
stochastic quantization in order to lay out the foundation for later considerations.

For a quantum field theory of a real field $\phi$ governed by a real action $S[\phi]$, stochastic quantization provides an intuitive way to understand path integrals of the form
\begin{equation}
  \label{Eq:PathIntegralSQ}
  \CZ = \int\CD\phi\ {\rm e}^{-S[\phi]}.
\end{equation}
 As a first step, we introduce a purely fictitious time variable $t$, which represents the direction of the stochastic evolution. This fictitious time evolution is then governed by a stochastic differential equation, namely the Langevin equation:
 \begin{equation}
   \label{Eq:Langevin_continuum}
   \frac{{\rm d}\phi}{{\rm d}t} = -\frac{\delta S[\phi]}{\delta \phi} + \tilde{\eta}.
 \end{equation}
The first term on the right hand side is called the drift term, also sometimes referred to as the classical flow as it constitutes the deterministic part of the time-propagation of the fields. The second term on the left hand encodes the random nature of the equation and is given by a white-noise with zero autocorrelation, i.e. $\tilde\eta \sim \delta(t-t')$.

For practical purposes it is convenient to rewrite this equation in a discrete form which can be done by integrating both sides over the time interval $\Delta t$. This leads to the discrete Langevin equation
\begin{equation}
  \label{Eq:Langevin}
  \Delta \phi = K[\phi] \Delta t + \eta,
\end{equation}
where $\eta$ is typically chosen to be a Gaussian random variable with $\langle \eta \rangle = 0$ and $\langle \eta^2 \rangle = 2\Delta t$. The angle brackets denote an
average over $\eta$. This random process will produce configurations $\phi$ that follow a certain probability distribution. The key ingredient of stochastic quantization
is the realization that the equilibrium distribution (if it exists) of the $d+1$ dimensional random process in \equref{Langevin} corresponds to the probability measure in the $d$-dimensional path integral \equref{PathIntegralSQ}. The extra dimension is simply the fictitious time $t$.

It is instructive to consider the above stochastic differential equation without the noise term. In that case, \equref{Langevin_continuum} reduces to an ordinary differential equation and its form is nothing but that of a gradient descent. Starting out at a random (non-pathological) state, this implies that the solution will converge to a stationary point of the action, which is the ``mean-field'' or classical solution. The simple interpretation of the noise term is that it represents quantum fluctuations around this classical solution. In order to reproduce the correct physics, we have to ``add the correct amount of fluctuations" which is set by the Fluctuation-Dissipation theorem. Thus, stochastic quantization can be viewed as a very explicit form of quantization.


The random process results in a sequence of time-dependent configurations $\phi_\eta(t)$ distributed according to a
time-dependent probability measure $P[\phi,t]$. The expectation value of a given observable $\mathcal O[\phi]$ is then
given by
\begin{equation}
  \label{Eq:EVOprob}
  \ev{\CO[\phi_\eta(t)]} = \int \mathcal D\phi\ P[\phi,t] \CO[\phi].
\end{equation}
To establish the validity of such a Langevin average, we need to investigate the temporal behavior of the expectation value and show that the time-dependent probability distribution depends on $S[\phi]$ in the way dictated by \equref{PathIntegralSQ}, at least at large enough $t$.
Such a property will justify the use of temporal averages along the Langevin evolution to estimate the true expectation values of the theory.
An instructive discussion can be found in \cite{PTPS1993CLSimulation}, which we will follow closely.

As a first step, we take the fictitious-time derivative of the expectation value
  \bea
  \label{Eq:PhiAvgSide}
  \frac{{\rm d} \ev{\CO[\phi_\eta(t)]}}{{\rm d}t} = \int \mathcal D\phi\ \frac{{\rm d} P[\phi,t]}{{\rm d}t} \CO[\phi].
\eea
Note, that only the probability distribution carries a fictitious-time dependence. Alternatively, we may perform the same fictitious-time derivative by expanding the observable to second order in its $\phi$ dependence
\begin{equation}
  \label{Eq:obs_expansion}
  {\rm d} \CO[\phi] = \frac{\delta \CO[\phi]}{\delta\phi} {\rm d} \phi + \frac{1}{2} \frac{\delta^2 \CO[\phi]}{\delta\phi^2} ({\rm d} \phi)^2.
\end{equation}
According to \equref{Langevin} we may write
\begin{equation}
  \label{Eq:LEdiff}
  {\rm d}\phi = -\frac{\delta S[\phi]}{\delta\phi} {\rm d}t + {\rm d}w\, ,
\end{equation}
where ${\rm d}w$ is the so-called {\it Wiener increment} with the property
\begin{equation}
  \langle {\rm d}w^2 \rangle = \int_{t}^{t+dt} {\rm d}\tau\ \int_{t}^{t+dt} {\rm d}\tau'\ \langle \eta(\tau)\eta(\tau') \rangle = 2 \, {\rm d}t \, ,
\end{equation}
and vanishing mean $\langle{\rm d}w\rangle = 0$.
Substituting in \equref{obs_expansion} and using the properties of ${\rm d}w$ yields
\bea
  \langle {\rm d} \CO[\phi] \rangle = \left\langle  -\frac{\delta \CO[\phi]}{\delta\phi}\frac{\delta S[\phi]}{\delta\phi} + \frac{\delta^2 \CO[\phi]}{\delta\phi^2}  \right\rangle{\rm d}t\, ,
\eea
which allows us to write
\begin{equation}
  \frac{{\rm d} \langle \CO [\phi_\eta(t)] \rangle}{{\rm d} t} =
  \left\langle  -\frac{\delta \CO[\phi]}{\delta\phi}\frac{\delta S[\phi]}{\delta\phi} + \frac{\delta^2 \CO[\phi]}{\delta\phi^2}  \right\rangle
  \equiv \ev{L_r O}\, ,
\end{equation}
where we defined the so-called Langevin operator
\begin{equation}
  L_{r} = \int{\rm d}\tau{\rm d}^dx\ \left(\frac{\delta}{\delta\phi} + K[\phi]\right)\frac{\delta}{\delta\phi}\, ,
\end{equation}
with the drift $K[\phi] = -\frac{\delta S[\phi]}{\delta\phi}$, according to \equref{Langevin}. We may again write this as an integral over configurations
\begin{align}
  \frac{{\rm d}  \langle \CO [\phi_\eta(t)] \rangle}{{\rm d} t} &= \int \mathcal D \phi \left( -\frac{\delta \CO[\phi]}{\delta\phi}\frac{\delta S[\phi]}{\delta\phi} + \frac{\delta^2 \CO[\phi]}{\delta\phi^2}\right)P[\phi,t] \\
  &= \int \mathcal D\phi \ \CO[\phi] \left(\frac{\delta}{\delta \phi}\frac{\delta S[\phi]}{\delta \phi}  + \frac{\delta^{2}}{\delta \phi ^{2}}\right) P[\phi,t].
\end{align}
where the second line results from partial integration. Here we made the important assumption that the probability vanishes at the boundaries (or decays fast enough if the integration region is non-compact). These assumptions are in fact crucial and will be discussed in more detail below.

Comparing the above equation with \equref{PhiAvgSide} yields the Fokker-Planck (FP) equation:
\bea
  \label{Eq:FPE}
  \frac{\rm d}{{\rm d}t}P[\phi,t] = L_{r}^{\rm T} P[\phi,t]\, ,
\eea
with the formal adjoint of the above Langevin operator
\begin{equation}
L_{r}^{\rm T} \equiv \int{\rm d}\tau{\rm d}^dx\ \frac{\delta}{\delta \phi}\left(\frac{\delta}{\delta \phi} - K[\phi]\right) ,
\end{equation}
which is also referred to as the FP operator or FP Hamiltonian. To show that the stationary solution of this equation is indeed our desired probability distribution we perform similarity transformation
\begin{equation}
\label{Eq:PtildeP}
  \widetilde{P}[\phi,t] = {\rm e}^{S[\phi]/2} P[\phi,t] \, ,
\end{equation}
to rewrite the FP equation
\begin{equation}
  \frac{{\rm d}}{{\rm d}t}\widetilde{P}[\phi,t] =  \widetilde{L}_{r}^{\rm T}\widetilde{P}[\phi,t] \, ,
\end{equation}
with
\begin{equation}
  \widetilde{L}_{r}^{\rm T} = {\rm e}^{S[\phi]/2}\,L_{r}^{\rm T}\,{\rm e}^{-S[\phi]/2} =
  \int{\rm d}\tau{\rm d}^dx\ \left(- \frac{\delta}{\delta \phi} + \frac{1}{2}K[\phi]\right)
    \left(\frac{\delta}{\delta \phi} + \frac{1}{2}K[\phi]\right).
\end{equation}
This last equation reveals that, with a real action $S[\phi]$, our modified FP Hamiltonian is a self-adjoint and positive semidefinite operator, with a unique FP ground state $\psi_0 = {\rm e}^{-S[\phi]/2}$ and vanishing FP energy $E_0 = 0$.
We can therefore project our probability over the complete set of eigenfunctions and non-negative eigenvalues of
$\widetilde{L}_{r}^{\rm T}$, and see that our probability collapses to the ground state in the long time limit:
\bea
  \label{Eq:ProbFPE}
  \widetilde{P}[\phi,t] = \sum_{n = 0}^{\infty}a_{n} \psi_{n} {\rm e}^{-E_{n} t}\  \xrightarrow{t\to\infty}\ a_{0} {\rm e}^{-S[\phi]/2}.
\eea
Upon performing the back-transformation according to \equref{PtildeP} we obtain
\begin{equation}
  \lim_{t\to\infty} P[\phi,t] \sim {\rm e}^{-S[\phi]} \, ,
\end{equation}
which shows that the Langevin equation produces field configurations distributed according to the Boltzmann weight ${\rm e}^{-S[\phi]}$ in the limit of large fictitious time \footnote{We have assumed here that the spectrum of $\widetilde{L}_{r}^{\rm T}$ is discrete, which may not be true in practice. This assumption can be relaxed but it is important that the $E_0 = 0$ eigenvalue be non-degenerate.}.

The above justifies the use of temporal averages to estimate equilibrium expectation values.
In fact, expectation values are obtained in practice by integrating over a time $T$:
\bea
  \langle \mathcal{O}\rangle \approx \frac{1}{T} \int_{t_{\rm th}}^{t_{\rm th}+T} {\rm d}t\ \mathcal{O}[\phi_\eta(t)],
\eea
where $t_{\rm th}$ reflects the equilibration time that is needed to approach the stationary probability distribution.

\subsection{A practical guide to real Langevin}
The above procedure is a well-established method for real-valued fields $\phi$ on a real manifold $\mathcal{M}$. In the following, we connect the concept with conventional Monte Carlo approaches based on Markov chains and highlight the similarities of these approaches in a practical way.

Generally, we are interested in expectation values of a given (Euclidean) field theory of the form of \equref{EVOprob}:
\begin{equation}
  \langle \CO \rangle = \frac{1}{\mathcal Z}\int{\mathcal D \phi\ \CO[\phi] {\rm e}^{-S[\phi]}} \equiv \int{\mathcal D \phi\ \CO[\phi] P[\phi]}.
  \label{Eq:langevin_toy_expectation}
\end{equation}
Typically, the evaluation of such high-dimensional path integrals is achieved by stochastic sampling, i.e. by producing a sequence of random states $\phi$ (interchangeably called samples or configurations). However, rather than producing a random state from scratch at every step (which might be expensive), new states are obtained by changing or {\it updating} an existing one. This can be written in the generic form
\begin{equation}
  \label{Eq:markov_chain}
  \phi_{n+1} = F[\phi_n]\, ,
\end{equation}
where the sample $\phi$ may be any representation of a physical state\footnote{In fact, finding suitable and efficient updates is a crucial part in devising any useful Monte Carlo method.}. If the next state is only dependent on the current one and not on any previous states the process is called ``memoryless" and the sequence of random samples represents a so-called ``Markov chain". This is the basis of the vast majority of most conventional Monte Carlo methods (often dubbed Markov-Chain Monte Carlo (MCMC) methods).

The Langevin method can be understood as such an approach, as we can see by recasting \equref{Langevin} into the form of \equref{markov_chain}:
\begin{equation}
  \label{Eq:discrete_langevin}
  \phi_{n+1} = \phi_{n} - \frac{\delta S[\phi]}{\delta\phi}\bigg|_{\phi=\phi_n}\Delta t + \sqrt{2\Delta t}\, \eta \, .
\end{equation}
By virtue of the discussion in the previous section, we know that in the long-time limit the samples $\{\phi_n\}$ follow the desired probability distribution $P[\phi]$ from \equref{langevin_toy_expectation}. Therefore, the strategy to evaluate the random sequence is identical to the one in regular Monte Carlo approaches: after starting out from a randomly produced configuration, we let the sampling process equilibrate for a certain thermalization time (typically a few multiples of the autocorrelation time -- see below) before we start to collect samples. An unbiased estimator of the expectation value with a total number of samples $N$ is then given by
\begin{equation}
  \label{Eq:mc_average}
  \langle \CO \rangle \approx \frac{1}{N} \sum_{i=1}^{N}{\CO[\phi_i]}\, ,
\end{equation}
and the statistical uncertainty (i.e., the variance of the mean) is estimated by
\begin{equation}
  \label{Eq:mc_error}
  \sigma_L = \sigma \sqrt{\frac{1 + 2\tau_a}{N}}.
\end{equation}
Here, $\sigma$ denotes the standard deviation of the set of values $\{ \CO[\phi_i]\}$ and $\tau_a$ is the integrated autocorrelation time, which reflects the (statistical) dependence of samples over the their Langevin-time history\footnote{For any Markov chain method, estimation of $\tau_a$ might be a challenging task depending on the number of accessible samples. Techniques such as bootstrap and jackknife can be used to obtain a reliable error estimate that considers the statistical dependence of the samples. An educational summary can be found in Ref.~\cite{Troyer2012}.}. From this expression we can also deduce that the number of samples is proportional to the elapsed Langevin time and the same conclusion holds: a longer Langevin evolution will result in better statistics and hence a smaller error bar.

It is important to note that the statistical uncertainty is not the only source of error, as we have introduced a systematic bias through the discretization of the Langevin equation in \equref{Langevin}. Results must then be extrapolated to the limit of $\Delta t\to 0$~\footnote{Notably, the hybrid Monte Carlo algorithm, mentioned above as a close cousin of RL, avoids this extrapolation in fictitious time by using Metropolis accept/reject steps. This property, however, does not imply that hybrid Monte Carlo can operate at arbitrarily large step sizes.}. The discretization of the Langevin equation presented above, which corresponds to Euler integration, leads to a linear dependence on the integration step $\Delta t$. Higher-order integrators can be designed if the noise term is appropriately accounted for, as investigated within RL in Refs.~\cite{DRUMMOND1983119, HOROWITZ1987510, CATTERALL1991177} and later extended to the complex case in Ref.~\cite{Aarts2012SU3}. Within the studied model, these extensions showed great success in reducing computational cost at equal systematic bias as well as to bring finite-step dependence below the statistical uncertainty. It is noted, however, that the majority of stochastic quantization studies still rely on the linear discretization as it is often sufficient to obtain useful results at modest computational effort.
\begin{figure}[t!]
  \centering
  \includegraphics[width=\columnwidth]{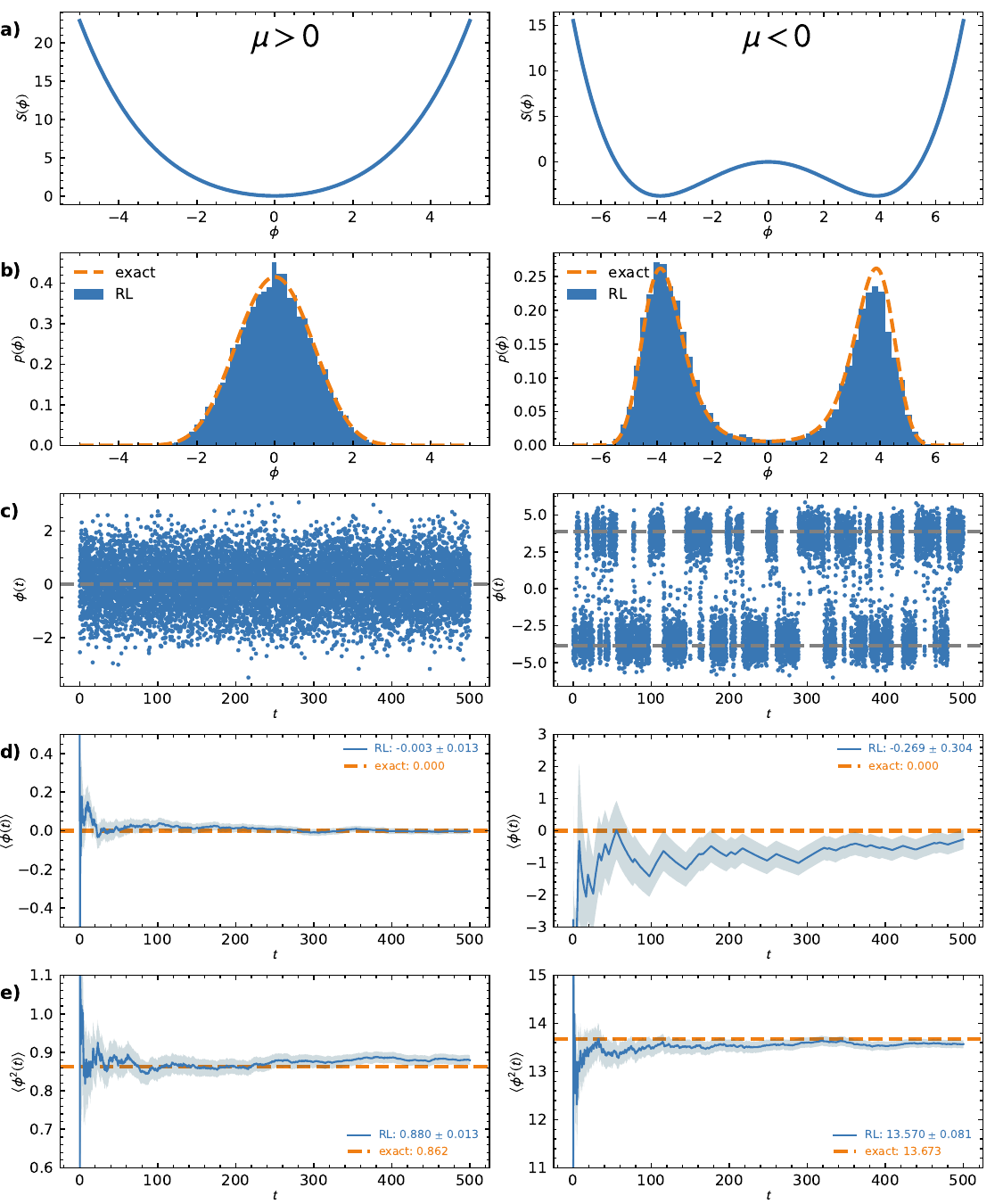}
  \caption{\label{fig:sq_toy}(a) Action as a function of the real variable $\phi$. (b) Probability distribution ${\rm e}^{-S[\phi]}$ (dashed lines) along with the sampled histogram (bins). (c) Measured values of $\phi$ as a function of Langevin time. (d) Running average of $\langle \phi \rangle$ compared to the exact solution (orange dashed line) (e) Running average of $\langle \phi^2 \rangle$ compared to the exact solution (orange dashed line).}
\end{figure}
%

\subsubsection{Toy problem I: a pedagogical example of real Langevin \label{sect:sq_toy_real}}
In order to illustrate the RL method in a concrete numerical setting, we consider a simple integral as a toy problem for a $0 + 0$-dimensional field theory (i.e., the field $\phi$ depends neither on space nor time). Note that here we are not interested in a detailed study of the specific model at hand but rather aim to investigate the behavior of the RL method in a straightforward case. Of course it is highly inefficient to use RL for the solution of this simple problem, however, the section serves as a basic example of an application of the method. Furthermore, conclusions that generalize to more involved problems can be drawn by these simple considerations.

In this case, we consider the action
\begin{equation}
  S(\phi)\ =\ \frac{\mu}{2}\phi^2\ + \ \frac{\lambda}{4!}\phi^4,
  \label{Eq:langevin_toy_continuum}
\end{equation}
with real couplings $\mu$ and $\lambda$. In the spirit of a real field theory, we will keep $\lambda$ positive and thus end up with two distinct scenarios: one in which $\mu >0$ (the single well anharmonic potential) and one in which $\mu <0$ (the double well potential).

We readily derive the discrete Langevin equation for our toy problem, according to \equref{discrete_langevin}:
\begin{equation}
  \phi_{n+1} = \phi_{n} - \left(\mu\phi_{n} + \frac{\lambda}{6}\phi_{n}^3\right)\Delta t\ +\ \sqrt{2\Delta t} \;\eta,
  \label{Eq:langevin_toy_discrete}
\end{equation}
where $\eta$ denotes a standard Gaussian white noise. In principle, this is everything needed to calculate expectation values of the form \equref{mc_average}.

In \figref{sq_toy}, a detailed analysis of two simulations at fixed $\lambda = 0.4$ and $\mu = \pm 1$ is presented (left and right columns, respectively). The second row from the top shows the histograms of the sampled field values, which should follow the distribution ${\rm e}^{-S[\phi]}$ (exact solution shown with dashed lines) in the limit of large Langevin time $t \rightarrow \infty$. While in the single-well system (left column) this is the case to a good approximation, it is apparent that the double-well scenario still suffers from a slight asymmetry. This can occur when the random process gets ``stuck" in an area of configuration space and does not easily move to another high probability area of configuration space (i.e. the other well).

This behavior can be further elucidated by the measured field values as a function of $t$ [row (c)]: on the left we see white noise centered around the expected value of $0$ while on the right we observe several correlated plateaus, corresponding to either the negative or the positive well. Ultimately, this behavior leads to a signal-to-noise issue in the calculation of the expectation value $\langle \phi \rangle$, reflected by large statistical uncertainties for the double well case [right column of row (d)].  Due to the symmetry of the problem, however, the running average for $\langle \phi^2 \rangle$ converges to the exact value relatively smoothly in both cases. Thus, by investigating one observable no profound statements can be made about a different one. While the autocorrelation of observable $A$ may be small and its statistical errors under control, observable $B$ could display erratic behavior and suffer from extremely slow convergence.

By tuning the parameters of the model, one could even study the extreme case where the two wells are separated by a barrier that cannot be surmounted by the random walk
(signaling a breakdown of ergodicity, a topic that we will return to below). In such a situation, the expectation value $\langle \phi \rangle$ would indicate that the discrete symmetry is broken, which certainly is not a physical result for our model. This reflects the problem of meta-stability of any Markov chain method, which is often very hard to detect {\it a priori}. Generally, one needs to address this issue carefully in real simulations, for example by sweeping numerical parameters in a systematic manner.

\begin{figure}[t]
  \centering
  \includegraphics[width=\columnwidth]{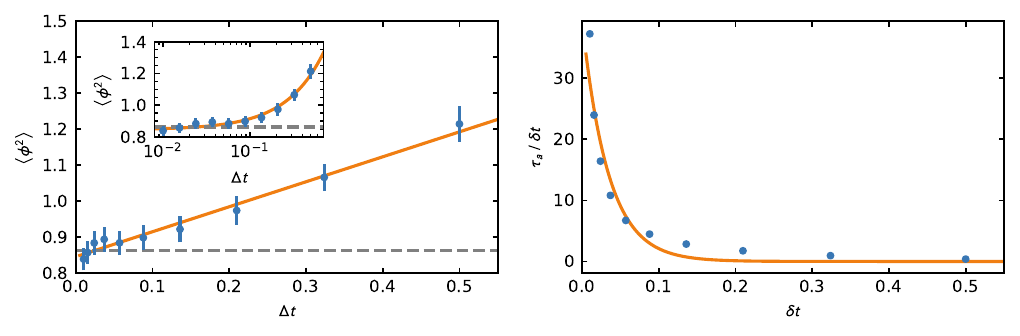}
  \caption{\label{fig:sq_stepsize} Real Langevin results for the simulation parameters $\mu=1$, $\lambda=0.4$ and a total Langevin-time of $10^3$. (Left) Symbols reflect RL results for  $\langle\phi^2\rangle$ along with the statistical errorbars as a function of integration time step $\Delta t$. The dashed line represents a linear extrapolation to the limit $\Delta t\rightarrow 0$ (horizontal dashed line). (Inset) Same data on a semi-log scale. (Right) Integrated autocorrelation time of $\phi^2$ in units of the time step as a function of $\Delta t$.}
\end{figure}
Finally, by inspecting the Langevin equation in \equref{langevin_toy_discrete} it becomes apparent that the autocorrelation time between samples $\tau_a$ should be inversely proportional to the Langevin time step $\Delta t$, i.e., statistically independent samples will be more expensive as the integration step decreases. On the other hand, a coarser integration step will yield a larger systematic error and thus a balance must be found where both the computational effort as well as the precision are within reasonable bounds.
This behavior is illustrated in \figref{sq_stepsize}, where we show the dependence for $\langle\phi^2\rangle$ as well as $\tau_a$ on the integration stepsize $\Delta t$. Indeed, we observe a systematic linear behavior of $\langle\phi^2\rangle$, which is expected due to the order of the Langevin equation (see discussion in the previous section). Further, also the integrated autocorrelation $\tau_a$ shows the expected behavior: it increases as the integration step approaches $0$.

\subsection{A practical guide to complex Langevin}

Thus far we have seen that the concept of stochastic quantization works well, given that the action $S[\phi]$ is real, i.e. when there is no sign problem. But what about the much more general and interesting case of complex-valued actions? In that case, the probability distribution in \equref{EVOprob} becomes a {\it complex} distribution
\begin{equation}
  \label{Eq:complex_p}
\rho[\phi] = \frac{{\rm e}^{-S[\phi]}}{\mathcal Z},
\end{equation}
while the field $\phi$ is still a real quantity. With such an action, a single step in the Langevin process according to \equref{discrete_langevin} would result in an imaginary component for $\phi$. At least from a practical perspective, as remarked by Parisi, ``nothing forbids to write a Langevin equation also for complex actions'' \cite{ParisiWu}.
Formal aspects aside (we return to those below), the idea is appliedthat one may perform calculations by extending the real stochastic process described above to a complex one. For a theory governed by a complex action $S[\phi]$, CL extends the target manifold of the field $\phi(x)$ to the complex plane by setting
\begin{equation}
  \phi \to \phi_{\rm R} + \i \phi_{\rm I},
\end{equation}
and analytically extending the domain of the action functional:
\begin{equation}
S[\phi] \to S[\phi_{\rm R} + \i \phi_{\rm I}].
\end{equation}
Naturally, a necessary condition for this extension to be valid is that $S[\phi]$ must be a holomorphic function of $\phi$.

The two Langevin methods -- real and complexified -- are compared side-by-side in \tabref{SQCL}.
With such an extension, the CL method proceeds very much in the same way as the real Langevin method, but now with a double system of coupled stochastic differential equations:
\bea
  \label{Eq:complex_langevin_equations}
  \Delta \phi_{\rm R} &=& K_{\rm R}\Delta t + \eta_{\rm R}(t),\\
  \Delta \phi_{\rm I} &=& K_{\rm I}\Delta t + \eta_{\rm I}(t) ,
\eea
where the real and imaginary drift functions $K_{\rm R}$ and $K_{\rm I}$ are found by taking the real and imaginary parts of the functional derivative of the complex action:
\bea
  K_{\rm R} &= -\textnormal{Re}\left[\frac{\delta S[\phi]}{\delta \phi}\right],\\
  K_{\rm I} &= -\textnormal{Im}\left[\frac{\delta S[\phi]}{\delta \phi}\right].
\eea
The real and imaginary noise obey the properties shown in \tabref{SQCL}.

It is important to note that, while the amplitudes of the real and imaginary noise terms are related, the two Wiener processes are completely independent. Additionally it should be pointed out that, beyond the complexification of each real degree of freedom, the above (coupled) Langevin processes are themselves real.
In practice, the imaginary noise is usually set to zero, which satisfies the constraints shown in \tabref{SQCL} and it has been found to have the best numerical properties~\cite{AartsPRD81054508}.

Although a rigorous derivation of this empirical ``rule'' is not available, some intuition may be gained as follows: For overly broad distributions of the imaginary noise (i.e. large $N_I$) the CL trajectory may be pushed into regions of the complex plane which are far away from the physical configuration space (which necessarily needs to be somewhat local in the imaginary directions as discussed further below). If the trajectory wanders off too far in the imaginary direction, it could take a long time to return and thus a long time to restore the proper sampling ratios required to reproduce physical results (in that sense, the strategy may be viewed as a variance reduction technique in a similar spirit to importance sampling). Eventually such a behavior will interfere with the decay of the sampled histogram in the imaginary directions, with possible implications for the formal validity of the algorithm, as already pointed out in Ref.~\cite{AartsPRD81054508}. Therefore, in order to avoid these potential complications, the noise term is typically chosen to be real, which seems to be the most convenient choice within the formal requirements.

We return to the justification for CL and related challenges after giving an illustrative example.
%
%

\begin{table*}[]
  \centering
  \begin{small}
    \begin{tabular}{@{}L{0.2\textwidth}M{0.35\textwidth}M{0.35\textwidth}M{0.02\textwidth}@{}N}\toprule
       & {\bf Real Langevin} & {\bf Complex Langevin} &\\ \toprule
     Fields & $\phi(\vec{x},\tau)$  & $\phi_{\rm R}(\vec{x},\tau) + \i\phi_{\rm I}(\vec{x},\tau)$ &\\[12pt] \hdashline
     Action & $S[\phi]$ & $S[\phi_{\rm R} + \i\phi_{\rm I}]$ &\\[12pt] \hdashline
     Field update & $\Delta\phi = K[\phi]\Delta t +\eta(t)$ &
     $\begin{array} {rcl}
       \Delta\phi_{\rm R} &=& K_{\rm R}[\phi]\Delta t +\eta_{\rm R}(t) \\
       \Delta\phi_{\rm I} &=& K_{\rm I}[\phi]\Delta t +\eta_{\rm I}(t)
     \end{array}$
     &\\[24pt] \hdashline
     Drift & $K[\phi] = -\frac{\delta S[\phi]}{\delta\phi(\vec x,\tau)}$&
     $\begin{array} {rcl}
       K_{\rm R}[\phi] &=& -{\rm Re}\left[\frac{\delta S[\phi]}{\delta\phi(\vec x,\tau)}\right] \\
       K_{\rm I}[\phi] &=& -{\rm Im}\left[\frac{\delta S[\phi]}{\delta\phi(\vec x,\tau)}\right]
     \end{array}$
     &\\[32pt] \hdashline
     Noise & $\ev{\eta}$ = 0 & $\ev{\eta_{\rm R}} = \ev{\eta_{\rm I}} = 0$&\\[10pt]
     & $\ev{\eta(t)\eta(t')} = 2\Delta\delta(t-t')$ &
     $\begin{array} {rcl}
       \ev{\eta_{\rm R}(t)\eta_{\rm R}(t')} &=& 2N_{\rm R}\Delta\delta(t-t') \\
       \ev{\eta_{\rm I}(t)\eta_{\rm I}(t')} &=& 2N_{\rm I}\Delta\delta(t-t') \\
       N_{\rm R} - N_{\rm I} &=& 1
     \end{array}$
     & &\\[36pt]
     \toprule

    \end{tabular}    
  \end{small}
  \caption{\label{tab:SQCL} Properties and expressions for the Langevin method with real-valued fields versus complex-valued fields.}
\end{table*}

%

\subsubsection{Toy problem II: a pedagogical example of complex Langevin\label{sect:sq_toy_complex}}
  In order to see the CL machinery at work, we build on the toy problem in \secref{sq_toy_real}. Indeed, from a computational standpoint, CL is largely just the Langevin process of stochastic quantization with complex variables. Of course we could just turn our toy problem into a CL problem by using complex noise but, as mentioned above, real noise has preferable numerical properties. Instead we turn our toy problem into a complex field theory which results in complex drift terms.

  The most natural route towards a complex field theory would of course be to consider complex-valued couplings $\mu$ and $\lambda$ in \equref{langevin_toy_continuum}, which in fact has been considered before, see e.g. \cite{PTPS1993CLSimulation}. However, this would amount to solving a different theory as the couplings necessarily take on different values. Alternatively, we may rewrite the above problem with a suitable Hubbard-Stratonovich transformation which merely amounts to an alternative representation of the same physical scenario (note, however, that the representation given below is by no means unique). Moreover, this is very much in the spirit of real-world problems, where a HS-transform is often used to integrate out fermionic degrees of freedom.
  To achieve this, we insert a suitable factor of $1$ into the partition function in terms of an auxiliary variable $\sigma$:
  \bea
    \CZ &=& \int_{-\infty}^{\infty} {\rm d}\phi\ {\rm e}^{-S(\phi)} \\
        &=& \sqrt{\frac{\lambda}{24\pi}}\ \int_{-\infty}^{\infty} {\rm d}\sigma\ \exp\left( -\frac{\lambda}{24} \sigma^2 \right)\
          \int_{-\infty}^{\infty} {\rm d}\phi\ \exp\left[ - \left( \frac{\mu}{2}\phi^2 + \frac{\lambda}{24}\phi^4\right) \right].
  \eea
  A shift $\sigma\rightarrow\sigma + \i\phi^2$ allows us to write
  \begin{equation}
      \mathcal{Z} = \sqrt{\frac{\lambda}{24\pi}} \int_{-\infty}^{\infty} \d\sigma \int_{-\infty}^{\infty}\d\phi\
          \exp\left[ -\left( \phi^2 \left( \frac{\mu}{2} + \frac{\i\lambda\sigma}{12} \right) + \frac{\lambda}{24}\sigma^2 \right) \right]\, ,
  \end{equation}
  and subsequently to integrate out the dependence on the old field $\phi$ (note that this is only possible in the case $\textnormal{Re}[\mu] > 0$). Ultimately we obtain the ``bosonized" version
  \begin{equation}
    \label{Eq:cl_bosonized_z}
    \mathcal{Z} = \int_{-\infty}^{\infty} \d\sigma \exp\left[ - \left( \frac{\lambda}{24}\sigma^2 - \frac{1}{2}\log\frac{\lambda}{12\mu + 2\i\lambda\sigma} \right) \right] \equiv \int_{-\infty}^{\infty} \mathrm{d}\sigma\, {\rm e}^{- S_\mathrm{B}(\sigma)}\, ,
  \end{equation}
  where we defined the ``bozonized action"
  \begin{equation}
    S_\mathrm{B}(\sigma) = \frac{\lambda}{24}\sigma^2 - \frac{1}{2}\log\frac{\lambda}{12\mu + 2\i\lambda\sigma}\, ,
  \end{equation}
  which is, by construction, a complex quantity. According to the discussion in the previous section we can still evaluate expectation values stochastically by using the complex Langevin equation
  \bea
    \label{Eq:cl_toy2_langevin}
    \sigma_{\rm R}^{n+1} &=& \sigma_{\rm R}^{n} - \Delta t\ \textnormal{Re}\left[ \frac{\lambda}{12}\sigma^n + \i\frac{\lambda}{12\mu + 2\i\lambda\sigma^n} \right] + \eta\, ,\\
    \sigma_{\rm I}^{n+1} &=& \sigma_{\rm I}^{n} - \Delta t\ \textnormal{Im}\left[ \frac{\lambda}{12}\sigma^n + \i\frac{\lambda}{12\mu + 2\i\lambda\sigma^n} \right].
    %
  \eea
  Specifically, we write for the second moment of the initial field $\phi$ in terms of the new field $\sigma$:
  \begin{equation}
    \langle \phi^2 \rangle = \left\langle \frac{6}{6\mu+\i\lambda\sigma} \right\rangle_\sigma\, ,
  \end{equation}
  where the subscript $\sigma$ denotes averaging over different realizations of $\sigma$.

\begin{figure}[t]
   \centering
   \includegraphics[width=\columnwidth]{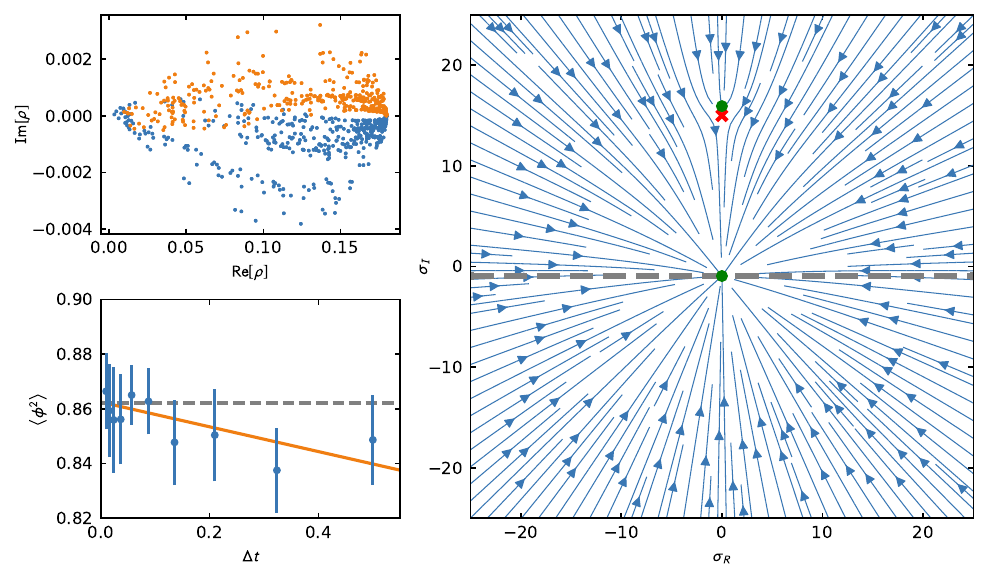}
   \caption{\label{fig:cl_stepsize}
   CL analysis for the action \equref{langevin_toy_continuum} with $\mu = 1.0$ and $\lambda = 0.4$, corresponding to the single well potential.
   (Upper left) Imaginary vs. real part of the complex weight in \equref{cl_bosonized_z}. Samples with ${\rm Re}[\sigma] > 0 $ and ${\rm Re}[\sigma] < 0$ are shown in orange and blue symbols respectively.
  (Lower left) Integration step dependence of the second moment $\langle \phi^2 \rangle$ as obtained with CL (symbols). The solid line represents a linear fit to the data in order to extrapolate to $\Delta t\to 0$ and the dashed line shows the exact result.
   (Right) Classical flow diagram with attractive fixed points (green dots) and the pole associated with the branch point of the action (red cross). The gray dashed line represents the domain of the equilibrium probability distribution.}
 \end{figure}

 From this point on, we proceed exactly as in the real case, with the exception that we now have to deal with a complex variable $\sigma$. The qualitative dependence of the numerical results on the integration step size $\Delta t$ should still be the same. This is indeed the case, as apparent from the lower left panel of \figref{cl_stepsize}, where we show CL results for the model given by \equref{langevin_toy_continuum} with parameters  $\mu = 1.0$ and $\lambda = 0.4$. We observe that the CL correctly reproduces the exact result. Interestingly the CL values show a much milder dependence on the integration step, which  is likely a consequence of the specific representation and in any way should not to be interpreted as a general result.

 It is instructive to investigate the sampled configurations by looking at the complex weight according to \equref{cl_bosonized_z}. This is shown in the upper left panel of \figref{cl_stepsize} where it becomes clear that the Langevin process now samples a complex quantity which is necessary to represent the correct answer. The coloring highlights the sampled configurations of $\sigma$ corresponding to regions with ${\rm Re}[\sigma] > 0 $ and ${\rm Re}[\sigma] < 0$.
 These two sets are connected through a point in configuration space which corresponds to a stationary point in the ``classical flow", i.e. the vanishing of the drift terms in \equref{cl_toy2_langevin}. {(The entire flow diagram is shown on the right panel of \figref{cl_stepsize}}). The attractive fixed point (marked by the lower green dot) pulls the field towards the connection point, while the noise induces fluctuations around that point, which occur mainly in the real component, as we have chosen to use only real noise in \equref{cl_toy2_langevin} (such that the only way the imaginary part can change is through the complex part of the drift). Thus, the imaginary component stays approximately constant during the Langevin evolution.

 The situation might change drastically if poles are encountered inside the domain of the distribution, which could lead to a breakdown of ergodicity, due in turn to a breakdown of holomorphicity of the action stated above (see also the discussion below). In the model considered here, we can find a pole at the point $\sigma = 6\tfrac{\mu}{\lambda}$, which corresponds to the branch-point of the action (marked by the red cross). At first glance this looks very dangerous as there is also an attractive stationary point (upper green dot) right next to the pole which would suggest faulty behavior. However, the imaginary part of the drift points away from the pole and even if a trajectory approaches this area of configuration space, fluctuations (in the real direction) will kick the process back into a stable trajectory that decays towards the attractive fixed point below. In equilibrium, the distribution will thus be confined to the gray dashed line far away from the pole, ensuring correct behavior (i.e. it is approximately shifted from the real axis by a constant offset). In fact, the existence of such an attractive fixed point is a necessary condition for the existence of an equilibrium distribution of the Langevin process \cite{CLZeroesFermionDet,Seiler2017StatusOfCL}. Luckily, this appears to be the case for systems of physical interest.


\subsection{Formal aspects and justification}

The CL process defines a random walk in a complexified manifold, such that for a given
configuration $\phi =  \phi_{\rm R} + \i\phi_{\rm I}$ there is a well-defined probability $P[\phi_{\rm R},\phi_{\rm I},t]$ at time $t$.
For a given observable $\mathcal O$, there will be an expectation value
\begin{equation}
\langle \mathcal O \rangle_{P(t)}^{} \equiv
\int \mathcal D\phi_{\rm R} \mathcal D\phi_{\rm I}\ P[\phi_{\rm R},\phi_{\rm I},t] \CO[\phi_{\rm R}+ \i \phi_{\rm I}].
\end{equation}
By virtue of the CL process, the real probability $P[\phi_{\rm R},\phi_{\rm I},t]$ obeys the FP equation:
\begin{equation}
\frac{\partial P}{\partial t} = L^{T} P \, ,
\end{equation}
where
\begin{equation}
\label{Eq:complex_langevin_operator}
L^T = \int {\rm d}\tau{\rm d}^dx\left\{ \frac{\delta}{\delta{\phi_{\rm R}}}\left [N_{\rm R} \frac{\delta}{\delta{\phi_{\rm R}}} - K_{\rm R} \right]
+ \frac{\delta}{\delta{\phi_{\rm I}}}\left [N_{\rm I} \frac{\delta}{\delta{\phi_{\rm I}}} - K_{\rm I} \right]\right\}\,.
\end{equation}
It is not obvious {\it a priori} whether this process would reproduce the desired expectation values of the physical observables,
i.e., whether $\langle \mathcal O \rangle_{P(t)}^{}$ actually corresponds to the physical expectation value of the theory,
at least in the large-$t$ limit. In fact, it is not even clear that the process would converge and if it does, whether it converges to the
correct answer. Indeed, following the steps outlined above for the case of real Langevin, one finds that the resulting FP Hamiltonian
is neither self-adjoint nor positive semidefinite, such that the proof of convergence to the desired probability distribution is spoiled.

The fundamental question underlying the validity of the CL approach is the relation between the CL distribution $P[\phi_{\rm R},\phi_{\rm I},t]$
and the desired complex distribution $\rho[\phi]$ in \equref{complex_p}. The latter defines the physics of interest and is a fixed point of its own FP equation
\begin{equation}
  \label{Eq:LTrho}
  \frac{\partial \rho}{\partial t} = L^{T}_0 \rho\, ,
\end{equation}
where
\begin{equation}
  L^T_0 = \int {\rm d}\tau{\rm d}^dx\ \frac{\delta}{\delta{\phi_{\rm R}}}\left [ \frac{\delta}{\delta{\phi_{\rm R}}} +  \frac{\delta S}{\delta{\phi_{\rm R}}}\right],
\end{equation}
which is obtained by temporal differentiation of
\begin{equation}
  \langle \mathcal O \rangle_{\rho(t)}^{}  \equiv \int \mathcal D\phi_{\rm R}\ \rho[\phi_{\rm R},t] \CO[\phi_{\rm R}].
\end{equation}
Again, the absence of boundary terms at infinity when integrating by parts was assumed. More specifically, the crucial question is whether
\begin{equation}
  \label{Eq:realcomplexprobabilities}
  \langle \mathcal O \rangle_{P(t)}^{}  = \langle \mathcal O \rangle_{\rho(t)}^{}\, ,
\end{equation}
holds. Stated explicitly, the question arises whether the expectation value of a given observable with respect to the complex measure $\rho[\phi_{\rm R}]$ may be obtained by instead sampling the positive semi-definite probability measure $P[\phi_{\rm R} +\i\phi_{\rm I}]$.

In Refs.~\cite{AartsPRD81054508, 2011EurPhysJC711756} it was shown how the desired relationship \equref{realcomplexprobabilities} can be proven for holomorphic observables, as long as the action and the associated drift are holomorphic functions of $\phi$. The proof relies on analyzing the behavior of
\begin{equation}
  \label{Eq:interpolating_function}
  F(t,\tau) = \int \mathcal D \phi_{\rm R} \mathcal D\phi_{\rm I}\ P[\phi_{\rm R}, \phi_{\rm I},t-\tau]\CO [\phi_{\rm R} + \i\phi_{\rm I},\tau],
\end{equation}
where $0 \leq \tau \leq t$. The function $F(t, \tau)$ interpolates between the two expectation values of interest:
\bea
F(t,0) = \langle \mathcal O \rangle_{P(t)}^{}, &  &  F(t,t) = \langle \mathcal O\rangle_{\rho(t)}^{}.
\eea
where we have assumed that the initial conditions are chosen as
\begin{equation}
  \label{Eq:InitialConditions}
  P (\phi_{\rm R}, \phi_{\rm I}, 0) = \rho[\phi_{\rm R}, 0]\,\delta(\phi_{\rm I} - \phi_{{\rm I},0}).
\end{equation}
We find that
\begin{equation}
  F(t,0) = \int \mathcal D \phi_{\rm R} \mathcal D\phi_{\rm I}\ P[\phi_{\rm R}, \phi_{\rm I}, t]\CO [\phi_{\rm R} + \i\phi_{\rm I},0] = \langle \mathcal O \rangle_{P(t)}^{},
\end{equation}
while, using the initial conditions of \equref{InitialConditions},
\begin{equation}
  F(t,t) = \int \mathcal D \phi_{\rm R} \mathcal D\phi_{\rm I}\ P[\phi_{\rm R}, \phi_{\rm I},0] \CO [\phi_{\rm R} + \i\phi_{\rm I},t] =
  \int \mathcal D \phi_{\rm R}\ \rho[\phi_{\rm R}, 0]\CO [\phi_{\rm R} + \i\phi_{{\rm I},0},t] =  \langle \mathcal O \rangle_{\rho(t)}^{},
\end{equation}
where we have used \equref{LTrho} to shift the Langevin evolution operator from $\mathcal O$ to $\rho$ by transposition
(which involves integration by parts).

Now the crucial observation is that if $F(t,\tau)$ is independent of $\tau$, then \equref{realcomplexprobabilities} holds, and
the Langevin method is formally shown to be valid for complex-valued variables, i.e. to converge
to the correct physical answers (assuming it converges). Naturally, this statement assumes that the expectation values in \equref{realcomplexprobabilities} agree at $t=0$, which can be ensured by choosing the initial condition of the Langevin process as in~\equref{InitialConditions}. The $\tau$ derivative of $F(t,\tau)$ again involves an integration by parts:
\begin{align}
\label{Eq:CL_IBP}
  \frac{\partial}{\partial \tau} F(t,\tau) =
  \int \mathcal D \phi_{\rm R} \mathcal D\phi_{\rm I}\ \bigg\{ &P[\phi_{\rm R}, \phi_{\rm I},t-\tau]  L \CO [\phi_{\rm R} + \i\phi_{\rm I},\tau]- \nonumber \\
  & \qquad\qquad L^{T} P[\phi_{\rm R}, \phi_{\rm I},t-\tau]\CO [\phi_{\rm R} + \i\phi_{\rm I},\tau]\bigg\} ,
\end{align}
where $L$ is the Langevin operator and $L^{T}$ its adjoint, as defined in~\equref{complex_langevin_operator}. If integration by parts is carried out and -- importantly --
the boundary terms are zero, then $\frac{\partial}{\partial \tau} F(t,\tau) = 0$. If the decay of
\begin{equation}
  \label{Eq:product_decay}
  P[\phi_{\rm R}, \phi_{\rm I},t-\tau] \CO [\phi_{\rm R} + \i\phi_{\rm I},\tau] \, ,
\end{equation}
and its derivatives is not fast enough to ensure that the boundary terms will vanish, then it cannot be guaranteed that the expectation values of the quantities of interest obtained via a Langevin process will converge to the correct
values~\cite{AartsPRD81054508, 2011EurPhysJC711756}. It is this property that ultimately determines the applicability of CL to a given theory.

While the condition of fast decay of \equref{product_decay} was recognized in~\cite{AartsPRD81054508, 2011EurPhysJC711756}, the precise rate was not immediately clear. In Ref.~\cite{CLJustificationPRD94114515}, the above arguments were reviewed by considering a finite step-size in Langevin time. It was then found that the above integration by parts is valid if the probability distribution of the drift term falls off faster than any power at large drift magnitude.
In practice, it is very difficult to establish the behavior of \equref{CL_IBP}, but it is perfectly possible to study the probability distribution of the drift and establish whether the decay is exponential.

\subsubsection{Practical monitoring of the boundary terms}
In practice, it would be useful to have an accessible diagnostic device which tells us whether the results of the CL evolution are trustworthy. Such a tool may be derived by re-examining the interpolating function in \equref{interpolating_function}, which must be independent of the interpolating time $\tau$ in order for the boundary terms to vanish. It has been found~\cite{2011EurPhysJC711756} that the maximal violation, if there is any, of this condition occurs at $\tau = 0$. Thus, we focus on the more straightforwardly accessible quantity
\begin{equation}
  \label{Eq:simplified_constraint}
  \frac{\partial}{\partial \tau} F(t,\tau)\bigg|_{\tau=0} = 0,
\end{equation}
which is weaker than the expression without the restriction to $\tau = 0$, however, still a sufficient criterion to ensure correctness~(up to some technical conditions, as shown in Ref.~\cite{2011EurPhysJC711756}).

Evaluating~\equref{CL_IBP} at $\tau = 0$ we observe that the first term vanishes, since $L^T P[\phi_{\rm R},\phi_{\rm I},t=\infty] = 0$ by virtue of the Fokker-Planck equation. Therefore, only the second term contributes and we finally arrive at the so-called consistency conditions
\begin{equation}
  \label{Eq:consistency_condition}
  \langle{\tilde{L}\CO}\rangle = 0,
\end{equation}
with the corresponding Langevin operator\footnote{
Note that this operator agrees with $L^{T}$ defined in \equref{complex_langevin_operator} for holomorphic observables because of the Cauchy-Riemann equations.}
\begin{equation}
  \tilde{L}^T = \int {\rm d}\tau{\rm d}^dz\ \frac{\delta}{\delta\phi}
  \left[ \frac{\delta}{\delta\phi} +  \frac{\delta S}{\delta{\phi}}\right].
\end{equation}
These conditions need to be fulfilled in order for the complex Langevin process to be correct. Technically, \equref{consistency_condition} represents an infinite tower of identities, since the condition would need to hold for all observables. However, it was argued that the check for a finite number of observables suffices in order to establish the validity of the results~\cite{AartsPRD81054508}.

In fact, for practical purposes, it was realized that a slight re-definition of the interpolating function is more useful. Following Ref.~\cite{PhysRevD.99.014512} we introduce a cutoff $Y$ in the imaginary direction, such that~\equref{interpolating_function} reads
\begin{equation}
  F(Y;t,\tau) = \int_{|\phi_{\rm I}| \le Y} \mathcal D \phi_{\rm R} \mathcal D\phi_{\rm I}\ P[\phi_{\rm R}, \phi_{\rm I},t-\tau]\CO [\phi_{\rm R} + \i\phi_{\rm I},\tau],
\end{equation}
and the original interpolating function $F(t,\tau)$ is recovered in the limit $Y\to\infty$. Inserting this expression into the condition~\equref{simplified_constraint} finally yields
\begin{align}
  \label{Eq:boundary_term_cutoff}
  \frac{\partial}{\partial \tau} F(Y;t,\tau)\bigg|_{\tau=0} \equiv B_{\CO}(Y,t,\tau)\bigg|_{\tau=0}
  =  \int\CD\phi_{\rm R}\ \bigg\{
    &K_{\rm I}[\phi_{\rm R},Y]P[\phi_{\rm R},Y,t]\CO[\phi_{\rm R}+\i Y,0] -\nonumber \\
    &\qquad K_{\rm I}[\phi_{\rm R},-Y]P[\phi_{\rm R},-Y,t]\CO[\phi_{\rm R}-\i Y,0]
  \bigg\}
\end{align}
where the temporal evolution of the observables $\tfrac{\partial}{\partial\tau} \CO[\phi,\tau] = L_0O[\phi,\tau]$ has been inserted.\footnote{For the sake of simplicity, it was assumed that the integration along the real direction is benign and boundary terms are absent. In order to accommodate potential issues, a cutoff in the real direction may also be introduced, see, e.g.,~\cite{aarts2013}. An explicit derivation may be found in the appendix of~\cite{scherzer2018}.} The form of this quantity also makes immediately apparent that a correct CL simulation requires a sufficient decay of the product $K_{\rm I}P\CO$ in the imaginary direction.

Just like with regular observables, we must evaluate~\equref{boundary_term_cutoff} in the limit $t\to\infty$, such that the relevant boundary term is given by $B_{\CO}(Y) \equiv \lim_{t\to\infty} B_{\CO}(Y,t,0)$. Note, that this evaluation itself does not require separate simulations at fixed values of the cutoff, but merely a certain post-processing of the sampled data. In practice, however, it may be challenging to assess the value of $B_{\CO}(Y)$ for large cutoffs due to excessive noise. Therefore, it is convenient to investigate the boundary term as a function of $Y$, which typically allows one to identify a converged plateau region before the growing influence of the noise makes an extrapolation $Y\to\infty$ impossible. The value of the plateau may then be interpreted as the value of the boundary terms, which, if consistent with zero, marks the validtiy of the CL results. For non-zero values, the results suffer from incorrect convergence. Interestingly, the magnitude of the boundary term may be used to assess the systematic error of the CL values, as recently exploited in Refs.~\cite{Scherzer2020,Scherzer2020b}.

Finally, the issue of correct convergence and boundary terms was very recently independently revisited in Ref.~\cite{cai2020validity}, where the understanding of these shortcomings was put on more solid mathematical footing.

\subsection{\label{sect:CLchallenges} Challenges and some solutions}
Although our understanding of CL has progressed considerably during the past few years, the approach still faces some challenges that remain to be fully understood. These can be roughly divided into two kinds: mathematical and practical, which naturally have some overlap. In this subsection we attempt to summarize the current understanding of these issues.

\subsubsection{Mathematical shortcomings}
Without a doubt, the biggest challenge for CL is the lack of general mathematical proofs, the previous section notwithstanding. More specifically, as pointed out most recently in Ref.~\cite{Seiler2017StatusOfCL}, it remains unknown whether the Langevin operators defined above as $L$, $L_0$, $L^T$, $L_0^T$ properly define unique stochastic processes, although in practice this is not typically an issue. Crucially, it remains unknown under what conditions the positive measure $P[\phi_{R}, \phi_{I},t]$ converges to an equilibrium measure, although (again) there is substantial numerical evidence that such an equilibrium
measure exists in many cases of interest.

\subsubsection{Instabilities}
One of the issues recognized early on (in fact, since the 1980s; see e.g.~\cite{Ambjorn1985,Ambjorn1986}) is the appearance of instabilities in the form of runaway trajectories along the CL evolution. These can become frequent enough to completely spoil a calculation performed at fixed step size. Only with the availability of increased computational resources was it realized that too coarse an integration step would yield problematic trajectories. Finally, in Ref.~\cite{AARTS2010154}, the need for adaptive step size integration of the complex Langevin equations was identified. It was found that such an approach provides a full solution to the problem of instabilities arising from the coarse integration step, by moderating the change between subsequently sampled configurations. In fact, this strategy has become standard in the field and virtually all modern applications of CL in both relativistic and non-relativistic physics implement the adaptive step (which is possible in several ways, see, e.g., \cite{AARTS2010154}).

\subsubsection{Convergence to incorrect values}
The above issues aside, CL has been shown to fail in certain cases (due either to failure to converge or convergence to the wrong answer), but also appears to work in scenarios that lie outside the holomorphic-action regime mentioned above. In cases of failure, the behavior has been traced back to insufficient {\it decay at infinity}, or to a breakdown of ergodicity due to {\it poles in the action} (which should be expected in fermionic systems as the fermion determinant will vanish at specific points, thus leading to meromorphic drifts).


\paragraph{Boundaries at infinity}
As outlined above, the behavior of boundaries at infinity is a relevant question, and a possible failure to meet this condition plagues models in both relativistic and non-relativistic physics. In particular, for gauge theories the complexification of the link variables leads to non-compact groups, e.g. SU(3) becomes SL(3,C). As we explain in \secref{NRQFT}, a similar effect is seen in nonrelativistic physics when using compact HS transformations. In either case, merely assuming that the derivative of $F(t,\tau)$ in \equref{CL_IBP} vanishes is a bad idea. For that to happen, the solutions to the FP equation must fall off sufficiently quickly along non-compact directions in the (complexified) space of field configurations (see in particular Refs.~\cite{PRDSalcedoCL2016, PhysRevD.99.014512} for a recent and insightful discussion of an exactly solvable case). That property is very difficult to determine {\it a priori}, but can be checked {\it a posteriori} following the arguments of Ref.~\cite{CLJustificationPRD94114515}.
Case studies show that in many instances, while the solutions fall off faster than exponentially in the real directions, the decay in the imaginary directions may be insufficient~\cite{AartsPRD81054508}. The prime source of the non-vanishing boundary terms are unphysical excursions far away from the real line, dubbed ``the excursion problem." Below we present several practical ways to mitigate such a problematic behavior.



\paragraph{Poles, ergodicity and singular drift}
Besides the behavior near the boundary at large imaginary direction, another issue which could potentially spoil the formal justification of the CL method is the non-analyticity of the action. This is intimately connected to the occurrence of zeros of the measure, due to the relation
\begin{equation}
  S[\phi] = \e^{-\ln P[\phi]},
\end{equation}
such that the action diverges when $P[\phi]$ vanishes. Such potentially problematic points appear, for instance, for fermionic actions as zeroes of the determinant of the fermion matrix. Naively, one could argue that the problematic point is never sampled since it has zero probability measure. However, as with the boundaries at infinity, the success of a CL simulation crucially depends on the behavior of the probability distribution as the poles are approached. Formally, this can be elucidated by manually cutting out a region around the singularity, thereby extending the boundary of the integration domain. This procedure is justified as long as the probability measure vanishes around those poles sufficiently fast; in other words, one has to address the boundaries around the poles potentially alongside those at infinity mentioned above~\cite{CLZeroesFermionDet,aarts2016b,aarts2016c}.

A detailed study of incorrect convergence due to poles in the drift function showed that the location of these poles, the decay of the drift function, and the behavior of the observables in the region near the poles all played a role in whether the method would return correct results~\cite{CLZeroesFermionDet, CLJustificationPRD94114515}. To be precise, there are three distinct scenarios, depending on the location of the poles with respect to the support of the probability measure:
\begin{description}
  \item {\it Poles on the outside -- } These are unproblematic, as the influence of these singularities will be lost after thermalization such that the equilibrium distribution is not affected. Note that this is also the case for the toy problem discussed in \secref{sq_toy_real} (visible as the red dot in \figref{cl_stepsize}, whereas the equilibrium distribution lies near the gray dashed line)
  \item {\it Poles at the edge -- } This type of singularity may potentially spoil the correctness due to a singular drift and therefore is a violation of the above derived rules for the decay of the histograms. Indeed, a careful investigation of several toy problems has elucidated such a behavior~\cite{CLZeroesFermionDet}.
  \item {\it Poles on the inside -- } These are potentially problematic in two ways: As in the above case, a singular drift probem could appear, however, the additional complication is given by the formation of bottlenecks, which arise because of the vanishing probability distribution at these points~\cite{Seiler2017StatusOfCL}. As a consequence, the equilibrium distribution separates into disjoint patches and the Langevin process may fail to tunnel between them, resulting in a correct but restricted sampling of the configuration space and therefore a failure of ergodicity. If the relative weight of the avoided regions is small, the associated bias might be indiscernible and useful values may be obtained for practical purposes. Conversely, if the contributions of the avoided configurations may not be neglected, the CL process converges to erroneous values. Such a behavior is closely related to the concept of meta-stability in conventional Monte Carlo approaches.
\end{description}

For simple toy models, some of the problems connected to the occurence of zeroes of the measure can be inferred from the study of so-called classical flow-patterns, as also shown in the toy model studied above. For real-world applications, on the other hand, these are impossible to generate such that other diagnostic tools are necessary.

As mentioned above, a re-examination of the conditions for correctness in Refs.~\cite{shimasaki2016,PhysRevD.92.011501,CLJustificationPRD94114515} revealed that a failure of CL in some cases has been attributed to the excursion and singular drift problems are actually due to the drift function falling off too slowly. This motivates the drift histogram as a diagnostic tool and it has indeed been shown to be a useful indicator for wrongfully converged values~\cite{Nagata2018b}. In fact, this type of analysis covers both types of boundary terms, the ones at infinity and around poles, in the same manner. It should be noted, however, that the observation of the drift histogram alone may be insufficient to distinguish the validity of multiple observables, as recently discussed in the context of the XY-model in Ref.~\cite{Scherzer2020}.

Finally, it was argued in Ref.~\cite{Hayata:2015lzj}, using a semiclassical analysis, that when more than one saddle point in the complex plane contributes to the ensemble averages, the CL method can lead to incorrect answers due to the different complex phases associated with each saddle point. The interference of these complex phases is essential in phenomena such as the Silver Blaze phenomenon and real-time dynamics.

\subsubsection{Practical solutions}
The preceding discussion lays out the possible issues of a CL simulation with the tentative takeaway that certain theories seem to be intractable with this method. On the practical side, several practical strategies have been devised to enable CL simulations in otherwise problematic models and parameter regimes. These ideas try to either prevent uncontrolled excursions in non-compact complex direction which are ultimately responsible for the violation of the criteria for correctness, try to circumvent the issues of a diverging drift term, or a combination of both. In the following we provide an overview on the most successful strategies in this regard.

\begin{description}

\item {\it Gauge cooling (GC) -- } By now, this is a standard technique to prevent long excursions of the CL trajectory in the complex plane when simulating gauge theories~\cite{SeilerGaugeCooling,Bongiovanni:2013nxa,Aarts2013b,PhysRevD.92.085020,Nagata:2015uga,Nagata:2016alq,Zhenning2020,dong2020alternating}.
The underlying idea is that at each Langevin step, one can make a gauge transformation to keep the link variables, which live in the unbounded complexified manifold, close to the compact subgroup without interfering with the correctness of the simulation.
For full QCD simulations, for instance, gauge transformations are performed in the extended non-compact group $SL(3,C)$ such that the sampled configurations are closed to the original compact $SU(3)$ manifold, measured by the unitary norm.
Among all the strategies to mitigate the excursion problem, GC represents the best understood approach from a mathematical perspective and is, in fact, mathematically exact~\cite{Nagata:2015uga}. Finally, it is worthwhile to note that GC has been found to be necessary, however, sometimes not sufficient to bring CL simulations of gauge theories under control.

\item {\it Dynamic stabilization (DS) -- } This approach was developed to further aid with
the excursion problem~\cite{Aarts:2016qhx, Lattice2016AttanasioJager, Lattice2017AttanasioJager, attanasio2018, Attanasio2019}.
The essential idea is to add a term to the Langevin drift $K[\phi]$ in the schematic form
\begin{equation}
K[\phi] \to K[\phi] + \i \alpha_{\rm DS} M,
\end{equation}
where $\alpha_{\rm DS}$ is a control parameter and $M$ acts only in the non-$SU(3)$ directions. The latter is chosen such that it grows rapidly with the distance from the $SU(3)$ manifold, thereby suppressing contributions with large deviations from the target manifold. Although this modification to the Langevin evolution cannot be derived from the action itself, it can be shown to vanish in the continuum limit.

\item {\it Regulators -- } The idea behind a regulator, similar to the above presented strategies, is to suppress terms far from the original configuration space. The approach is similar in nature to DS and first appeared in a nonrelativistic application, namely Ref.~\cite{PRD95094502}, and was further discussed more recently in Ref.~\cite{PhysRevD.99.014512}.

As opposed to DS, the resulting modifications on the Langevin equations are directly derived from the action, which is modified by a term of the form
\begin{equation}
  S[\phi] \to S[\phi] + \int\d\tau\d^dx\ \xi \phi(x,\tau)^2,
\end{equation}
which effectively adds a harmonic oscillator trapping potential, i.e. a restoring force that prevents the field from running away.\footnote{In a relativistic setting, the modification of the action may be understood as a a mass-like term for the auxiliary field $\phi$.} The modification, strictly speaking, amounts to solving a different theory and therefore introduces a systematic bias in the computed expectation values. To remove this bias, it is necessary to either show that the results are insensitive to the regulator strength $\xi$ or, equivalently, extrapolate $\xi \to 0$. Stated differently, the idea is to simulate a series of well-behaved systems with decreasing deviation from the physical system of interest in order to extrapolate to the physically correct point.

While the advantages of such a practical solution are clear, it is by no means a full solution and
in many cases -- especially at strong coupling or low temperatures -- it is not possible to make $\xi$ small and obtain a converging calculation. Nevertheless, it has been found that this strategy is essential for simulations of non-relativistic fermions via CL and it is by now a standard approach in this context (see~\secref{fermions_ft_1d} for a detailed discussion).

\item{\it Deformation technique -- } In order to mitigate the occurrence of singular drift values, a viable option is to deform the action by adding a fermion bilinear term to the action~\cite{Ito2016,Ito2018b,2018EPJWC17507017N}, ideally in such a way that the vacuum of the system is minimally affected~\cite{Ito2016}. For a suitable choice of this term, the effect is to suppress (near-)zero eigenvalues of the fermion matrix (in this context also often referred to as the Dirac operator) which are ultimately responsible for exceedingly large drift values~\cite{Splittorff2015,Ichihara2016}. In analogy to the regulator prescription discussed above, it is necessary to extrapolate the coefficient of this additional coupling to zero to obtain results for the physical theory of interest. For gauge theories, this strategy is typically used in combination with GC in order to achieve stable and correct CL simulations, most recently for a full QCD simulation~\cite{2018EPJWC17507017N}. For non-relativistic theories, this approach remains unexplored to the best of our knowledge.

\end{description}

\subsubsection{Schwinger-Dyson representation}
Quantum field theories can be cast in terms of differential equations rather than path integrals; this is called the Schwinger action
principle. A specific set of differential equations resulting from this principle is the Schwinger-Dyson equations, and these involve
variations with respect to the source fields. Of the many solutions to this set of differential equations, only one at most corresponds
to the Feynman path integral.  It appears that the CL method is able to converge to multiple stationary distributions, and while at
most one of those distributions corresponds to the desired distribution, it has been shown that these stationary distributions of the
complex Langevin equation satisfy the Schwinger-Dyson equations~\cite{PhysRevD75045007, xue1986}.

The initial conditions appear to play a role in this behavior: CL converges to certain solutions of the Schwinger-Dyson equation with
one set of initial conditions and to another with a different set, and for certain conditions CL does not converge at all
~\cite{GuralnikNPB200905503213, Salcedo1993SpuriousCLSolutions}. The role of the Schwinger-Dyson equations in the presence of
zeros of the complex density $\rho$ was analyzed in detail in Ref.~\cite{Salcedo:2018fvt}. Further study of the convergence of CL to incorrect solutions
must be an integral part of the understanding and application of this method to systems with complex actions.

%
%
%
%
%
%
%


\section{\label{sect:RQFT}Applications in relativistic physics: from toy models to finite density QCD}
As already mentioned above, the first applications of CL were undertaken in the context of lattice gauge theories in the 1980s. Following several successful applications in this regard \cite{PhysRevLett.55.2242,Flower1986,Ambjorn1986,Haymaker1988} the initial wave of excitement simmered down due to observation of incorrect convergence and a poor understanding of the causes. After a period of reduced activity, the CL method resurfaced in the 2000s in the context of relativistic physics (see Refs.~\cite{PhysRevLett95202003,PhysRevD75045007,JHEP200809018,AartsPRL102131601,JHEP200905052,Lattice2012Aarts}).
Motivated by the revived interest, much work has gone into the testing of the methodology, its capabilities, and its limitations. While CL has successfully circumvented the sign problem in a few key areas, it still suffers from some of the problems discussed in \secref{formalism}. As a result, recent emphasis has been on determining regions of applicability and adjusting the method in order to prevent uncontrolled excursions of the Langevin evolution into the complex plane and ensure convergence to the correct solution.

While the main focus of this review is on applications of CL to nonrelativistic quantum many-body problems, such an overview would be incomplete without accounting for the developments that enabled these advances. Therefore, we provide a brief summary of selected applications of CL to relativistic field theories. For a more in-depth overview on the use of CL in relativistic physics, we refer the reader to Refs.~\cite{Lattice2012Aarts,aarts2014,Seiler2017StatusOfCL,attanasio2020b} for various status reports.

\subsection{QCD-inspired toy models}
In Lattice QCD, the main promise of CL is its potential to explore regions of the QCD phase diagram which are currently inaccessible due to a sign problem. One of the primary goals is to reliably treat the region of non-zero quark chemical potential (see, e.g., Refs.~\cite{Muroya2003,forcrand2010,Aarts2015,Aarts2016}), which eludes efficient treatment with conventional methods due to a severe complex phase problem.
In addition, theories with coupling to a topological charge (see Ref.~\cite{Seiler2017StatusOfCL}) or computation in Minkowski space to obtain real-time dynamics~\cite{PhysRevLett95202003,PhysRevD75045007,
BERGES2008306,deAguiar:2010ue} are also of great interest.

While the CL method has not yet been able to produce detailed solutions for these problems (some attempts notwithstanding, see below), a number of simpler models, which partly contain the phenomenology of QCD in certain limits, have been successfully studied. The findings in these reduced models allow deep insights into the behavior of the CL method in these limits and can help elucidate where problems may arise with such a treatment. In the following we intend to highlight some of these advances.

\subsubsection {\it XY model}
The XY model is an extremely useful benchmark case for CL as the sign problem can be circumvented in other ways, for example by using imaginary asymmetries or dual variables in a world-line formulation.. Results from both of these alternate approaches have been compared to calculations done with CL for the three-dimensional XY model at non-zero chemical potential~\cite{aarts2010,JHEP20100820}.
The CL results turn out to be very promising for the ordered phase at low temperatures but fail to reproduce known answers for the disordered (high-temperature) phase. From this failure, first criteria for the correctness of CL results have been extracted, as already discussed above. Similar conclusions on the applicability of the CL method have been reached in a recent study for the $O(3)$ model~\cite{Katz2017}.

Moreover, the model is very sensitive to instabilities in the algorithm -- just as heavy-dense QCD is, see also below.
The instabilities in the algorithm can be eliminated by the addition of adaptive stepsize to the Langevin evolution, which was first shown to work in the three-dimensional XY model at non-zero chemical potential. As remarked above, the adaptive step size mitigates the issue of unstable CL trajectories by preventing overly large changes in the Markov process.
Another way to improve the convergence of CL in the XY model is by dynamic stabilization, which forces the Langevin trajectory to remain near the real axis by means of unitary transformations~\cite{Lattice2017AttanasioJager}. These transformations modify the drift function to prevent large excursions in the imaginary direction~\cite{Seiler2017StatusOfCL,Lattice2016AttanasioJager}.

\subsubsection {Polyakov and SU(3) spin models}
Polyakov models at non-zero density were among the first theories under investigation with CL, and their exploration has led to methodological breakthroughs.
Most notably, it was realized that the complexification of the group elements led to an enlarged manifold that needs to be constrained via gauge cooling in order to prevent unstable trajectories~\cite{SeilerGaugeCooling}. This strategy is key to studying more sophisticated theories.

\begin{figure}
  \includegraphics[width=0.95\textwidth]{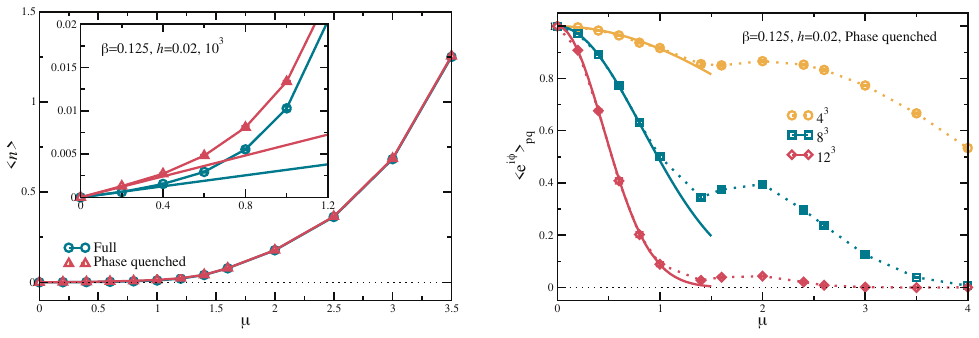}
  \caption{(Left) SU(3) spin-model in the full and the phase-quenched theory as a function of $\mu$. The inset shows a close-up of the small $\mu$ region. The lines are the predicted linear dependence for small $\mu$ in excellent agreement with the CL results. (Right) Average phase factor as a function of $\mu$ for the SU(3) model, highlighting the severeness of the sign problem that the CL is able to solve. See Ref.~\cite{Aarts2012SU3}.}
  \label{fig:su3_plots}
\end{figure}

The three-dimensional SU(3) spin model is an effective Polyakov loop model for QCD at nonzero temperature and density. It suffers from a sign problem at nonzero chemical potential and typically reweighting or the phase-quenched approximation are used to study its dynamics. Complex Langevin was used to successfully address this model, originally in Refs.~\cite{PhysRevLett.55.2242,Bilic1988} and more recently in Refs.~\cite{Aarts2012SU3,Aarts:2012ft}. The latter studies found remarkable accuracy across a phase transition. Some of these results are shown in \figref{su3_plots} along with the average phase factor in the phase-quenched theory, which vanishes quickly with increasing system size. The success of CL within this model, despite the smallness of the average phase, underlines its ability to solve severe sign problems.
Moreover, a comparison with the 3D XY model, for which CL fails for such a phase transition, sheds light on the justification of the approach and sets the stage for the derivation of further criteria of correctness.

Later, CL results for the SU(3) spin-model were also compared with the relative weights method across a wide range of chemical potentials~\cite{Greensite2014,greensite2014b}. While values for the densities agreed perfectly in all cases, results for the Polyakov loop displayed some discrepancies at large chemical potentials. The shortcoming that was brought in connection with the appearance of a logarithmic term (and hence a gauge cooling) in the action, as already encountered in~\cite{Mollgaard2014}.
branch cut

\subsubsection {Random matrix theory (RMT)}
Random matrix theory has been proven generally useful for the analysis of QCD~\cite{Verbaarschot:2000dy}. Perhaps most importantly from a CL perspective, just like full QCD, RMT models are ``constructed" with complex fermion determinants and both exhibit a severe sign problem for small masses. As opposed to QCD, however, analytic solutions for RMT models are available, making them a valuable playground to assess the ability of CL to deal with complex-valued fermion determinants.

Some work has been done for models without a finite-density transition~\cite{PRD201311116007}, where it was found that CL works well for large quark masses but incorrect convergence occurs for smaller values of the mass. The failure was linked to CL paths that wind around the origin of the complex plane, therefore encountering the ambiguity in the logarithmic part of the action, i.e., a branch cut. The latter criterion was initially believed to be a general marker of incorrectly converged CL results and indeed had some success in explaining the failure encountered in other models. However, a later study found that CL also fails to deliver reliable results for cases where such a point is on the edge of the domain of the distribution function where no full winding around the pathological point is possible~\cite{CLZeroesFermionDet}. In a follow-up study, CL was shown to be able to fully capture the chiral limit of RMT which was achieved by shifting the integration variables and changing to a polar representation~\cite{Mollgaard2014}.

In addition, an extended gauge cooling procedure was proposed and shown to deliver exact results in the regime of light quark masses~\cite{Nagata:2016alq,Nagata2016b}. Here, the general idea is to avoid small eigenvalues of the Dirac operator which could possibly spoil the correctnes of CL~\cite{Splittorff2015,Ichihara2016}.

Moreover, models with finite-density phase transition, where many problems with CL may arise, has been addressed. A
recent work on one of these models suggests that with suitably designed reweighting methods, CL is able to fully reproduce the known analytical solution of the model at the finite-density phase transition~\cite{2017EPJWC13707030B}.

\subsubsection{Thirring model}
The Thirring model model constitutes one of the rare cases of an exactly solvable quantum field theory. CL studies of this model have been conducted in the $0+1$ dimensional case where it was found that CL is able to reproduce the exact results in the weak coupling regime~\cite{Pawlowski:2013pje}.
At larger couplings the method is only able to reproduce exact results at small and large values of the chemical potential but fails for intermediate $\mu$. Similar findings have been reported for the $2+1$ dimensional scenario~\cite{Pawlowski:2013gag}. A later comparison (for both dimensions) with the fermion-bag method (see \secref{fermion_bags}), underscored this behavior and showed that CL does not perform sufficiently well in this parameter regime~\cite{li2016, li2016b}.

In an independent study of the $0+1$ dimensional Thirring model, essentially the same behavior was observed and it was found that the CL trajectories were sampled in the vicinity of zeroes of the fermion determinant~\cite{fujii2017}. Consequently, large values of the drift occur in such studies and the distributions of the drift term are no longer localized, thereby violating certain criteria for correctness. A solution in terms of reweighting and deformation was achieved. However, the scaling of these ``remedies" is exponenphase diagramtial in system size.

\subsubsection{Heavy-dense QCD (HDQCD) and hopping parameter expansion}
A very prominent simplification of QCD is the limit of large quark masses and large densities, called heavy-dense QCD (HDQCD).
In this limit, the kinetic term of the quarks is suppressed to the point that they become essentially static. In this sense, the limit may also be
used as a starting point for an expansion in the hopping parameter, which approaches the dynamic limit with increasing order, see below.

Studies of HDQCD have been among the first objectives since CL was revived~\cite{JHEP200809018} and have led to profound insights and methodological advances, most notably in the light of adaptive step-size control~\cite{AARTS2010154} and the gauge cooling procedure~\cite{SeilerGaugeCooling}. These advances made it possible to map out the phase diagram in the full $T-\mu$ plane at several values of the inverse coupling $\beta$, and excellent agreement with known results was found in the large-$\beta$ regime~\cite{aarts2014d} (this includes, for instance, the Silver Blaze behavior that was observed to sharpen with decreasing temperature).
More recent studies have refined these findings~\cite{Aarts:2016qrv,Attanasio2016,aarts2016e,Aarts:2013nja,POSInsightsintoHDQCDusingCL}. However, it was also found that for $\beta \lesssim 5.9$, instabilities emerged where the histograms of the drift term developed skirts and sometimes even shifted the mean. These shortcomings were accompanied by an increasing unitary norm (i.e., the trajectories were far away from the $SU(3)$ submanifold), despite the use of gauge cooling.
Close monitoring was found sufficient to get reliable {\em ab initio} insights in the entire $T-\mu$ plane, albeit with reduced precision in different parameter regimes. With advances in the assessment of systematic errors of CL, the failure was linked to boundary terms which significantly increase as $\beta$ is decreased~\cite{Scherzer2020b}. In fact, such boundary terms are also present in the ``correct'' parameter regime, but they are sufficiently small such that the estimated systematic error is well below the statistical accuracy.
\begin{figure}[t]
  \includegraphics[width=0.475\textwidth]{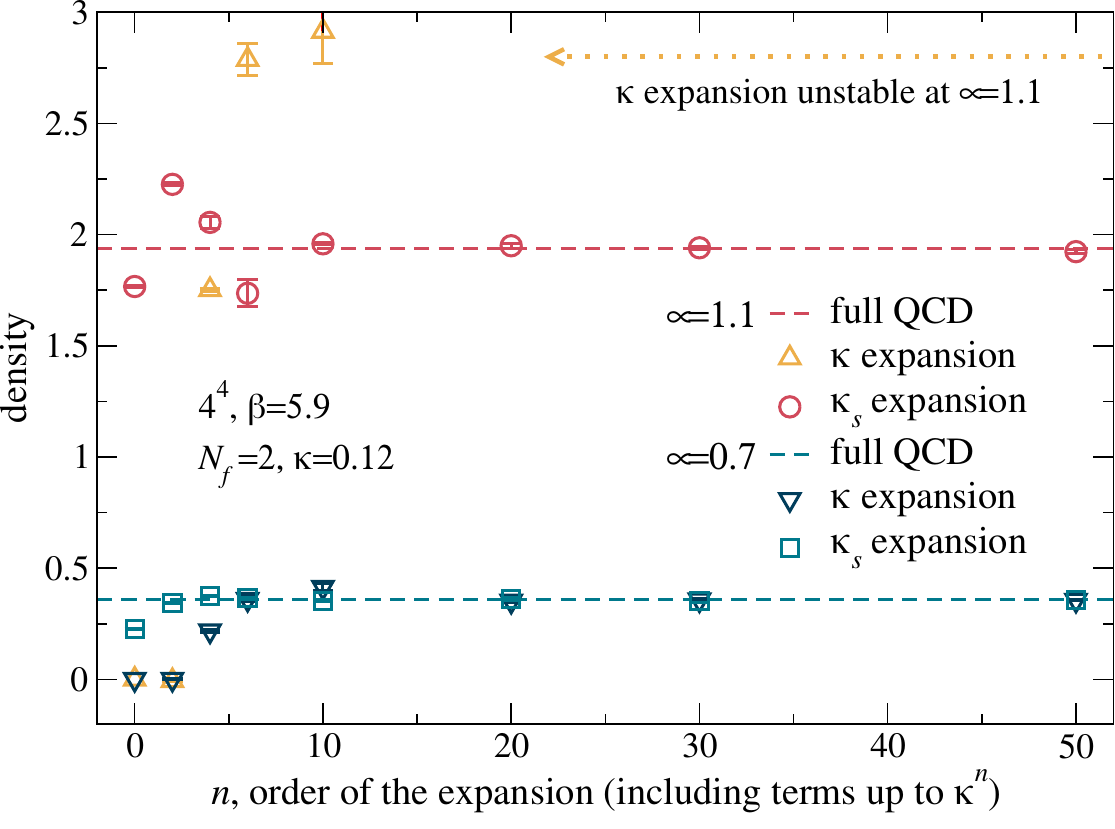}
  \includegraphics[width=0.505\columnwidth]{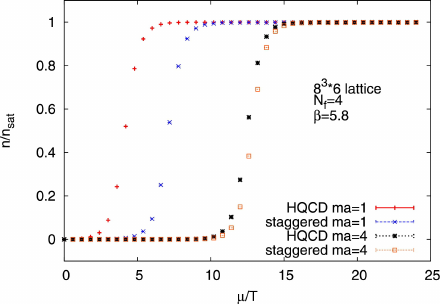}
  \caption{
    \label{fig:hopping_results} (Left) Dependence of the quark density (lattice units) on the order of the hopping expansion at $\beta=5.9$ on a $4^4$ lattice for two distinct types of expansions~\cite{Aarts2014b}.  The lower data shows results for $\mu = 0.7$, where both expansions and full QCD results (dashed lines) agree well at large orders. For $\mu = 1.1$ (upper data) only one type of expansion gives consistent results while the other breaks down.
    (Right) Comparison of the average quark densities as obtained from HQCD and a CL study of full QCD with staggered fermions, see Ref.~\cite{sexty2014b}.
  }
\end{figure}

In addition to the phase diagrams in the heavy limit, CL studies have also been performed in combination with a hopping parameter expansion. It has been mentioned~\cite{Aarts2014b} that the effective theories resulting from such an expansion exhibit a much milder sign problem, sometimes allowing a comparison with reweighting. Low-order expansions have been used in Refs.~\cite{Fromm2013,Langelage2014}, where excellent agreement between Metropolis-based algorithms and CL has been observed, and it was reported that CL is vastly superior at larger lattices. Moreover, all-order hopping expansions have been simulated with CL~\cite{aarts2014e,Aarts2014b,aarts2015b} and were found to agree well with CL results for the full theory~\cite{sexty2014b} (see below), which is considered to be a validation. Results for the density are shown in~\figref{hopping_results}. It is worthwhile to note that two different expansions -- one that includes poles in the measure and one that does not -- agree, indicating the absence of the influence of poles in the drift term in this case, providing us with an optimistic outlook for the application of CL to full QCD.

\subsubsection{Low-dimensional QCD at non-zero chemical potential}
Another strategy to benchmark the applicability of CL in the context of QCD is to reduce the spatial dimensionality of the lattice, thereby considerably reducing the numerical effort. Applications include $0+1$ dimensional QCD, where the quark density and chiral condensate have been computed using CL, with gauge cooling used to prevent the simulation from deviating in the imaginary direction~\cite{Bloch:2015coa}. Simulations in $1+1$-dimensional QCD have also successfully provided us with results for the chiral condensate in the thermodynamic limit~\cite{JHEP0820101017}.
In recent years, work on combining reweighting techniques with CL have expanded the range of applications of the method to previously inaccessible parameter space in QCD, allowing for greater ranges of values for the mass, temperature, and chemical potential in $1+1$-dimensional QCD~\cite{Bloch:2017sfg,PhysRevD95054509,Schmalzbauer2017}.

\subsection{$3+1$ dimensional QCD at finite chemical potential}
As mentioned already above, one of the big goals in the development of CL is the application to full QCD on the lattice at finite chemical potential, i.e., the consideration of dynamical quarks with physical masses. The earliest attempts to calculate the properties of full QCD at finite chemical potential is the work of Refs.~\cite{Sexty:2013ica,sexty2013,Sexty2014,sexty2014b} which employed a staggered-fermion formulation. As shown in~\figref{hopping_results}, the CL approach using gauge cooling agrees with the HQCD results in the valid region, i.e., for large quark masses. For values of $\beta \gtrsim 5.9$ the simulations are well under control and reliable results may be obtained. Similar to previous model studies, however, smaller values of $\beta$ exhibit large unitary norms despite gauge cooling and, as a consequence, develop skirts in the observable distributions.
Another early application to the full theory was performed in Ref.~\cite{Fodor2015}, where an extensive comparison with data from studies based on reweighting techniques at $\mu=0$ was performed. Both methods agree in a wide parameter range, however, the conclusion was similar to the parallel studies of QCD, namely that lower values of $\beta$ remain problematic.

Later on, the results of Ref.~\cite{Lattice2015KogutSinclair} suggested CL with gauge cooling produces trustworthy solutions for $3+1$-dimensional QCD at zero temperature and finite chemical potential. Work on this same system using a combination of adaptive step size and gauge cooling to prevent runaways in the complex plane for weak coupling showed promise. In fact, the results share some important physical traits with the system. However, known results were not accurately recovered and CL results seem to be influenced by the phase-quenched theory.
To be more specific, the observation of the transition from hadronic to nuclear matter (at $\mu \approx m_{N}/3$) is visible, as is evidence of saturation at large chemical potential~\cite{LATTICE2016PROCSinclairKogut}. The zero chemical potential limit disagreed with known results, but to a very small degree. With larger lattices and smaller coupling, the known results are still not accurately recovered. However, it was argued that the agreement becomes better in these cases~\cite{LATTICE2016PROCSinclairKogut}.
This has been traced back to the fact that these calculations suffer from the appearance of zeroes in the fermion determinant in this regime~\cite{Lattice2017KogutSinclair} (see also~\cite{Sinclair:2018rbk, Kogut2019}). Very recent extensions of these investigations employed modified actions with additional irrelevant terms (for the continuum theory) that are designed to separate the physics between different mass scales~\cite{Sinclair2019}. This strategy seemed to improve the location of the phase transition, however, further computations, in particular towards the continuum limit, are required.

A parallel development started with investigations of full QCD at small $4^3 \times 8$ lattices~\cite{Nagata2016} where encouraging results have been obtained with gauge cooling alone. While problems associated with a singular drift have not been observed at these lattice sizes, somewhat increased autocorrelation times have been argued to be the harbinger of potential issues in that regard on larger lattices. A subsequent study on the same lattice size but with the deformation technique yielded well-controlled results and was able to extend the applicable parameter range~\cite{2018EPJWC17507017N,Nagata2018}. In fact, it was found that the deformation is a necessary tool to circumvent the increasingly problematic influence of the singular drift and it was argued that when $\mu$ is chosen within the region of validity of CL, the eigenvalue distribution of the Dirac operator exhibits a gap along the real axis~\cite{Ito2018b}.
Most strikingly, a delayed onset of the influence of the chemical potential on the quark density was observed, a signal that was argued to be in agreement with the Silver Blaze property. Later, the results were extended to larger lattices~\cite{ito2018}, where the formation of a plateau in the quark density has been found. This was interpreted as the onset of nuclear matter (i.e., the formation of a Fermi surface) which was observed to persist even for lattices up to $16^3 \times 32$, which to date is the largest calculation of full QCD with CL~\cite{tsutsui2019,ito2020}.

Although progress has been made in the exploration of full QCD in $3+1$ dimensions, many aspects of the role of CL in this regard remain to be investigated. One of the pressing questions is the general applicability of CL in the confined phase, which was argued to be questionable at least due to the structure of the Dirac operator~\cite{tsutsui2018,Bloch2018}. Note that this was suspected, on more practical grounds, already in Ref.~\cite{Fodor2015}.
In addition, the issue of erroneous  convergence and the absence of a systematic error estimate constitutes a potential roadblock. To this end, recent advances in the analysis of boundary terms may constitute a way to handle this situation and, in fact, have also been applied to full QCD recently~\cite{Scherzer2020,Scherzer2020b}.
Finally, it remains to be seen if the use of certain improved operators, a strategy that was recently employed in a precise calculation of the equation of state in the quark gluon plasma~\cite{Sexty2019}, will help to circumvent some of these issues. A very recent overview on the current status of full QCD simulations with CL can be found in Ref.~\cite{attanasio2020b}.

\subsection{Other applications in relativistic field-theories}
Although most of the applications and development of CL have been achieved in the context of lattice QCD and related models, several other attempts have been made in the past. Here, we only provide a short overview of such studies.

\subsubsection{Real-time dynamics}
As mentioned above, a very intriguing application of the CL method is the direct simulation of QFTs in Minkowski spacetime, which suffers from a so-called ``dynamical'' sign-problem, originating from the complex exponent in the path integral (and typically avoided by going to the imaginary-time formulation). This potential was realized immediately after the CL prescription was originally proposed~\cite{Hueffel1984} which led to several applications in the context of high-energy physics and related  models~\cite{Hueffel1984,Gozzi1985,Hiromichi1986,Fukuda1988,Kolley1988} (with early applications even in the field of gravity~\cite{Rumpf1986}), see Ref.~\cite{PhysicsReportsStochasticQuantization} for an overview on the formalism and an early review.

After a period of relatively low activity, stochastic quantization once again was employed to study non-equilibrium QFT in the context of non-relativistic physics in the mid 2000s. Due to the non-perturbative nature of these non-equilibrium systems, standard approximation techniques fail. heavy ionulations may potentially be applied here but a Minkowski formulation suffers from a sign problem. In 2005, Berges and Stamatescu demonstrated the viability of CL to treat non-equilibrium QFT using first-principles simulations~\cite{PhysRevLett95202003,BERGES2008306}, albeit, only close to the equilibrium case as the long-time limit is prone to erroneous convergence. Later progress built on these results to examine how CL could lead to breakthroughs in our understanding of QCD plasmas in heavy ion collisions, early thermalization, and other open questions in quantum field theory~\cite{PhysRevD75045007}.

More recently, CL in combination with non-Markovian Langevin equations was used to address non-equilibrium physics for simple non-interacting field theories~\cite{deAguiar:2010ue}, leading to first sucesses and a potential route to future applications for interacting theories. Moreover, non-equilibrium aspects of scalar $0+1$ dimensional field theories (or, equivalently, quantum mechanics) have been studied within the CL algorithm and found promising results but also potentially troublesome, unphysical fixed points in the Langevin propagation~\cite{Anzaki2015}. To deal with the latter issue, a restriction in phase space was introduced, yielding correctly converged, yet potentially approximate, results. Finally, Ref.~\cite{Anzaki2015} also contains an elucidating comparison of real-time stochastic quantization with several state-of-the-art methods for the determination of dynamical properties.

\subsubsection{Theories with topological terms}
To (potentially) address the question of the strong CP problem, theories need to contain a so-called $\theta$-term, generically written as $S_\theta = -\i\theta Q$ with the topological charge $Q$. Such a term is purely imaginary, thereby introducing a phase problem. Besides results obtained by reweighting~\cite{Vicari2009}, a few CL studies have been conducted to examine the effect of such terms.

In Ref.~\cite{Bongiovanni:2013nxa,Bongiovanni:2014rna} the $SU(3)$ Yang-Mills theory was investigated, for the first time directly with a purely real $\theta$ - for imaginary $\theta$ the sign problem is absent and HMC may therefore be used as a benchmark in this regime. It was found that gauge cooling is essential for an efficient implementation and that CL is well controlled above an inverse coupling of $\beta \gtrsim 5.9$, below which the conditions for correctness were observed to be violated.

Additionally, the two-dimensional $U(1)$ gauge theory in the presence of a $\theta$-term was investigated~\cite{hirasawa2020}. It was shown that a naive implementation fails because of the crossing between different ``topology sectors'' (which implies the crossing of a branch cut). Reassuringly, the failure was observed to be accompanied by the violation of correctness markers. By removing a plaquette from the torus, i.e. ``piercing'' the manifold, the CL model was observed to deliver reliable results even at large values of $\theta$, previously inaccessible with methods based on reweighting.

\subsubsection{Matrix models for superstring theory}
While the overwhelming application of CL in the context of relativistic field theories has been in the context of QCD and related simplified models, several studies have explored quantum field theories beyond this scope. Recent progress was specifically made in the realm of matrix models, which are conjectured to provide a non-perturbative formulation of certain superstring theories~\cite{Banks1997} (in a similar fashion as lattice gauge theories constitute a numerically accessible non-perturbative formulation of QCD).

Amongst others~\cite{Pallab2018,Anosh2019}, particular interest was paid to the type IIB matrix model, which is conjectured to be a nonperturbative definition of type IIB superstring theory~\cite{Nishimura2019}. In its Euclidean version, a severe sign problem arises upon integrating out the fermions, leaving a complex-valued Pfaffian. The complex-valued measure was found to be responsible for the central phenomenon in these theories, namely a dynamical compactification of spacetime from ten to four dimensions, which rules out phase-quenched approaches to investigate these models~\cite{ito2016b}. Initial CL studies focused on the low-dimensional sector~\cite{Ito2016} and found that a singular drift problem arises which is treatable with the deformation technique.
Subsequently, the study of this model was extended to six dimensions~\cite{anagnostopoulos2019,Anagnostopoulos2018}, where the a breakdown from $SO(6)$ to $SO(3)$ was observed, marking a major improvement over reweighting-based methods.
Most recently, these investigations have been extended to the original ten dimensions of the model in the Lorentzian~\cite{Nishimura2019} as well as the Euclidean version~\cite{anagnostopoulos2020, anagnostopoulos2020b} of the model, where clear signals of the spontaneous symmetry breaking have been observed.


\section{\label{sect:NRQFT}Applications in non-relativistic matter: ultracold atomic gases}
In contrast to relativistic theories, applications of CL to non-relativistic matter remain largely in their infancy (notwithstanding notable early work such as Ref.~\cite{PRC2001026303}), perhaps not least due to an early finding that conventional MC approaches are superior to CL in this regard because of uncontrolled runaway trajectories~\cite{Lin1986}. Recent advances concerning these issues made largely in the high-energy community, however, induced progress also for non-relativistic models.
As described below, there have been several attempts to characterize bosons as well as fermions in a variety of situations~\cite{Drut2018}, but as of this writing only one calculation has been successfully carried out for a strongly coupled fermionic system in 3D~\cite{2018UFGviaCL} (see the discussion in \secref{ufg}).
Non-relativistic systems have difficulties of their own in the form of phase transitions and strongly coupled regimes, but they do not feature quintessential (and technically challenging) QCD elements such as non-abelian gauge fields. Nevertheless, as we shall see, some of the challenges faced by CL calculations are universal, as are the ideas to diagnose and tackle them. The different viewpoints taken in relativistic and non-relativistic applications leads to an important interplay between these two subfields and ultimately progresses the exploration of CL in both communities (see \secref{formalism}).

Among the vast array of non-relativistic systems that remain to be explored, some pressing candidates remain across different areas. In the condensed matter context, for example, the leading candidate is undoubtedly the repulsive 2D Hubbard model which faces a severe sign problem upon introducing a finite doping; a single exploratory study was conducted for the repulsive 2D Hubbard-model at half-filling~\cite{yamamoto2015}). Other highly interesting candidates include the spin-polarized electron gas, spin-isospin polarized nuclei, and neutron and nuclear matter, to name only a few. The list of applications outlined in this section may potentially also lead to further advances in those areas.

\subsection{Non-relativistic bosons}
One of the pioneering applications of CL to non-relativistic systems in the modern era (roughly within the last decade) concerned the study of a 3+1 dimensional bosonic quantum field theory in a rotating frame~\cite{yamamoto2015,PhysRevA92043628}. The number density and condensate fraction with no rotation computed via CL was found to agree with mean-field calculations, and when rotation was introduced, the CL results showed quantized circulation for high condensate fraction. This is consistent with vortex formation in rotating superfluids, a phenomenon which has been directly observed using rotating ultracold bosonic atoms. The results for the circulation using CL do not agree with mean-field calculations for small condensate fraction, as expected due to the breakdown of mean-field theory to describe a system with strong quantum fluctuations. Besides this study in three spatial dimensions, an exploration in two spatial dimensions, where quantum fluctuations exhibit an even greater impact, was conducted~\cite{Berger2019}. However, in the latter study, the parameter regime where a quantized vorticity is expected remained out of reach, requiring further CL studies on the topic.

In addition to these single component systems, the effect of spin-orbit coupling in a two-dimensional Bose gas with two pseudospin components interacting via a delta function potential was explored~\cite{Attanasio2020}. In this case, the sign problem arises due to a first-order derivative in the imaginary time which was well controlled with the CL approach. Mean-field predictions were found to overestimate thermodynamic equations of state, in particular at large spin-orbit coupling.

Aside from these brief explorations, the application of CL to circumvent the sign problem in systems of non-relativistic bosons remains scarce.

\subsection{Non-relativistic fermions in one dimension}
\begin{figure}[t]
  \centering
  \includegraphics[width=\columnwidth]{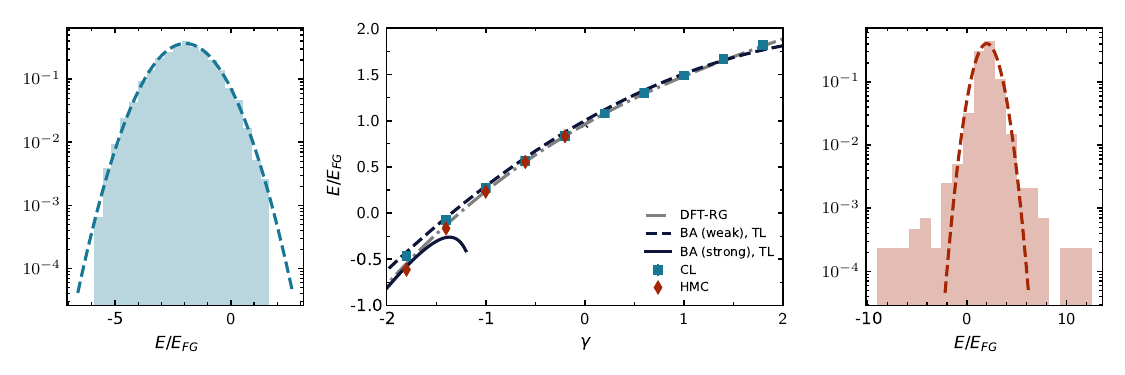}
  \caption{\label{fig:1d_bal_eos}  (Left) Log-histogram for sampled ground-state energies at strong attractive coupling $\gamma = -3.0$. Perfect gaussian behavior is observed (dashed line). (Center) Ground-state energy of $N = 5 + 5$ fermions as a function of the dimensionless coupling $\gamma$. The CL results (blue squares) are compared to HMC results (red diamonds) as well as to the Bethe-ansatz solutions for strong and weak coupling (solid and dashed lines, respectively) and results obtained by a DFT-RG approach \cite{0954-3899-44-1-015101} (dashed-dotted line). (Right) Log-histogram for sampled ground-state energies at strong repulsive coupling $\gamma = 3.0$. Heavy tails of the distributions (as compared to a Gaussian, dashed line) spoil the correctness criterion.}
\end{figure}
One-dimensional (1D) Fermi gases have drawn considerable interest in the past, when exact solutions for various models have been derived by the use of the so-called Bethe ansatz (see e.g., Ref.~\cite{RevModPhys.85.1633} for an extensive review). The availability of these exact solutions, paired with a relatively modest computational cost, make these systems excellent benchmark scenarios for any newly devised method, such as the CL approach for non-relativistic quantum matter.  Moreover, 1D systems have become accessible in experiments in recent years, which provides yet another motivation to study these exotic and intrinsically strongly-interacting systems.

It is worthwhile to note here that in the special case of 1D it is often possible to re-write relativistic and non-relativistic models using dual variables in a way that avoids sign problem. One would be able to then compute quantities by conventional Monte Carlo methods. While this is an option in 1D, those approaches do not generalize to higher dimensions, in contrast to the CL method. In fact, the dimensionality of the problem is mainly a question of computational effort, and all insights gained on the numerical behavior of CL simulations carry over to higher dimensions.

In this section, we will review recent advances that have been made in applying CL to one-dimensional fermionic systems that suffer from a sign problem. Specifically, we will discuss systems at zero temperature featuring attractive and repulsive interactions, finite spin-polarization as well as asymmetry in the masses of the fermions in \secref{fermions_gs_1d}. Furthermore, we recall results obtained at finite temperature in \secref{fermions_ft_1d}, where repulsively interacting fermions are shown as well as systems at asymmetric chemical potential.

\subsubsection{1D fermions in the ground state \label{sect:fermions_gs_1d}}
In the following, we will consider a system of $N = N_\uparrow + N_\downarrow$ fermions at fixed linear lattices of $N_x$ sites, which corresponds to working in the canonical ensemble. The physics in 1D is set by the dimensionless interaction parameter $\gamma = g/n$ where $g$ is the bare interaction and $n$ is the linear particle density.

\paragraph{Balanced systems}
\begin{figure}[t]
  \centering
  \begin{subfigure}[t]{0.56\columnwidth}
    \includegraphics[width=\columnwidth]{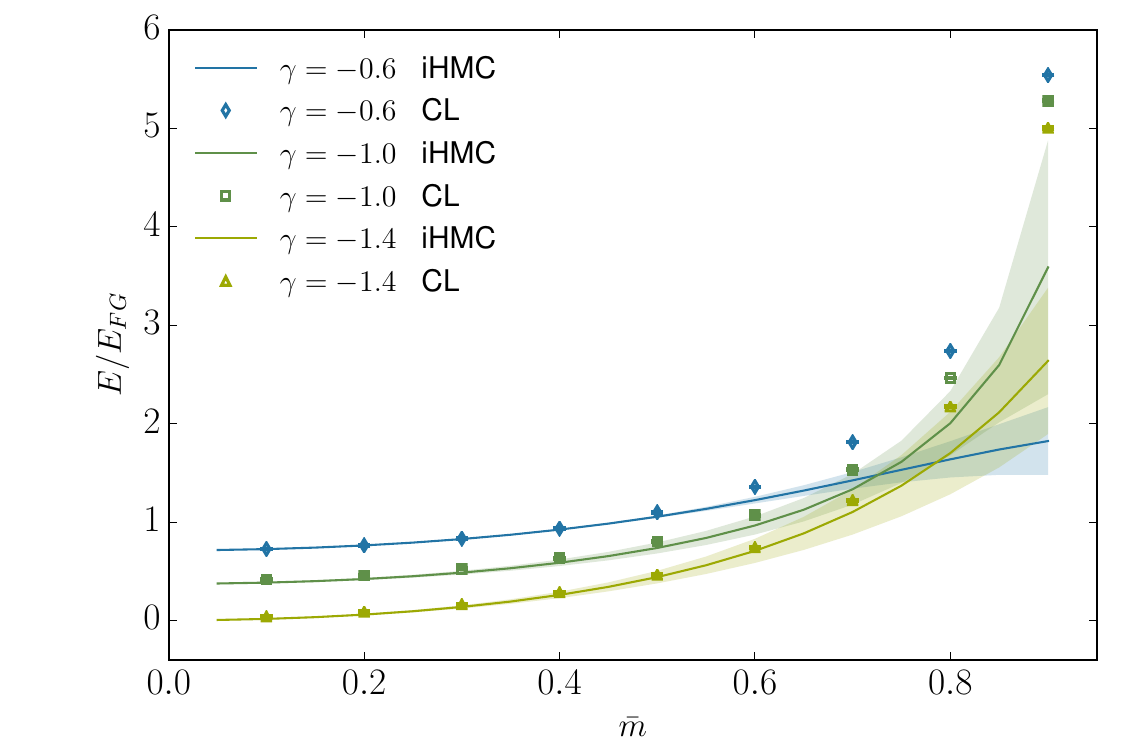}
  \end{subfigure}
  \begin{subfigure}[t]{0.4\columnwidth}
    \includegraphics[width=\columnwidth]{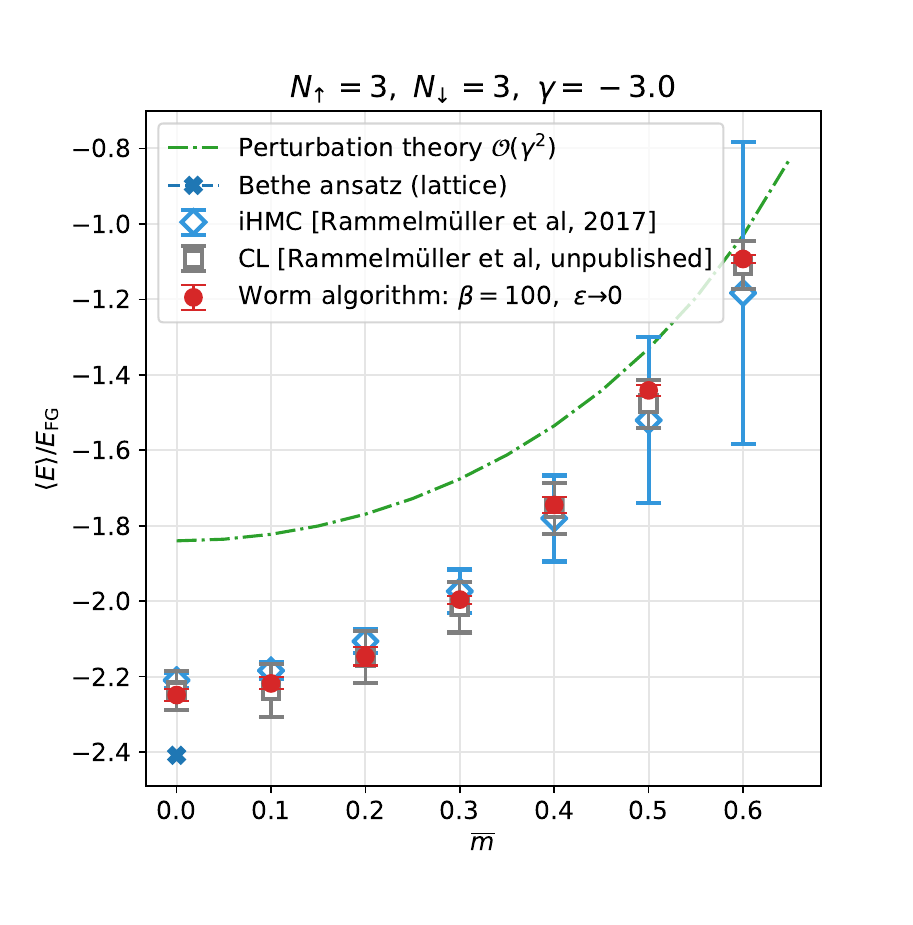}
  \end{subfigure}
  \caption{\label{fig:1d_mib_eos} (Left) Comparison of iHMC and CL as a function of relative mass imbalance $\bar m$ for $N_\uparrow+N_\downarrow = 5+5$ particles at fixed interaction streghts $\gamma$. (Right) Ground-state energy of $N = 3 + 3$ fermions as a function of $\bar m$ as obtained with the Worm algorithm \cite{PhysRevD.99.074511} in comparison with CL results as well as iHMC values and perturbation theory. The plot is taken from Ref.~\cite{PhysRevD.99.074511}.}
\end{figure}
As a first benchmark of the CL method in the ground state, the spin- and mass-balanced Fermi gas was investigated~\cite{PRD96094506,Rammelmueller2018}. A comparison to various other methods is shown in \figref{1d_bal_eos}, where generally good agreement is apparent across the entire range of interactions shown. While the CL results shown are at $N = 5+5$ particles and a finite lattice of $N_x = 40$, the curves for the Bethe ansatz expansions (strong~\cite{1742-5468-2007-06-P06011} and weak~\cite{doi:10.1063/1.4964252} coupling) correspond to the thermodynamic limit (TL) of taking particle number and volume to infinity at constant density.
The agreement on the repulsive sides indicates sufficiently large box sizes. On the attractive side agreement is observed; however, larger spatial lattices are needed as is suggested by the slight discrepancy between CL and the volume-extrapolated HMC values of Ref.~\cite{PhysRevA.96.033635}, as well as results obtained by a DFT-RG approach in Ref.~\cite{0954-3899-44-1-015101}. Another possible source of systematic bias is the finite (adaptive) integration step of $\Delta t = 0.01$, which was used throughout this study. Although the $\Delta t$ dependence was checked in Ref.~\cite{PRD96094506}, where it was found that $\Delta t$ was sufficiently small (within the statistical uncertainty), the influence of $\Delta t$ could vary in other areas of parameter space, in particular with varying coupling strength.

While the agreement among the methods looks excellent, a closer look reveals interesting technical subtleties. During stochastic calculations it is instrumental to study histograms of all measured quantities to ensure correct behavior. As can be appreciated in the outer panels of \figref{1d_bal_eos}, two distinct behaviors were observed in different regimes: the strongly attractive case ($\gamma = -3.0$) displays well-behaved, localized distributions; the strongly repulsive case ($\gamma=3.0$), on the other hand, exhibits a large amount of outliers with respect to an assumed Gaussian. The existence of so-called ``fat tails" renders the simulation problematic, as this violates the conditions for correct behavior and thus undermines the validity of the CL approach. Therefore, it was conjectured that the CL values are possibly faulty in this parameter regime, which was later confirmed by a different few-body method \cite{PhysRevD.99.074511} based on dual variables and the worm algorithm. A more precise understanding of the behavior at strong repulsive interaction, and eventually its resolution, would be very useful.

\paragraph{Mass-imbalanced Fermi-Fermi mixtures}
Mixtures of two different fermion species, i.e., systems where the constituents have unequal masses, are challenging to address theoretically. While many 1D models are integrable via the Bethe ansatz, there is no such solution in the mass-imbalanced case. Nevertheless, these systems are of great interest as there are a number of experimental realizations available for these mixtures. The CL method can be straightforwardly extended to cover fermion systems with unequal masses by the use of spin-dependent dispersion relations, i.e. a spin-dependent mass $m_\sigma$ (or, equivalently, anisotropic hopping parameters). To quantify the mass asymmetry between two species, it is convenient to introduce the dimensionless relative mass imbalance $\bar m = {(m_\uparrow-m_\downarrow)/(m_\uparrow+m_\downarrow)}$,
which is restricted to the interval $[0,1]$. While the use of spin-dependent dispersion relations is unproblematic from a conceptual viewpoint, numerical difficulties might occur due to an increasing separation of scales. Thus, one may use different parametrizations of the mass imbalance depending on the specific implementation of the algorithm.

\begin{figure}[t!]
  \includegraphics{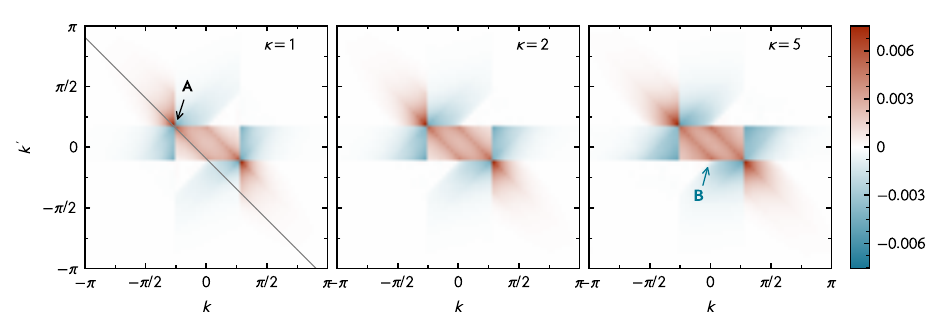}
  \caption{Shot-noise correlator $G_{\uparrow\downarrow}$ for $17\uparrow +\, 9\downarrow$ particles and mass rations of $\kappa = 1, 2$, and $5$ (corresponding to a heavy-majority setup). The gray dashed line in the left panel corresponds to fixed total momentum of $k_{\rm tot} = -q$, which is the expected FFLO momentum.
  The peaks labeled ``A'' and ``B'' are the leading FFLO peak and the emergent pairing peak, respectively.}
  \label{fig:1d_shotnoise}
\end{figure}

In Refs.~\cite{PRD96094506,Rammelmueller2018} the mass-imbalanced few-body problem was studied and the average mass was fixed to $m_0 = 1$. In this way, a comparison is made possible to results obtained with the iHMC method (imaginary asymmetry -- introduced in \secref{ImaginaryAsymmetry}). A comparison is shown in the left panel of \figref{1d_mib_eos} for fixed values of the interaction strength $\gamma$. Excellent agreement of CL and iHMC is reported across a wide range of mass imbalances. Starting at $\bar{m} \sim 0.6$ the iHMC results start to deviate as a consequence of the instability of the analytic continuation. The CL results, on the other hand, remain smooth and precise up to high mass imbalances. The agreement between these different methods provides confidence that the CL method is suitable to study this problem.

Recently, a worldline algorithm was adapted to study mass-imbalanced few-body systems without a sign problem \cite{PhysRevD.99.074511, Singh:2018pci} in 1D and its results have been compared to the CL values obtained in \cite{PRD96094506} (shown in the right panel of \figref{1d_mib_eos}). On the attractive side, excellent agreement is observed for all mass imbalances, which further validates the CL method in this setting. A comparison on the repulsive side, however, revealed that the CL values deviate from the wordline results, which is connected to the occurrence of heavy tails mentioned above and in Ref.~\cite{PRD96094506}. It remains to be shown whether mass-imbalanced Fermi mixtures with repulsive interactions in the ground state are accessible with CL. This question is also intimately related to the applicability to the repulsive Hubbard model.

\paragraph{Spin- and mass-imbalanced systems}
Very recently, the investigation of mass-imbalanced fermionic systems was extended to study pairing phenomena based on two-body correlation functions~\cite{rammelmueller2020} which, to date, is the first determination of correlation functions with the CL approach. Concretely, the effect of unequal masses on the behavior of a spin-polarized system of two-component fermions was addressed. While it is known (theoretically) that such a system exhibits Cooper pairing at finite pair-momentum (referred to as Fulde-Ferell-Larkin-Ovchinnikov or FFLO pairing), the effect of mass imbalance in this situation remained relatively little explored. Notably, no exact solutions for these mass-imbalanced system are known to date whereas exact solutions from the Bethe ansatz exist for the mass-balanced case, even for finite spin imbalance~\cite{1742-5468-2007-06-P06011}, see Ref.~\cite{RevModPhys.85.1633} for a review on the Bethe ansatz.

With CL, it was shown that the appropriate correlation functions can be extracted accurately. Interestingly, in addition to the expected FFLO-type pairing correlations, unprecedented correlations in the density-density correlation function in momentum space (also referred to as shot-noise correlations) have been observed in case of finite mass imbalance, see \figref{1d_shotnoise}. The extracted results for the mass-balanced case (leftmost panel in \figref{1d_shotnoise}) agree with related DMRG studies of the spin-imbalanced Hubbard model at low filling~\cite{Luescher2008}.
Specifically, dominant correlations are observed at the points $(\pm k_{{\rm F},\uparrow},\mp k_{{\rm F},\downarrow})$ which indicates the expected pairing at the opposite Fermi momenta. With increasing mass imbalance, on the other hand, an additional correlation maximum emerges, suggesting a previously unknown scattering of fermions far below (heavy majority) or far above (heavy minority) the Fermi point of the heavier spin species. It is worthwhile to note that the underlying mechanism is still associated with pairing of fermions with a pair-momentum also assumed in case of conventional FFLO correlations, i.e. $q = |k_{{\rm F},\uparrow} - k_{{\rm F},\downarrow}|$. However, the relative momentum of the bound fermions is shifted.

The two-body quantities obtained from CL provide an exceptionally clean signal of the pairing instability. Moreover, the shot noise is of direct experimental interest, as it may be extracted with readily available time-of-flight measurements.

\subsubsection{1D fermions at finite temperature \label{sect:fermions_ft_1d}}
\begin{figure}[t]
  \centering
  \includegraphics[width=0.49\columnwidth]{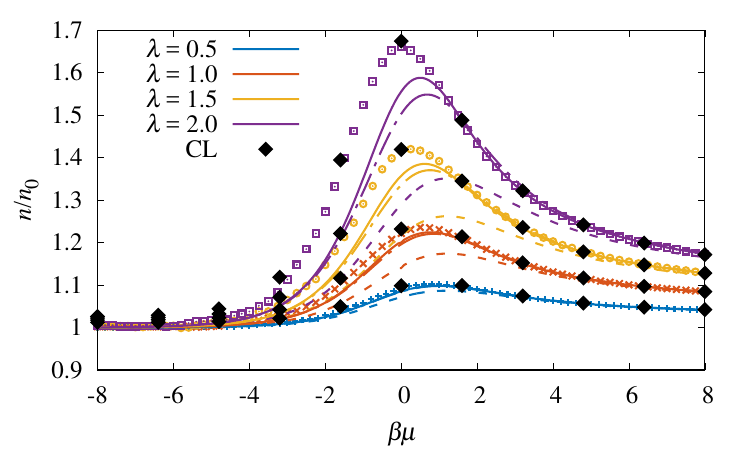}
  \includegraphics[width=0.49\columnwidth]{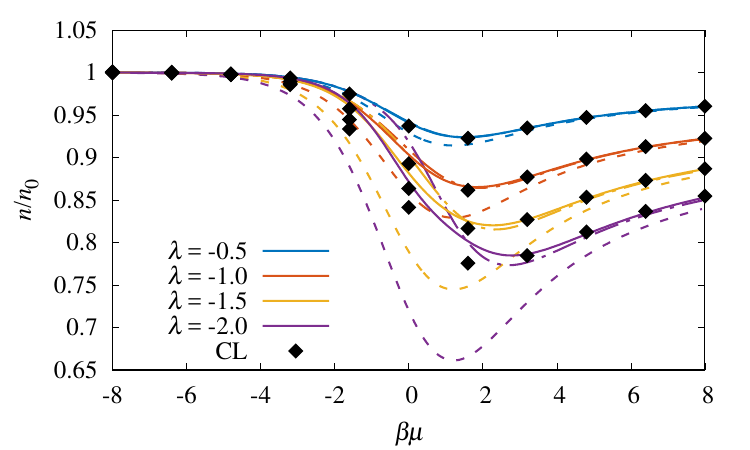}
  \caption{\label{fig:DensityEoSCL1DFiniteT} Density $n$ of the attractive (left) and repulsive (right) unpolarized Fermi gas in units of the density of the
	noninteracting system $n_0$, as shown for the dimensionless interaction strengths $\lambda = 0.5$, 1.0, 1.5, 2.0 (attractive), and $\lambda = -0.5$, -1.0, -1.5, -2.0
	(repulsive). The NLO (dashed line), N2LO (dash-dotted line), and N3LO (solid line) results of perturbation theory are displayed for each coupling and are
	compared with HMC results (see Ref.~\cite{PhysRevA.91.033618}) in the attractive case. For both plots, the black diamonds show CL results
	(RL for the attractive case), regulated with $\xi=0.1$ as described in the main text. Results were computed on a spatial lattice of $N_x = 80$ for CL and HMC, and of $N_x = 100$
  for perturbation theory. The statistical uncertainty of the CL results is estimated to be on
	the order of the size of the symbols, or less, as supported by the smoothness of those results.}
\end{figure}
One of the first attempts to use CL for non-relativistic fermions was made in Refs.~\cite{PRD95094502,Loheac2018}, where the thermodynamics of a repulsive\footnote{In fermionic systems with a repulsive coupling, the sign problem is present even for spin-balanced systems, as the Hubbard-Stratonovich transformation necessarily requires the determinants (and hence the probability measure) to be complex quantities.} 1D Fermi gas was calculated using that method as well as third-order lattice perturbation theory (see \figref{DensityEoSCL1DFiniteT}).  The density and pressure were computed for a range of interaction strengths and temperatures. The calculations were also compared with prior hybrid Monte Carlo results for attractive interactions, where the sign problem is absent and the convergence of the perturbative expansion can be assessed. On the repulsive side, CL was found to agree with the third-order perturbative expansion at weak coupling and away from the virial region (where the fugacity is small), beyond which perturbation theory was expected to break down. In the virial regime, a related study employed particle projection in combination with CL to obtain high-order virial coefficients in 1D~\cite{Shill2018}. This approach, while so far only applied in 1D, is applicable in any spatial dimension and could therefore be beneficial in extending the range of accessible temperatures of the virial expansion.

\begin{figure}[t]
  \floatbox[{\capbeside\thisfloatsetup{capbesideposition={right,top},capbesidewidth=0.4\textwidth}}]{figure}[\FBwidth]
  {\caption{\label{fig:XiComparison}The normalized density $n/n_0$, where $n_0$ is the noninteracting result,
  	for $\lambda=-1.0$ and $\beta\mu = 1.6$, as a function of the Langevin time $t$ for
  	several values for the regulating parameter $\xi$ [see \equref{CLModifiedEqs}]. The result was computed on a spatial lattice of $N_x = 80$
    and a temporal lattice of $N_\tau = 160$.
  	For a choice of $\xi = 0$, where the regulating term is removed, CL tends toward an incorrect value for the density. When $\xi \simeq 0.1$,
  	the additional term provides a restoring force and the stochastic process converges to a different value consistent with perturbation theory.
  	For $\xi < 0$, the solution diverges, as expected.
    for a given $\xi$.}}
  {\includegraphics[width=0.55\textwidth]{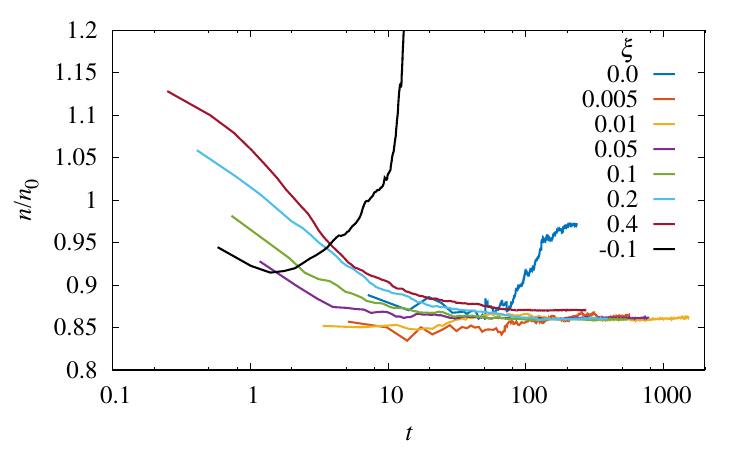}}
\end{figure}

At the same time, the aforementioned work~\cite{PRD95094502} explored for the first time the use of a regulator to avoid uncontrolled excursions of the auxiliary field into the complex plane. In those calculations, the excursions into the complex plane were highly problematic because the dependence of the action and the drift on the auxiliary field $\sigma$ involved hyperbolic functions.  The HS transformation used in that work depends on $\sin \sigma$, and the complexification $\sigma \to \sigma_R + i \sigma_I$ results in
\begin{equation}
  \sin \sigma = \sin (\sigma_R) \cosh(\sigma_I) + i \cos (\sigma_R) \sinh(\sigma_I).
\end{equation}
As a consequence, a growing magnitude of $\sigma_I$ amounted to increasing the fermion-fermion coupling at an exponential rate. In such a situation, the calculation would either stall or result in a converged but incorrect answer. This exponential growth is similar to the problem found in gauge theories, as the complexified link variables representing the gauge field become unbounded in the same fashion in those theories.
For lattice calculations of gauge theories, Refs.~\cite{Lattice2016AttanasioJager,Lattice2017AttanasioJager} tackled the problem of large excursions by dynamic stabilization (see \secref{CLchallenges}). Following a similar idea, Ref.~\cite{PRD95094502} added a regulating term to the CL dynamics controlled by a parameter $\xi$, such that the new CL equations became
\bea
  \label{Eq:CLModifiedEqs}
  \Delta \sigma_R &=& -\textrm{Re}\left[\frac{\delta S[\sigma]}{\delta \sigma}\right] \Delta t  - 2 \xi \sigma_R \Delta t + \eta \sqrt{\Delta t}, \\
  \Delta \sigma_I &=& -\textrm{Im}\left[\frac{\delta S[\sigma]}{\delta \sigma}\right] \Delta t - 2 \xi \sigma_I \Delta t.
\eea
As remarked above, the new term in the action can be understood as a harmonic oscillator trapping potential, i.e. a restoring force that prevents the field from wandering too far in the imaginary direction. That modification introduces a systematic effect that needs to be studied for each quantity of interest as a function of $\xi$ as $\xi \to 0$. \figref{XiComparison} shows the running average of the density as a function of the Langevin time $t$ for several values of $\xi$ in the neighborhood of $0$. As is evident in that figure, there is a sizable window of small values of $\xi$ where CL converges.

\paragraph{Spin-polarized systems}
Encouraged by the success of CL in calculating 1D systems with repulsive interactions, the follow-up study of Refs.~\cite{PhysRevD.98.054507,Loheac2018} extended the results of Ref.~\cite{PRD95094502} to the spin-polarized case by introducing a non-vanishing asymmetry $\beta h = \beta (\mu_\uparrow - \mu_\downarrow)/2$. Such an asymmetry leads to a sign problem for the case of attractive interactions and to a phase problem for repulsive interactions. As mentioned above, such a sign or phase problem can be avoided in 1D using specific methods which, unfortunately, do not generalize to higher dimensions.

\begin{figure*}[t]
	\centering
	\includegraphics[width=0.49\columnwidth]{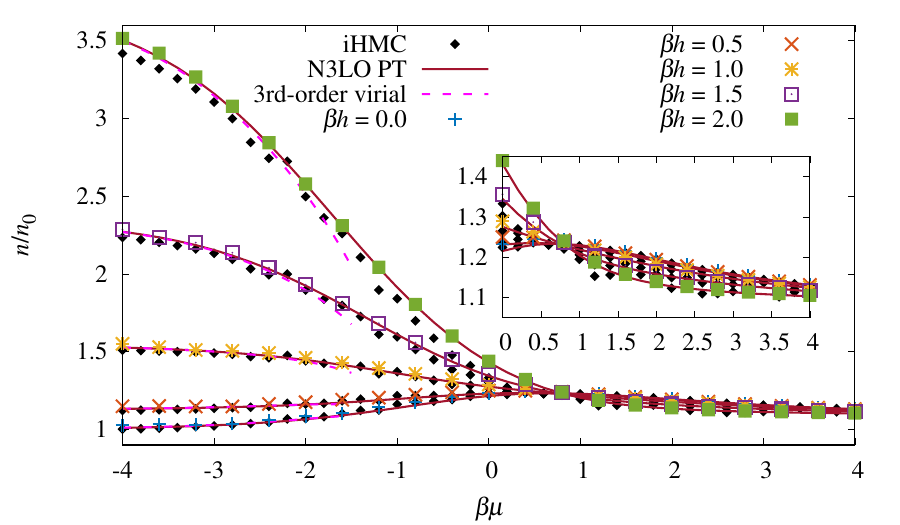}
	\includegraphics[width=0.49\columnwidth]{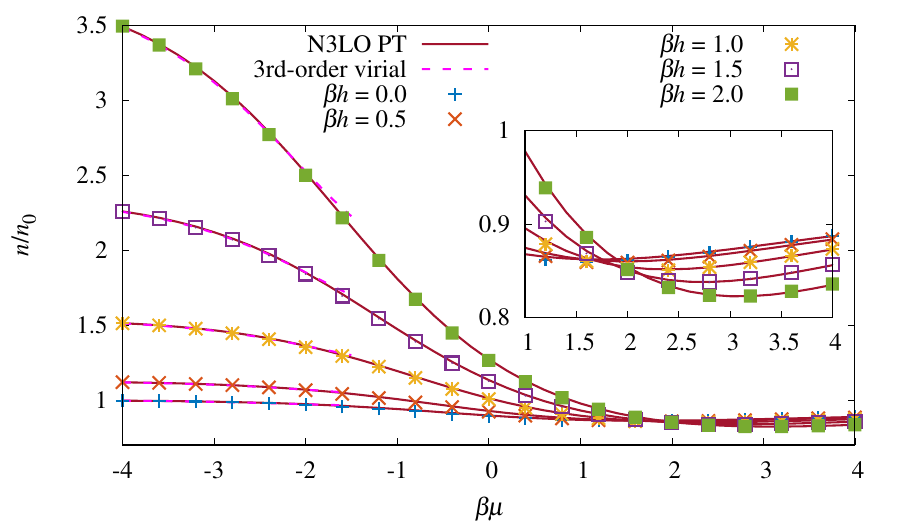}
	\caption{\label{fig:DensityLambda1} Density equation of state $n = n_\uparrow + n_\downarrow$ normalized by the non-interacting, unpolarized counterpart $n_0$,
	{for attractive (left) and repulsive (right) interactions of strength $\lambda = \pm 1$.
	Insets: Zoom in on the region $\beta\mu > 0$ (left) and $\beta\mu > 1$ (right).}
	In all cases, the CL results are shown with colored symbols, iHMC results (from Ref.~\cite{PhysRevA.92.063609}) appear with
	black diamonds, perturbative results at third order are shown with solid lines, and virial expansion results
	appear as dashed lines.}
\end{figure*}

In the study of Ref.~\cite{PhysRevD.98.054507}, the density and magnetization equations of state were calculated with CL and compared with perturbation theory, imaginary polarization approaches (iHMC), and the virial expansion. A sample of those results is reproduced in \figref{DensityLambda1} where the remarkable agreement among all of these approaches is apparent. However, at the largest chemical potential asymmetry (i.e. $\beta h =2.0$ in the figure), the results obtained with imaginary asymmetry start to deviate from the ones obtained by other means, including CL. This trend is not surprising, as the iHMC method is expected to be less robust in this limit (see also the discussion above). Reassuringly, as in the mass-imbalanced case discussed above, the CL values continue to agree with benchmark data in this regime, confirming the robustness of CL for spin-polarized systems. Interestingly, a study based on Lefschetz thimbles (see~\secref{Thimbles}) finds a better agreement with iHMC in the virial regime for this value of the imbalance~\cite{Alexandru2018}. At larger chemical potential, on the other hand, excellent agreement between thimbles and CL was reported in the same study.

\subsection{Non-relativistic fermions in three dimensions: the polarized unitary Fermi gas \label{sect:ufg}}
One of the most studied many-body systems in recent years is the so-called unitary Fermi gas (UFG) which is associated with a diverging s-wave scattering length. Due to this dominant length scale, the microscopic information about the interaction between the spin-up and spin-down particles becomes effectively irrelevant, and all observed quantities may be written as universal functions of the fermion density and temperature (as these are the only scales left in the system). Its intricate behavior is linked to various many-body phenomena, and moreover it is realized to an excellent precision in numerous cold atoms experiments, having led to a plethora of precise measurements of its unique behavior (see, e.g., \cite{Zwerger2012} for a comprehensive review).

For SU(2) symmetric fermions, i.e. for equal numbers of spin-up and spin-down particles, it is possible -- albeit challenging -- to study the UFG with conventional Monte Carlo approaches. Past stochastic studies of the UFG have been conducted using methods such as hybrid Monte Carlo (HMC) \cite{PhysRevA.85.051601, PhysRevLett.110.090401, PhysRevA.93.053604} and auxiliary field methods (AFQMC) on the lattice \cite{Jensen2019, PhysRevA.84.061602, PhysRevB.73.115112} as well as various flavors of diagrammatic Monte Carlo methods \cite{PhysRevLett.121.130405, VanHoucke2012} in the continuum.
Building on these (and many more) theoretical and experimental advances, it was found that the UFG undergoes a phase transition from a strongly correlated normal fluid to a BCS superfluid at the unusually high transition temperature $T/T_{\rm F} \approx 0.17$. Moreover, the Bertsch parameter, which relates the non-interacting energy to the ground-state energy of the UFG, was determined accurately to be $\xi = 0.37(2)$ in excellent agreement with high-precision experiments~\cite{He2020,Ku2012}.

For a broken SU(2) symmetry, in which the particle numbers of the up and down components differ, experiments have shed light on the phase diagram in the interesting regime of intermediate polarizations up to $p = (N_\uparrow - N_\downarrow)/(N_\uparrow + N_\downarrow) \approx 0.4$, above which a first-order phase transition (at, and very close to, zero temperature) from the superfluid to the normal state takes place \cite{Shin2006,PhysRevLett.101.070404,Shin2008,Nascimbene2010,Navon2010}.
On the theory side, progress was made in the highly-imbalanced limit of polarons through a variational ansatz for the $N+1$ wavefunction. However, for the full many-body problem with a finite spin-imbalance no sufficiently good ansatz is known. Diffusion MC calculations in the ground state, however, show good agreement with the experimental findings despite the fixed-node approximation~\cite{Lobo2006}. This simplification, which enables such calculations, unfortunately limits the predictive power in terms of possibly exotic pairing states (such as the FFLO mechanism) and was therefore mainly limited to the normal phase.

Nevertheless, several studies based on mean-field theory (see, e.g., \cite{Chevy2010,Radzihovsky2010} for a broad overview on such advances) and (semi-)analytic perturbative \cite{Strinati2018} as well as non-perturbative studies based on the functional renormalization group \cite{Gubbels:2007xc,Krippa:2014kra,Boettcher2014,Roscher:2015xha} and the Luttinger-Ward (LW) formalism~\cite{Frank2018} have led to insights into the phase diagram spanned by the temperature and polarization. Whereas these studies do not suffer from a sign problem, their application requires the construction of truncation schemes (e.g., in terms of a set of correlation functions that is taken into account). Therefore, complementary studies based on stochastic approaches allow us to enhance our understanding of the spin-imbalanced UFG.

\begin{figure}[t]
  \centering
  \includegraphics[width=\columnwidth]{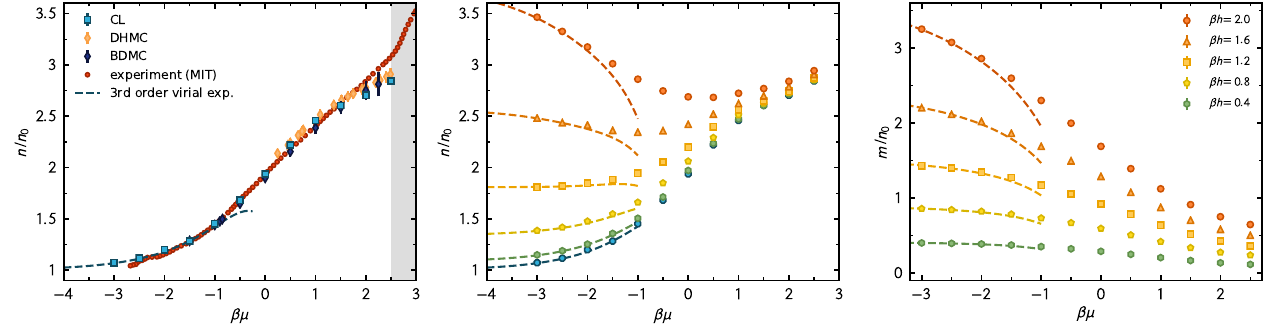}
  \caption{\label{fig:ufg_eos} (Left) Density in units of the non-interacting density as a function of the dimensionless parameter $\beta \mu$ for the balanced Fermi gas. Additionally, comparison to theoretical values (BDMC \cite{VanHoucke2012}, DHMC \cite{PhysRevA.85.051601}) and experimental values from the MIT group \cite{VanHoucke2012} are shown. (Center) Density in units of the non-interacting density as a function of $\beta \mu$ for the polarized Fermi gas, compared to the third order virial expansion at high temperatures. (Right) Magnetization in units of the non-interacting density as a function of $\beta \mu$ compared to the 3rd order virial expansion at high temperatures.}
\end{figure}

As stated above, a finite spin polarization introduces a sign problem in stochastic approaches which is especially severe at low temperatures, where the interesting physics takes place. To circumvent this sign problem, the CL method was applied to the spin-polarized unitary Fermi gas for the first time in Ref.~\cite{2018UFGviaCL}, finding excellent agreement with available benchmark results in the balanced limit. In the left panel of \figref{ufg_eos}, the density equation of state is shown for a balanced gas in comparison to previously obtained results from DHMC \cite{PhysRevA.85.051601} and bold diagrammatic Monte Carlo (BDMC) \cite{VanHoucke2012} studies, as well as experimental values from the MIT group \cite{VanHoucke2012}. Across a wide range of $\beta\mu$ values, excellent agreement was reported down to low temperatures.
In the vicinity of the phase transition to superfluidity, however, the CL results, along with all other theoretical values, start to deviate from the experimental values, which is attributed to finite lattice effects and can, in principle, be mitigated by extending the box sizes beyond~$N_x = 11$.

In the central and right panels of \figref{ufg_eos} first predictions for the density and magnetic equations of state are shown, respectively. It is important to note here that these systems require essentially the same computational effort as in the spin-balanced case, whereas a reweighting approach would suffer from an exponential increase in computational effort. As can be appreciated from the figure, excellent agreement with the virial expansion (see e.g. \cite{LIU201337}) is achieved at high temperatures, which gives confidence on the reliability of the CL results in that regime. Moving towards lower temperatures, the virial expansion is expected to break down, whereas the CL results continue to evolve smoothly across the entire parameter range studied. With respect to the latter regime, a computation of the pressure equation of state of the polarized UFG (beyond the mean-field approximation) based on the LW formalism~\cite{Frank2018} is found to agree well with the CL results up to small positive values of $\beta\mu$. Above this regime, the LW study appears to overestimate the density already in the spin-balanced case, as opposed to the CL values.

This precise treatment of the spin-polarized UFG at finite temperature may considered one of the biggest successes of the CL method in the context of non-relativistic physics so far and sets the stage for the search of possibly inhomogeneous superfluid sectors in the phase diagram both at finite and zero temperatures. Such phenomena, most notably FFLO phases with a spatially inhomogeneous order parameter, are not only of great current experimental interest but also reminiscent of the dynamics underlying the QCD phase diagram at finite baryon chemical potential and low temperature, where such inhomogeneous phases may also exist~\cite{Anglani_2014,Buballa_2015}. This highlights a phenomenological similarity between high-energy physics and cold-dilute quantum matter with the perspective that success in one of these fields likely will trigger progress in the~other.


\section{\label{sect:outlook}Summary and Outlook}

In this report we have discussed the origin and methods to address the sign problem as it appears across a wide range of
systems in relativistic and nonrelativistic quantum many-body physics. We have proposed a broad classification of the various methods into
{\it new variables}, {\it statistical}, and {\it complex plane} approaches. After a brief overview of each of those categories,
we focused on the complex form of stochastic quantization, namely CL.

Complex Langevin first appeared in the 1980s as a natural generalization of stochastic quantization for cases with a complex action.
Operationally, stochastic quantization does not require a probability to be defined (there is no Metropolis accept/reject step), which made field
complexification (itself needed in the presence of a sign or complex phase problem) appear more like a feature than a bug. From the
mathematical standpoint, however, the challenges seemed daunting. It took decades for the community to begin to understand the properties and
behavior of CL, to clarify the origin of its problems and limitations, and to propose solutions. Some of that progress was enabled by advances in
hardware, as modern personal computers are powerful enough to run small but useful quantum field theory calculations with little wait; 
that was certainly not the case in the 1980s, nor in the 1990s when initial explorations of this method were underway.

As computer power continues to grow, and given the overall progress made during the last decade, there is good reason for optimism.
Furthermore, there are now several groups and international collaborations around the world applying CL methods to many systems, which
generate a wider range of situations than ever before from which insight can be gained. The most remarkable step forward, in the latest chapter
in the history of CL, is the derivation of conditions for correctness and how they relate to behavior at the boundaries of the integration region (at
infinity and at zeros of the complex weight). This new understanding has spawned new practical solutions such as gauge cooling, dynamic
stabilization, and modified actions (regulators), which have enabled more applications than previously thought possible.
Continued studies of CL will help illuminate when the method is reliable and when it is not. A more detailed understanding of the structure of
the problems in CL might help develop new methods to ameliorate or solve those problems.

\section*{Acknowledgements}

We are especially grateful to G. Aarts, F. Attanasio, G. Basar, S. Chandrasekharan, P. de Forcrand, C. Gattringer, E. Huffman, K. Langfeld, and E. Seiler for their extremely useful comments on earlier versions of this manuscript. We would also like to acknowledge G. Aarts, C. Gattringer, C. Ratti, D. Sexty, and H. Singh for kindly allowing us to reproduce their figures.

This material is based upon work supported by the National Science Foundation under Grant No. PHY{1452635} (Computational Physics Program). C.E.B. acknowledges support from the United States Department of Energy through the Computational Science Graduate Fellowship (DOE CSGF) under grant number DE-FG02-97ER25308. J.B. acknowledges support by the Deutsche Forschungsgemeinschaft (DFG, German Research Foundation) under grant BR 4005/4-1 (Heisenberg program). J.B. and F.E. acknowledge support by the DFG grant BR 4005/5-1. J.B. and L.R. acknowledge support by HIC for FAIR within the LOEWE program of the State of Hesse. J.B. acknowledges support by the DFG -- Projektnummer 279384907 -- SFB 1245.

\bibliographystyle{mybst}
\bibliography{CLReview_bib.bib}

\begin{thebibliography}{100}
\providecommand{\url}[1]{\texttt{#1}}
\providecommand{\urlprefix}{URL }
\providecommand{\eprint}[2][]{\url{#2}}

\bibitem{BetheSalpeterBook}
H.~A. {Bethe}, E.~E. {Salpeter}, \emph{Quantum Mechanics of One- and
  Two-Electron Atoms} (1957).

\bibitem{DiracQuote}
P.~A.~M. Dirac, Proceedings of the Royal Society of London. Series A,
  Containing Papers of a Mathematical and Physical Character 123 (1929).

\bibitem{BARRETT2013131}
B.~R. Barrett, P.~Navratil, J.~P. Vary, \emph{Ab initio no core shell model},
  Progress in Particle and Nuclear Physics 69 (2013) 131 .

\bibitem{Johnson:2018hrx}
C.~W. Johnson, W.~E. Ormand, K.~S. McElvain, H.~Shan, \emph{{BIGSTICK: A
  flexible configuration-interaction shell-model code}}  (2018),
  \eprint{1801.08432}.

\bibitem{RevModPhys.79.291}
R.~J. Bartlett, M.~Musia\l{}, \emph{Coupled-cluster theory in quantum
  chemistry}, Rev. Mod. Phys. 79 (2007) 291.

\bibitem{Hagen:2013nca}
G.~Hagen, T.~Papenbrock, M.~Hjorth-Jensen, D.~J. Dean, \emph{{Coupled-cluster
  computations of atomic nuclei}}, Rept. Prog. Phys. 77 (2014) 096302,
  \eprint{1312.7872}.

\bibitem{Burgess:2007pt}
C.~P. Burgess, \emph{Introduction to Effective Field Theory}, Ann. Rev. Nucl.
  Part. Sci. 57 (2007) 329, \eprint{hep-th/0701053}.

\bibitem{RevModPhys.81.1773}
E.~Epelbaum, H.-W. Hammer, U.-G. Mei\ss{}ner, \emph{Modern theory of nuclear
  forces}, Rev. Mod. Phys. 81 (2009) 1773.

\bibitem{Machleidt:2011zz}
R.~Machleidt, D.~R. Entem, \emph{{Chiral effective field theory and nuclear
  forces}}, Phys. Rept. 503 (2011) 1, \eprint{1105.2919}.

\bibitem{Shankar:1996vk}
R.~Shankar, \emph{{Effective field theory in condensed matter physics}}, in:
  \emph{{Conceptual foundations of quantum field theory. Proceedings, Symposium
  and Workshop, Boston, USA, March 1-3, 1996}} (1996), pp. 47--55,
  \eprint{cond-mat/9703210}.

\bibitem{doi:10.1142.6826}
A.~M.~J. Schakel, \emph{Boulevard of Broken Symmetries} (World Scientific,
  2008), \eprint{https://www.worldscientific.com/doi/pdf/10.1142/6826}.

\bibitem{Metropolis}
N.~Metropolis, A.~W. Rosenbluth, M.~N. Rosenbluth, A.~H. Teller, E.~Teller,
  \emph{Equation of State Calculations by Fast Computing Machines}, The Journal
  of Chemical Physics 21 (1953) 1087.

\bibitem{PhysRevB.41.9301}
E.~Y. Loh, J.~E. Gubernatis, R.~T. Scalettar, S.~R. White, D.~J. Scalapino,
  R.~L. Sugar, \emph{Sign problem in the numerical simulation of many-electron
  systems}, Phys. Rev. B 41 (1990) 9301.

\bibitem{KOONIN19971}
S.~Koonin, D.~Dean, K.~Langanke, \emph{Shell model Monte Carlo methods}, Phys.
  Rept. 278 (1997) 1 .

\bibitem{Alhassid:2016ojg}
Y.~Alhassid, \emph{{The shell model Monte Carlo approach to level densities:
  recent developments and perspectives}}, Eur. Phys. J. A51 (2015) 171,
  \eprint{1601.00107}.

\bibitem{Gattringer:2016kco}
C.~Gattringer, K.~Langfeld, \emph{{Approaches to the sign problem in lattice
  field theory}}, Int. J. Mod. Phys. A 31 (2016) 1643007, \eprint{1603.09517}.

\bibitem{Bongiovanni:2016ess}
L.~Bongiovanni, \emph{{Numerical methods for the sign problem in Lattice Field
  Theory}}, Ph.D. thesis, Swansea U. (2015), \eprint{1603.06458}.

\bibitem{Aarts:2015tyj}
G.~Aarts, \emph{{Introductory lectures on lattice QCD at nonzero baryon
  number}}, J. Phys. Conf. Ser. 706 (2016) 022004, \eprint{1512.05145}.

\bibitem{PhysRevLett.61.2635}
A.~M. Ferrenberg, R.~H. Swendsen, \emph{New Monte Carlo technique for studying
  phase transitions}, Phys. Rev. Lett. 61 (1988) 2635.

\bibitem{PhysRevB.28.4059}
J.~E. Hirsch, \emph{Discrete Hubbard-Stratonovich transformation for fermion
  lattice models}, Phys. Rev. B 28 (1983) 4059.

\bibitem{PhysRevC.78.024001}
D.~Lee, \emph{Ground state energy at unitarity}, Phys. Rev. C 78 (2008) 024001.

\bibitem{PhysRevB.48.589}
G.~G. Batrouni, P.~de~Forcrand, \emph{Fermion sign problem: Decoupling
  transformation and simulation algorithm}, Phys. Rev. B 48 (1993) 589.

\bibitem{PhysRevB.71.155115}
C.~Wu, S.-C. Zhang, \emph{Sufficient condition for absence of the sign problem
  in the fermionic quantum Monte Carlo algorithm}, Phys. Rev. B 71 (2005)
  155115.

\bibitem{doi:10.1146/annurev-conmatphys-033117-054307}
Z.-X. Li, H.~Yao, \emph{{Sign-Problem-Free Fermionic Quantum Monte Carlo:
  Developments and Applications}}, Ann. Rev. Condensed Matter Phys. 10 (2019)
  337, \eprint{1805.08219}.

\bibitem{PhysRevB.42.2282}
G.~G. Batrouni, R.~T. Scalettar, \emph{Anomalous decouplings and the fermion
  sign problem}, Phys. Rev. B 42 (1990) 2282.

\bibitem{PhysRevB.56.15001}
F.~F. Assaad, M.~Imada, D.~J. Scalapino, \emph{Charge and spin structures of a
  ${d}_{{x}^{2}\ensuremath{-}{y}^{2}}$ superconductor in the proximity of an
  antiferromagnetic Mott insulator}, Phys. Rev. B 56 (1997) 15001.

\bibitem{doi:10.1143/JPSJ.66.1872}
Y.~Motome, M.~Imada, \emph{A Quantum Monte Carlo Method and Its Applications to
  Multi-Orbital Hubbard Models}, J. Phys. Soc. Jpn. 66 (1997) 1872.

\bibitem{PhysRevLett.120.107201}
T.~Sato, F.~F. Assaad, T.~Grover, \emph{Quantum Monte Carlo Simulation of
  Frustrated Kondo Lattice Models}, Phys. Rev. Lett. 120 (2018) 107201.

\bibitem{PhysRevB.63.155114}
S.~Capponi, F.~F. Assaad, \emph{Spin and charge dynamics of the ferromagnetic
  and antiferromagnetic two-dimensional half-filled Kondo lattice model}, Phys.
  Rev. B 63 (2001) 155114.

\bibitem{PhysRevLett.83.796}
F.~F. Assaad, \emph{Quantum Monte Carlo Simulations of the Half-Filled
  Two-Dimensional Kondo Lattice Model}, Phys. Rev. Lett. 83 (1999) 796.

\bibitem{PhysRevLett.86.2050}
F.~Wang, D.~P. Landau, \emph{Efficient, Multiple-Range Random Walk Algorithm to
  Calculate the Density of States}, Phys. Rev. Lett. 86 (2001) 2050.

\bibitem{PhysRevD.85.056010}
A.~Bazavov, B.~A. Berg, D.~Du, Y.~Meurice, \emph{Density of states and Fisher's
  zeros in compact $U(1)$ pure gauge theory}, Phys. Rev. D 85 (2012) 056010.

\bibitem{PhysRevLett.109.111601}
K.~Langfeld, B.~Lucini, A.~Rago, \emph{Density of States in Gauge Theories},
  Phys. Rev. Lett. 109 (2012) 111601.

\bibitem{Fodor:2007vv}
Z.~Fodor, S.~D. Katz, C.~Schmidt, \emph{{The density of states method at
  non-zero chemical potential}}, JHEP 03 (2007) 121, \eprint{hep-lat/0701022}.

\bibitem{PhysRevD.90.094502}
K.~Langfeld, B.~Lucini, \emph{Density of states approach to dense quantum
  systems}, Phys. Rev. D 90 (2014) 094502.

\bibitem{1742-6596-631-1-012063}
K.~Langfeld, B.~Lucini, A.~Rago, R.~Pellegrini, L.~Bongiovanni, \emph{{The
  density of states approach for the simulation of finite density quantum field
  theories}}, J. Phys. Conf. Ser. 631 (2015) 012063, \eprint{1503.00450}.

\bibitem{PhysRevD.88.071502}
K.~Langfeld, J.~M. Pawlowski, \emph{Two-color QCD with heavy quarks at finite
  densities}, Phys. Rev. D 88 (2013) 071502.

\bibitem{PhysRevLett.61.2054}
A.~Gocksch, \emph{Simulating Lattice QCD at Finite Density}, Phys. Rev. Lett.
  61 (1988) 2054.

\bibitem{Schmidt:2006us}
C.~Schmidt, \emph{{Lattice QCD at finite density}}, PoS LATTICE2006 (2006) 021,
  \eprint{hep-lat/0610116}.

\bibitem{PhysRevD.77.014508}
S.~Ejiri, \emph{Existence of the critical point in finite density lattice QCD},
  Phys. Rev. D 77 (2008) 014508.

\bibitem{PhysRevD.78.074507}
S.~Ejiri, \emph{Canonical partition function and finite density phase
  transition in lattice QCD}, Phys. Rev. D 78 (2008) 074507.

\bibitem{10.1093/ptep/pts005}
S.~Ejiri, K.~Kanaya, T.~Umeda, \emph{{Ab initio study of QCD thermodynamics on
  the lattice at zero and finite densities}}, PTEP 2012 (2012) 01A104,
  \eprint{1205.5347}.

\bibitem{GATTRINGER2015545}
C.~Gattringer, P.~Torek, \emph{Density of states method for the Z3 spin model},
  Phys. Lett. B 747 (2015) 545 .

\bibitem{GIULIANI2016627}
M.~Giuliani, C.~Gattringer, P.~Torek, \emph{Developing and testing the density
  of states FFA method in the SU(3) spin model}, Nucl. Phys. B 913 (2016) 627 .

\bibitem{robbins1951}
H.~Robbins, S.~Monro, \emph{A Stochastic Approximation Method}, Ann. Math.
  Statist. 22 (1951) 400.

\bibitem{GIULIANI2017166}
M.~Giuliani, C.~Gattringer, \emph{Density of States FFA analysis of SU(3)
  lattice gauge theory at a finite density of color sources}, Phys. Lett. B 773
  (2017) 166 .

\bibitem{ROSSI1984105}
P.~Rossi, U.~Wolff, \emph{Lattice QCD with fermions at strong coupling: A dimer
  system}, Nucl. Phys. B 248 (1984) 105 .

\bibitem{PhysRevB.61.10725}
F.~H\'ebert, G.~G. Batrouni, H.~Mabilat, \emph{Exact duality and dual Monte
  Carlo simulation for the bosonic Hubbard model}, Phys. Rev. B 61 (2000)
  10725.

\bibitem{PhysRevD.75.065012}
M.~G. Endres, \emph{Method for simulating $O(N)$ lattice models at finite
  density}, Phys. Rev. D 75 (2007) 065012.

\bibitem{PhysRevLett.87.160601}
N.~Prokof'ev, B.~Svistunov, \emph{Worm Algorithms for Classical Statistical
  Models}, Phys. Rev. Lett. 87 (2001) 160601.

\bibitem{PhysRevE.66.046701}
O.~F. Sylju\aa{}sen, A.~W. Sandvik, \emph{Quantum Monte Carlo with directed
  loops}, Phys. Rev. E 66 (2002) 046701.

\bibitem{PhysRevD.81.125007}
D.~Banerjee, S.~Chandrasekharan, \emph{Finite size effects in the presence of a
  chemical potential: A study in the classical nonlinear $O(2)$ sigma model},
  Phys. Rev. D 81 (2010) 125007.

\bibitem{PhysRevLett.106.222001}
Y.~D. Mercado, H.~G. Evertz, C.~Gattringer, \emph{QCD Phase Diagram According
  to the Center Group}, Phys. Rev. Lett. 106 (2011) 222001.

\bibitem{GATTRINGER2011242}
C.~Gattringer, \emph{Flux representation of an effective Polyakov loop model
  for QCD thermodynamics}, Nucl. Phys. B 850 (2011) 242 .

\bibitem{Fromm:2011qi}
M.~Fromm, J.~Langelage, S.~Lottini, O.~Philipsen, \emph{{The QCD deconfinement
  transition for heavy quarks and all baryon chemical potentials}}, JHEP 01
  (2012) 042, \eprint{1111.4953}.

\bibitem{MERCADO20121920}
Y.~D. Mercado, H.~G. Evertz, C.~Gattringer, \emph{Worm algorithms for the
  3-state Potts model with magnetic field and chemical potential}, Computer
  Physics Communications 183 (2012) 1920 .

\bibitem{MERCADO2012737}
Y.~D. Mercado, C.~Gattringer, \emph{Monte Carlo simulation of the SU(3) spin
  model with chemical potential in a flux representation}, Nucl. Phys. B 862
  (2012) 737 .

\bibitem{Gattringer:2012df}
C.~Gattringer, T.~Kloiber, \emph{{Lattice study of the Silver Blaze phenomenon
  for a charged scalar $\phi^4$ field}}, Nucl. Phys. B 869 (2013) 56,
  \eprint{1206.2954}.

\bibitem{Gattringer:2012ap}
C.~Gattringer, T.~Kloiber, \emph{{Spectroscopy in finite density lattice field
  theory: An exploratory study in the relativistic Bose gas}}, Phys. Lett. B
  720 (2013) 210, \eprint{1212.3770}.

\bibitem{PhysRevD.92.114508}
C.~Gattringer, T.~Kloiber, M.~M\"uller-Preussker, \emph{Dual simulation of the
  two-dimensional lattice U(1) gauge-Higgs model with a topological term},
  Phys. Rev. D 92 (2015) 114508.

\bibitem{Gattringer:2018dlw}
C.~Gattringer, D.~G{\"o}schl, T.~Sulejmanpasic, \emph{{Dual simulation of the
  2d U(1) gauge Higgs model at topological angle $\theta = \pi\,$: Critical
  endpoint behavior}}, Nucl. Phys. B 935 (2018) 344, \eprint{1807.07793}.

\bibitem{Sulejmanpasic:2019ytl}
T.~Sulejmanpasic, C.~Gattringer, \emph{{Abelian gauge theories on the lattice:
  $\theta$-terms and compact gauge theory with(out) monopoles}}, Nucl. Phys. B
  943 (2019) 114616, \eprint{1901.02637}.

\bibitem{Vairinhos:2014uxa}
H.~Vairinhos, P.~de~Forcrand, \emph{{Lattice gauge theory without link
  variables}}, JHEP 12 (2014) 038, \eprint{1409.8442}.

\bibitem{Vairinhos:2015ewa}
H.~Vairinhos, P.~de~Forcrand, \emph{{Integrating out lattice gauge fields}},
  PoS CPOD2014 (2015) 061, \eprint{1506.07007}.

\bibitem{WOLFF2009491}
U.~Wolff, \emph{Simulating the all-order strong coupling expansion I: Ising
  model demo}, Nucl. Phys. B 810 (2009) 491.

\bibitem{WOLFF2009549}
U.~Wolff, \emph{Simulating the all-order hopping expansion II: Wilson
  fermions}, Nucl. Phys. B 814 (2009) 549.

\bibitem{WOLFF2010254}
U.~Wolff, \emph{Simulating the all-order strong coupling expansion III: O(N)
  sigma/loop models}, Nucl. Phys. B 824 (2010) 254.

\bibitem{WOLFF2010520}
U.~Wolff, \emph{Simulating the all-order strong coupling expansion IV: CP(N-1)
  as a loop model}, Nucl. Phys. B 832 (2010) 520.

\bibitem{PhysRevLett.104.112005}
P.~de~Forcrand, M.~Fromm, \emph{Nuclear Physics from Lattice QCD at Strong
  Coupling}, Phys. Rev. Lett. 104 (2010) 112005.

\bibitem{Unger:2011it}
W.~Unger, P.~de~Forcrand, \emph{{Continuous Time Monte Carlo for Lattice QCD in
  the Strong Coupling Limit}}, J. Phys. G38 (2011) 124190, \eprint{1107.1553}.

\bibitem{PhysRevLett.83.3116}
S.~Chandrasekharan, U.-J. Wiese, \emph{Meron-Cluster Solution of Fermion Sign
  Problems}, Phys. Rev. Lett. 83 (1999) 3116.

\bibitem{PhysRevD.82.025007}
S.~Chandrasekharan, \emph{Fermion bag approach to lattice field theories},
  Phys. Rev. D 82 (2010) 025007.

\bibitem{Chandrasekharan:2013rpa}
S.~Chandrasekharan, \emph{{Fermion Bag Approach to Fermion Sign Problems}},
  Eur. Phys. J. A49 (2013) 90, \eprint{1304.4900}.

\bibitem{PhysRevD.96.114502}
E.~Huffman, S.~Chandrasekharan, \emph{Fermion bag approach to Hamiltonian
  lattice field theories in continuous time}, Phys. Rev. D 96 (2017) 114502.

\bibitem{Chandrasekharan2011}
S.~Chandrasekharan, A.~Li, \emph{Fermion bag approach to the sign problem in
  strongly coupled lattice QED with Wilson fermions}, JHEP 2011 (2011) 18.

\bibitem{PhysRevLett.108.140404}
S.~Chandrasekharan, A.~Li, \emph{Fermion Bags, Duality, and the Three
  Dimensional Massless Lattice Thirring Model}, Phys. Rev. Lett. 108 (2012)
  140404.

\bibitem{PhysRevD.86.021701}
S.~Chandrasekharan, \emph{Solutions to sign problems in lattice Yukawa models},
  Phys. Rev. D 86 (2012) 021701.

\bibitem{PhysRevB.89.111101}
E.~F. Huffman, S.~Chandrasekharan, \emph{Solution to sign problems in
  half-filled spin-polarized electronic systems}, Phys. Rev. B 89 (2014)
  111101.

\bibitem{PhysRevB.26.5033}
J.~E. Hirsch, R.~L. Sugar, D.~J. Scalapino, R.~Blankenbecler, \emph{Monte Carlo
  simulations of one-dimensional fermion systems}, Phys. Rev. B 26 (1982) 5033.

\bibitem{PhysRevA.85.063624}
M.~G. Endres, \emph{Lattice theory for nonrelativistic fermions in one spatial
  dimension}, Phys. Rev. A 85 (2012) 063624.

\bibitem{PhysRevLett.109.250403}
M.~G. Endres, \emph{Transdimensional Equivalence of Universal Constants for
  Fermi Gases at Unitarity}, Phys. Rev. Lett. 109 (2012) 250403.

\bibitem{PhysRevA.87.063617}
M.~G. Endres, \emph{Numerical study of unitary fermions in one spatial
  dimension}, Phys. Rev. A 87 (2013) 063617.

\bibitem{PhysRevD.97.074506}
C.~Gattringer, \emph{Baryon bags in strong coupling QCD}, Phys. Rev. D 97
  (2018) 074506.

\bibitem{PhysRevD.80.071503}
U.~Wenger, \emph{Efficient simulation of relativistic fermions via vertex
  models}, Phys. Rev. D 80 (2009) 071503.

\bibitem{PhysRevD.97.054501}
V.~Ayyar, S.~Chandrasekharan, J.~Rantaharju, \emph{Benchmark results in the 2D
  lattice Thirring model with a chemical potential}, Phys. Rev. D 97 (2018)
  054501.

\bibitem{Drut:2012md}
J.~E. Drut, A.~N. Nicholson, \emph{{Lattice methods for strongly interacting
  many-body systems}}, J. Phys. G 40 (2013) 043101, \eprint{1208.6556}.

\bibitem{PhysRevB.91.241117}
Z.-X. Li, Y.-F. Jiang, H.~Yao, \emph{Solving the fermion sign problem in
  quantum Monte Carlo simulations by Majorana representation}, Phys. Rev. B 91
  (2015) 241117.

\bibitem{PhysRevLett.115.250601}
L.~Wang, Y.-H. Liu, M.~Iazzi, M.~Troyer, G.~Harcos, \emph{Split Orthogonal
  Group: A Guiding Principle for Sign-Problem-Free Fermionic Simulations},
  Phys. Rev. Lett. 115 (2015) 250601.

\bibitem{PhysRevLett.116.250601}
Z.~C. Wei, C.~Wu, Y.~Li, S.~Zhang, T.~Xiang, \emph{Majorana Positivity and the
  Fermion Sign Problem of Quantum Monte Carlo Simulations}, Phys. Rev. Lett.
  116 (2016) 250601.

\bibitem{PhysRevLett.92.257002}
J.-W. Chen, D.~B. Kaplan, \emph{Lattice Theory for Low Energy Fermions at
  Nonzero Chemical Potential}, Phys. Rev. Lett. 92 (2004) 257002.

\bibitem{PhysRevLett.117.267002}
Z.-X. Li, Y.-F. Jiang, H.~Yao, \emph{Majorana-Time-Reversal Symmetries: A
  Fundamental Principle for Sign-Problem-Free Quantum Monte Carlo Simulations},
  Phys. Rev. Lett. 117 (2016) 267002.

\bibitem{Wei:2017wns}
Z.-C. Wei, \emph{{Semigroup Approach to the Sign Problem in Quantum Monte Carlo
  Simulations}}  (2017), \eprint{1712.09412}.

\bibitem{Li_2015}
Z.-X. Li, Y.-F. Jiang, H.~Yao, \emph{{Fermion-sign-free
  Majorana-quantum-Monte-Carlo studies of quantum critical phenomena of Dirac
  fermions in two dimensions}}, New J. Phys. 17 (2015) 085003,
  \eprint{1411.7383}.

\bibitem{PhysRevB.92.235129}
Y.-H. Liu, L.~Wang, \emph{Quantum Monte Carlo study of mass-imbalanced Hubbard
  models}, Phys. Rev. B 92 (2015) 235129.

\bibitem{s41467-017-00167-6}
Z.-X. Li, Y.-F. Jiang, S.-K. Jian, H.~Yao, \emph{{Fermion-induced quantum
  critical points}}, Nature Commun. 8 (2017) 314.

\bibitem{PhysRevB.96.155112}
S.-K. Jian, H.~Yao, \emph{Fermion-induced quantum critical points in
  three-dimensional Weyl semimetals}, Phys. Rev. B 96 (2017) 155112.

\bibitem{PhysRevB.96.035129}
T.~Hayata, A.~Yamamoto, \emph{Quantum Monte Carlo simulation of a
  two-dimensional Majorana lattice model}, Phys. Rev. B 96 (2017) 035129.

\bibitem{PhysRevB.41.811}
E.~Dagotto, A.~Moreo, R.~L. Sugar, D.~Toussaint, \emph{Binding of holes in the
  Hubbard model}, Phys. Rev. B 41 (1990) 811.

\bibitem{Alford:1998sd}
M.~G. Alford, A.~Kapustin, F.~Wilczek, \emph{{Imaginary chemical potential and
  finite fermion density on the lattice}}, Phys. Rev. D 59 (1999) 054502,
  \eprint{hep-lat/9807039}.

\bibitem{DEFORCRAND2003170}
P.~de~Forcrand, O.~Philipsen, \emph{The QCD phase diagram for three degenerate
  flavors and small baryon density}, Nucl. Phys. B 673 (2003) 170 .

\bibitem{PhysRevD.67.014505}
M.~D'Elia, M.-P. Lombardo, \emph{Finite density QCD via an imaginary chemical
  potential}, Phys. Rev. D 67 (2003) 014505.

\bibitem{PhysRevD.70.074509}
M.~D'Elia, M.-P. Lombardo, \emph{QCD thermodynamics from an imaginary
  ${\ensuremath{\mu}}_{B}$: Results on the four flavor lattice model}, Phys.
  Rev. D 70 (2004) 074509.

\bibitem{PhysRevLett.105.152001}
P.~de~Forcrand, O.~Philipsen, \emph{Constraining the QCD Phase Diagram by
  Tricritical Lines at Imaginary Chemical Potential}, Phys. Rev. Lett. 105
  (2010) 152001.

\bibitem{Philipsen:2012nu}
O.~Philipsen, \emph{{The QCD equation of state from the lattice}}, Prog. Part.
  Nucl. Phys. 70 (2013) 55, \eprint{1207.5999}.

\bibitem{PhysRevD.85.094512}
P.~Cea, L.~Cosmai, M.~D'Elia, A.~Papa, F.~Sanfilippo, \emph{Critical line of
  two-flavor QCD at finite isospin or baryon densities from imaginary chemical
  potentials}, Phys. Rev. D 85 (2012) 094512.

\bibitem{Bonati:2014kpa}
C.~Bonati, P.~de~Forcrand, M.~D'Elia, O.~Philipsen, F.~Sanfilippo,
  \emph{{Chiral phase transition in two-flavor QCD from an imaginary chemical
  potential}}, Phys. Rev. D 90 (2014) 074030, \eprint{1408.5086}.

\bibitem{Philipsen:2016hkv}
O.~Philipsen, C.~Pinke, \emph{{The $N_f=2$ QCD chiral phase transition with
  Wilson fermions at zero and imaginary chemical potential}}, Phys. Rev. D 93
  (2016) 114507, \eprint{1602.06129}.

\bibitem{Gunther:2016vcp}
J.~N. Guenther, R.~Bellwied, S.~Borsanyi, Z.~Fodor, S.~D. Katz, A.~Pasztor,
  C.~Ratti, K.~K. Szab{\'{o}}, \emph{{The QCD equation of state at finite
  density from analytical continuation}}, Nucl. Phys. A 967 (2017) 720,
  \eprint{1607.02493}.

\bibitem{Borsanyi:2020fev}
S.~Borsanyi, Z.~Fodor, J.~N. Guenther, R.~Kara, S.~D. Katz, P.~Parotto,
  A.~Pasztor, C.~Ratti, K.~K. Szabo, \emph{{The QCD crossover at finite
  chemical potential from lattice simulations}}, Phys. Rev. Lett. 125 (2020)
  052001, \eprint{2002.02821}.

\bibitem{Lombardo:2006yc}
M.~Lombardo, \emph{{Lattice QCD at finite density: Imaginary chemical
  potential}}, PoS CPOD2006 (2006) 003, \eprint{hep-lat/0612017}.

\bibitem{DElia:2007bkz}
M.~D'Elia, F.~Di~Renzo, M.~P. Lombardo, \emph{{Strongly interacting quark-gluon
  plasma, and the critical behavior of QCD at imaginary $\mu$}}, Phys. Rev. D
  76 (2007) 114509, \eprint{0705.3814}.

\bibitem{Karbstein:2006er}
F.~Karbstein, M.~Thies, \emph{{How to get from imaginary to real chemical
  potential}}, Phys. Rev. D 75 (2007) 025003, \eprint{hep-th/0610243}.

\bibitem{Roberge:1986mm}
A.~Roberge, N.~Weiss, \emph{{Gauge Theories With Imaginary Chemical Potential
  and the Phases of {QCD}}}, Nucl. Phys. B 275 (1986) 734.

\bibitem{Roscher:2013aqa}
D.~Roscher, J.~Braun, J.-W. Chen, J.~E. Drut, \emph{{Fermi gases with imaginary
  mass imbalance and the sign problem in Monte Carlo calculations}}, J. Phys. G
  41 (2014) 055110, \eprint{1306.0798}.

\bibitem{PhysRevLett.110.130404}
J.~Braun, J.-W. Chen, J.~Deng, J.~E. Drut, B.~Friman, C.-T. Ma, Y.-D. Tsai,
  \emph{Imaginary Polarization as a Way to Surmount the Sign Problem in Ab
  Initio Calculations of Spin-Imbalanced Fermi Gases}, Phys. Rev. Lett. 110
  (2013) 130404.

\bibitem{PhysRevA.92.063609}
A.~C. Loheac, J.~Braun, J.~E. Drut, D.~Roscher, \emph{Thermal equation of state
  of polarized fermions in one dimension via complex chemical potentials},
  Phys. Rev. A 92 (2015) 063609.

\bibitem{PhysRevLett.114.050404}
J.~Braun, J.~E. Drut, D.~Roscher, \emph{Zero-Temperature Equation of State of
  Mass-Imbalanced Resonant Fermi Gases}, Phys. Rev. Lett. 114 (2015) 050404.

\bibitem{PRD96094506}
L.~Rammelm\"uller, W.~J. Porter, J.~E. Drut, J.~Braun, \emph{Surmounting the
  sign problem in nonrelativistic calculations: A case study with
  mass-imbalanced fermions}, Phys. Rev. D 96 (2017) 094506.

\bibitem{Allton:2003vx}
C.~Allton, S.~Ejiri, S.~Hands, O.~Kaczmarek, F.~Karsch, E.~Laermann,
  C.~Schmidt, \emph{{Equation of state for two flavor QCD at nonzero chemical
  potential}}, Phys. Rev. D 68 (2003) 014507, \eprint{hep-lat/0305007}.

\bibitem{Ejiri:2005uv}
S.~Ejiri, F.~Karsch, E.~Laermann, C.~Schmidt, \emph{{Isentropic equation of
  state of 2-flavor QCD}}, Phys. Rev. D 73 (2006) 054506,
  \eprint{hep-lat/0512040}.

\bibitem{deForcrand:2007rq}
P.~de~Forcrand, S.~Kim, O.~Philipsen, \emph{{A QCD chiral critical point at
  small chemical potential: Is it there or not?}}, PoS LATTICE2007 (2007) 178,
  \eprint{0711.0262}.

\bibitem{Endrodi:2011gv}
G.~Endrodi, Z.~Fodor, S.~D. Katz, K.~K. Szabo, \emph{{The QCD phase diagram at
  nonzero quark density}}, JHEP 04 (2011) 001, \eprint{1102.1356}.

\bibitem{Bazavov:2017dus}
A.~Bazavov, et~al., \emph{{QCD equation of state to $\mathcal{O}(\mu_B^6)$ from
  lattice QCD}}, Phys. Rev. D 95 (2017) 054504, \eprint{1701.04325}.

\bibitem{Sharma:2017jwb}
S.~Sharma, \emph{{The QCD Equation of state and critical end-point estimates at
  $\mathcal O(\mu_B^6)$}}, Nucl. Phys. A 967 (2017) 728, \eprint{1704.05969}.

\bibitem{Brandt:2018omg}
B.~B. Brandt, G.~Endrodi, \emph{{Reliability of Taylor expansions in QCD}},
  Phys. Rev. D 99 (2019) 014518, \eprint{1810.11045}.

\bibitem{ROM1997382}
N.~Rom, D.~Charutz, D.~Neuhauser, \emph{Shifted-contour auxiliary-field Monte
  Carlo: circumventing the sign difficulty for electronic-structure
  calculations}, Chem. Phys. Lett. 270 (1997) 382 .

\bibitem{Witten:2010cx}
E.~Witten, \emph{{Analytic Continuation Of Chern-Simons Theory}}, AMS/IP Stud.
  Adv. Math. 50 (2011) 347, \eprint{1001.2933}.

\bibitem{Witten:2010zr}
E.~Witten, \emph{{A New Look At The Path Integral Of Quantum Mechanics}}
  (2010), \eprint{1009.6032}.

\bibitem{Aarts2013c}
G.~Aarts, \emph{Lefschetz thimbles and stochastic quantization: Complex actions
  in the complex plane}, Phys. Rev. D 88 (2013) 094501.

\bibitem{Alexandru:2015sua}
A.~Alexandru, G.~Ba\ifmmode~\mbox{\c{s}}\else \c{s}\fi{}ar, P.~F. Bedaque,
  G.~W. Ridgway, N.~C. Warrington, \emph{{Sign problem and Monte Carlo
  calculations beyond Lefschetz thimbles}}, JHEP 05 (2016) 053,
  \eprint{1512.08764}.

\bibitem{PhysRevLett.117.081602}
A.~Alexandru, G.~Ba\ifmmode~\mbox{\c{s}}\else \c{s}\fi{}ar, P.~F. Bedaque,
  S.~Vartak, N.~C. Warrington, \emph{Monte Carlo Study of Real Time Dynamics on
  the Lattice}, Phys. Rev. Lett. 117 (2016) 081602.

\bibitem{PhysRevD.95.114501}
A.~Alexandru, G.~Ba\ifmmode~\mbox{\c{s}}\else \c{s}\fi{}ar, P.~F. Bedaque,
  G.~Ridgway, \emph{Schwinger-Keldysh formalism on the lattice: A faster
  algorithm and its application to field theory}, Phys. Rev. D 95 (2017)
  114501.

\bibitem{PhysRevD.96.034513}
A.~Alexandru, G.~Ba\ifmmode~\mbox{\c{s}}\else \c{s}\fi{}ar, P.~F. Bedaque,
  N.~C. Warrington, \emph{Tempered transitions between thimbles}, Phys. Rev. D
  96 (2017) 034513.

\bibitem{PhysRevD.93.094514}
A.~Alexandru, G.~Ba\ifmmode~\mbox{\c{s}}\else \c{s}\fi{}ar, P.~F. Bedaque,
  G.~W. Ridgway, N.~C. Warrington, \emph{Fast estimator of Jacobians in the
  Monte Carlo integration on Lefschetz thimbles}, Phys. Rev. D 93 (2016)
  094514.

\bibitem{Bluecher:2018sgj}
S.~Bluecher, J.~M. Pawlowski, M.~Scherzer, M.~Schlosser, I.-O. Stamatescu,
  S.~Syrkowski, F.~P.~G. Ziegler, \emph{{Reweighting Lefschetz Thimbles}},
  SciPost Phys. 5 (2018) 044, \eprint{1803.08418}.

\bibitem{PhysRevD.86.074506}
M.~Cristoforetti, F.~Di~Renzo, L.~Scorzato, \emph{New approach to the sign
  problem in quantum field theories: High density QCD on a Lefschetz thimble},
  Phys. Rev. D 86 (2012) 074506.

\bibitem{PhysRevD.88.051501}
M.~Cristoforetti, F.~Di~Renzo, A.~Mukherjee, L.~Scorzato, \emph{Monte Carlo
  simulations on the Lefschetz thimble: Taming the sign problem}, Phys. Rev. D
  88 (2013) 051501.

\bibitem{PhysRevD.88.051502}
A.~Mukherjee, M.~Cristoforetti, L.~Scorzato, \emph{Metropolis Monte Carlo
  integration on the Lefschetz thimble: Application to a one-plaquette model},
  Phys. Rev. D 88 (2013) 051502.

\bibitem{Fujii:2013sra}
H.~Fujii, D.~Honda, M.~Kato, Y.~Kikukawa, S.~Komatsu, T.~Sano, \emph{{Hybrid
  Monte Carlo on Lefschetz thimbles - A study of the residual sign problem}},
  JHEP 10 (2013) 147, \eprint{1309.4371}.

\bibitem{PhysRevD.89.114505}
M.~Cristoforetti, F.~Di~Renzo, G.~Eruzzi, A.~Mukherjee, C.~Schmidt,
  L.~Scorzato, C.~Torrero, \emph{An efficient method to compute the residual
  phase on a Lefschetz thimble}, Phys. Rev. D 89 (2014) 114505.

\bibitem{Kanazawa2015}
T.~Kanazawa, Y.~Tanizaki, \emph{Structure of Lefschetz thimbles in simple
  fermionic systems}, JHEP 2015 (2015) 44.

\bibitem{PhysRevD.91.101701}
Y.~Tanizaki, H.~Nishimura, K.~Kashiwa, \emph{Evading the sign problem in the
  mean-field approximation through Lefschetz-thimble path integral}, Phys. Rev.
  D 91 (2015) 101701.

\bibitem{Tanizaki:2016cou}
Y.~Tanizaki, Y.~Hidaka, T.~Hayata, \emph{{Lefschetz-thimble approach to the
  Silver Blaze problem of one-site fermion model}}, PoS LATTICE2016 (2016) 030,
  \eprint{1610.00393}.

\bibitem{PhysRevD.95.014502}
A.~Alexandru, G.~Ba\ifmmode~\mbox{\c{s}}\else \c{s}\fi{}ar, P.~F. Bedaque,
  G.~W. Ridgway, N.~C. Warrington, \emph{Monte Carlo calculations of the finite
  density Thirring model}, Phys. Rev. D 95 (2017) 014502.

\bibitem{PhysRevD.98.034506}
A.~Alexandru, G.~Ba\ifmmode~\mbox{\c{s}}\else \c{s}\fi{}ar, P.~F. Bedaque,
  H.~Lamm, S.~Lawrence, \emph{Finite density ${\mathrm{QED}}_{1+1}$ near
  Lefschetz thimbles}, Phys. Rev. D 98 (2018) 034506.

\bibitem{Fukushima:2015qza}
K.~Fukushima, Y.~Tanizaki, \emph{{Hamilton dynamics for Lefschetz-thimble
  integration akin to the complex Langevin method}}, PTEP 2015 (2015) 111A01,
  \eprint{1507.07351}.

\bibitem{Hayata:2015lzj}
T.~Hayata, Y.~Hidaka, Y.~Tanizaki, \emph{{Complex saddle points and the sign
  problem in complex Langevin simulation}}, Nucl. Phys. B 911 (2016) 94,
  \eprint{1511.02437}.

\bibitem{Nishimura:2017eiu}
J.~Nishimura, S.~Shimasaki, \emph{{Unification of the complex Langevin method
  and the Lefschetz thimble method}}, EPJ Web Conf. 175 (2018) 07018,
  \eprint{1710.07027}.

\bibitem{Ulybyshev:2017hbs}
M.~V. Ulybyshev, S.~N. Valgushev, \emph{{Path integral representation for the
  Hubbard model with reduced number of Lefschetz thimbles}}  (2017),
  \eprint{1712.02188}.

\bibitem{Ulybyshev2019fte}
M.~Ulybyshev, C.~Winterowd, S.~Zafeiropoulos, \emph{{Lefschetz thimbles
  decomposition for the Hubbard model on the hexagonal lattice}}  (2019),
  \eprint{1906.07678}.

\bibitem{Ulybyshev2019hfm}
M.~Ulybyshev, C.~Winterowd, S.~Zafeiropoulos, \emph{{Taming the sign problem of
  the finite density Hubbard model via Lefschetz thimbles}}  (2019),
  \eprint{1906.02726}.

\bibitem{Fukuma2019wbv}
M.~Fukuma, N.~Matsumoto, N.~Umeda, \emph{{Applying the tempered Lefschetz
  thimble method to the Hubbard model away from half-filling}}  (2019),
  \eprint{1906.04243}.

\bibitem{Fukuma:2017fjq}
M.~Fukuma, N.~Umeda, \emph{{Parallel tempering algorithm for integration over
  Lefschetz thimbles}}, PTEP 2017 (2017) 073B01, \eprint{1703.00861}.

\bibitem{Alexandru:2020wrj}
A.~Alexandru, G.~Ba\ifmmode~\mbox{\c{s}}\else \c{s}\fi{}ar, P.~F. Bedaque,
  N.~C. Warrington, \emph{{Complex Paths Around The Sign Problem}}  (2020),
  \eprint{2007.05436}.

\bibitem{Ohnishi2018}
{Ohnishi, Akira}, {Mori, Yuto}, {Kashiwa, Kouji}, \emph{Path optimization
  method for the sign problem}, EPJ Web Conf. 175 (2018) 07043.

\bibitem{PhysRevD96.111501}
Y.~Mori, K.~Kashiwa, A.~Ohnishi, \emph{Toward solving the sign problem with
  path optimization method}, Phys. Rev. D 96 (2017) 111501.

\bibitem{MoriLattice2019}
Y.~Mori, K.~Kashiwa, A.~Ohnishi, \emph{{The path optimization for the sign
  problem of low dimensional QCD}}, PoS LATTICE2019  (2019) 183,
  \eprint{1912.12050}.

\bibitem{Ohnishi2019POMandNN}
A.~Ohnishi, Y.~Mori, K.~Kashiwa, \emph{{Path optimization method with use of
  neural network for the sign problem in field theories}}, PoS LATTICE2018
  (2019) 023.

\bibitem{MoriPOMandNN}
Y.~Mori, K.~Kashiwa, A.~Ohnishi, \emph{{Application of a neural network to the
  sign problem via the path optimization method}}, Progress of Theoretical and
  Experimental Physics 2018 (2018), 023B04,
  \eprint{https://academic.oup.com/ptep/article-pdf/2018/2/023B04/23956652/ptx191.pdf}.

\bibitem{Bursa2018}
F.~Bursa, M.~Kroyter, \emph{A simple approach towards the sign problem using
  path optimisation}, Journal of High Energy Physics 2018 (2018) 54.

\bibitem{PhysRevD102.014514}
W.~Detmold, G.~Kanwar, M.~L. Wagman, N.~C. Warrington, \emph{Path integral
  contour deformations for noisy observables}, Phys. Rev. D 102 (2020) 014514.

\bibitem{MoriPOMinQCD}
Y.~Mori, K.~Kashiwa, A.~Ohnishi, \emph{{Path optimization in $0+1$D QCD at
  finite density}}, Prog. Theor. Exp. Phys. 2019 (2019), 113B01,
  \eprint{https://academic.oup.com/ptep/article-pdf/2019/11/113B01/33450851/ptz111.pdf}.

\bibitem{PhysRevD99.014033}
K.~Kashiwa, Y.~Mori, A.~Ohnishi, \emph{Controlling the model sign problem via
  the path optimization method: Monte Carlo approach to a QCD effective model
  with Polyakov loop}, Phys. Rev. D 99 (2019) 014033.

\bibitem{PhysRevD99.114005}
K.~Kashiwa, Y.~Mori, A.~Ohnishi, \emph{Application of the path optimization
  method to the sign problem in an effective model of QCD with a repulsive
  vector-type interaction}, Phys. Rev. D 99 (2019) 114005.

\bibitem{PhysRevD.97.094510}
A.~Alexandru, P.~F. Bedaque, H.~Lamm, S.~Lawrence, \emph{Finite-density Monte
  Carlo calculations on sign-optimized manifolds}, Phys. Rev. D 97 (2018)
  094510.

\bibitem{PhysRevLett.121.191602}
A.~Alexandru, P.~F. Bedaque, H.~Lamm, S.~Lawrence, N.~C. Warrington,
  \emph{Fermions at Finite Density in $2+1$ Dimensions with Sign-Optimized
  Manifolds}, Phys. Rev. Lett. 121 (2018) 191602.

\bibitem{ParisiWu}
G.~Parisi, Y.-s. Wu, \emph{{Perturbation Theory Without Gauge Fixing}}, Sci.
  Sin. 24 (1981) 483.

\bibitem{PhysicsReportsStochasticQuantization}
P.~H. Damgaard, H.~H{\"u}ffel, \emph{Stochastic quantization}, Phys. Rept. 152
  (1987) 227 .

\bibitem{Klauder1983a}
J.~R. Klauder, \emph{{Stochastic Quantization}}, Acta Phys. Austriaca Suppl. 25
  (1983) 251.

\bibitem{Klauder1983b}
J.~R. Klauder, \emph{A Langevin approach to fermion and quantum spin
  correlation functions}, Journal of Physics A: Mathematical and General 16
  (1983) L317.

\bibitem{Parisi1983}
G.~Parisi, \emph{On complex probabilities}, Phys. Lett. B 131 (1983) 393.

\bibitem{Klauder1984}
J.~R. Klauder, \emph{Coherent-state Langevin equations for canonical quantum
  systems with applications to the quantized Hall effect}, Phys. Rev. A 29
  (1984) 2036.

\bibitem{Ambjorn1985}
J.~Ambjørn, S.-K. Yang, \emph{Numerical problems in applying the langevin
  equation to complex effective actions}, Phys. Lett. B 165 (1985) 140 .

\bibitem{Ambjorn1986}
J.~Ambjørn, M.~Flensburg, C.~Peterson, \emph{The complex langevin equation and
  Monte Carlo simulations of actions with static charges}, Nucl. Phys. B 275
  (1986) 375 .

\bibitem{Klauder1985}
J.~R. Klauder, W.~P. Petersen, \emph{Spectrum of certain non-self-adjoint
  operators and solutions of Langevin equations with complex drift}, J. Stat.
  Phys. 39 (1985) 53.

\bibitem{Gausterer1993}
H.~Gausterer, S.~Lee, \emph{The mechanism of complex Langevin simulations},
  Journal of Statistical Physics 73 (1993) 147.

\bibitem{Gausterer1998}
H.~Gausterer, H.~Thaler, \emph{Complex Langevin for semisimple compact
  connected Lie groups and U(1)}, J. Phys. A 31 (1998) 2541.

\bibitem{GaustererNPA1998}
H.~Gausterer, \emph{Complex Langevin: A numerical method?}, Nucl. Phys. A 642
  (1998) c239 , qCD at Finite Baryon Density.

\bibitem{Gausterer1986}
H.~Gausterer, J.~R. Klauder, \emph{Complex Langevin solution of the Schwinger
  model}, Phys. Rev. Lett. 56 (1986) 306.

\bibitem{Kieu1995}
T.~Kieu, P.~Hawkins, \emph{A numerical attempt on the chiral Schwinger model},
  Nuclear Physics B - Proceedings Supplements 42 (1995) 621 .

\bibitem{Haymaker1988}
R.~W. Haymaker, J.~Wosiek, \emph{Complex Langevin simulations of non-Abelian
  integrals}, Phys. Rev. D 37 (1988) 969.

\bibitem{PRC2001026303}
C.~Adami, S.~E. Koonin, \emph{{Complex Langevin equation and the many fermion
  problem}}, Phys. Rev. C 63 (2001) 034319, \eprint{nucl-th/0009021}.

\bibitem{Ganesan_2001}
V.~Ganesan, G.~H. Fredrickson, \emph{Field-theoretic polymer simulations}, EPL
  55 (2001) 814.

\bibitem{Duechs2003}
D.~D{\"u}chs, V.~Ganesan, G.~H. Fredrickson, F.~Schmid, \emph{Fluctuation
  Effects in Ternary AB + A + B Polymeric Emulsions}, Macromolecules 36 (2003)
  9237.

\bibitem{Fredrickson2002}
G.~H. Fredrickson, V.~Ganesan, F.~Drolet, \emph{Field-Theoretic Computer
  Simulation Methods for Polymers and Complex Fluids}, Macromolecules 35 (2002)
  16.

\bibitem{HOCHBERG200654}
D.~Hochberg, M.-P. Zorzano, F.~Morán, \emph{Complex reaction noise in a
  molecular quasispecies model}, Chem. Phys. Lett. 423 (2006) 54 .

\bibitem{DELOUBRIERE2002135}
O.~Deloubrière, L.~Frachebourg, H.~Hilhorst, K.~Kitahara, \emph{Imaginary
  noise and parity conservation in the reaction A+A{$\rightleftharpoons$}0},
  Physica A: Statistical Mechanics and its Applications 308 (2002) 135 .

\bibitem{Jonghoon2008}
J.~Lee, Y.~O. Popov, G.~H. Fredrickson, \emph{Complex coacervation: A field
  theoretic simulation study of polyelectrolyte complexation}, The Journal of
  Chemical Physics 128 (2008) 224908,
  \eprint{https://doi.org/10.1063/1.2936834}.

\bibitem{Popov2007}
Y.~O. Popov, J.~Lee, G.~H. Fredrickson, \emph{Field-theoretic simulations of
  polyelectrolyte complexation}, Journal of Polymer Science Part B: Polymer
  Physics 45 (2007) 3223,
  \eprint{https://onlinelibrary.wiley.com/doi/pdf/10.1002/polb.21334}.

\bibitem{Xingkun2014}
X.~Man, K.~T. Delaney, M.~C. Villet, H.~Orland, G.~H. Fredrickson,
  \emph{Coherent states formulation of polymer field theory}, The Journal of
  Chemical Physics 140 (2014) 024905,
  \eprint{https://doi.org/10.1063/1.4860978}.

\bibitem{Duechs2014}
D.~Düchs, K.~T. Delaney, G.~H. Fredrickson, \emph{A multi-species exchange
  model for fully fluctuating polymer field theory simulations}, The Journal of
  Chemical Physics 141 (2014) 174103,
  \eprint{https://doi.org/10.1063/1.4900574}.

\bibitem{PhysRevLett95202003}
J.~Berges, I.-O. Stamatescu, \emph{Simulating Nonequilibrium Quantum Fields
  with Stochastic Quantization Techniques}, Phys. Rev. Lett. 95 (2005) 202003.

\bibitem{PhysRevD75045007}
J.~Berges, S.~Bors\'anyi, D.~Sexty, I.-O. Stamatescu, \emph{Lattice simulations
  of real-time quantum fields}, Phys. Rev. D 75 (2007) 045007.

\bibitem{JHEP200809018}
G.~Aarts, I.-O. Stamatescu, \emph{{Stochastic quantization at finite chemical
  potential}}, JHEP 09 (2008) 018, \eprint{0807.1597}.

\bibitem{Aarts2009}
G.~Aarts, \emph{Two complex problems on the lattice: transport coefficients and
  finite chemical potential}, Nuclear Physics A 820 (2009) 57c , strong and
  Electroweak Matter.

\bibitem{AartsPRL102131601}
G.~{Aarts}, \emph{{Can Stochastic Quantization Evade the Sign Problem? The
  Relativistic Bose Gas at Finite Chemical Potential}}, Phys. Rev. Lett. 102
  (2009) 131601, \eprint{0810.2089}.

\bibitem{JHEP200905052}
G.~Aarts, \emph{Complex Langevin dynamics at finite chemical potential: mean
  field analysis in the relativistic Bose gas}, JHEP 2009 (2009) 052.

\bibitem{BERGES2008306}
{J\"urgen Berges}, {D\'enes Sexty}, \emph{Real-time gauge theory simulations
  from stochastic quantization with optimized updating}, Nucl. Phys. B 799
  (2008) 306 .

\bibitem{AARTS2010154}
G.~Aarts, F.~A. James, E.~Seiler, I.-O. Stamatescu, \emph{Adaptive stepsize and
  instabilities in complex Langevin dynamics}, Phys. Lett. B 687 (2010) 154 .

\bibitem{JHEP20100820}
G.~{Aarts}, F.~A. {James}, \emph{{On the convergence of complex Langevin
  dynamics: the three-dimensional XY model at finite chemical potential}}, JHEP
  8 (2010) 20, \eprint{1005.3468}.

\bibitem{PhysRevLett.89.240201}
D.~Weingarten, \emph{Complex Probabilities on ${R}^{N}$ as Real Probabilities
  on ${C}^{N}$ and an Application to Path Integrals}, Phys. Rev. Lett. 89
  (2002) 240201.

\bibitem{Salcedo:1996sa}
L.~L. Salcedo, \emph{{Representation of complex probabilities}}, J. Math. Phys.
  38 (1997) 1710, \eprint{hep-lat/9607044}.

\bibitem{Salcedo_2007}
L.~L. Salcedo, \emph{{Existence of positive representations for complex
  weights}}, J. Phys. A 40 (2007) 9399, \eprint{0706.4359}.

\bibitem{Wosiek:2015iwl}
J.~Wosiek, \emph{{Beyond complex Langevin equations I: two simple examples}}
  (2015), \eprint{1511.09083}.

\bibitem{Wosiek:2015bqg}
J.~Wosiek, \emph{{Beyond complex Langevin equations: from simple examples to
  positive representation of Feynman path integrals directly in the Minkowski
  time}}, JHEP 04 (2016) 146, \eprint{1511.09114}.

\bibitem{PhysRevD.94.074503}
L.~L. Salcedo, \emph{Gibbs sampling of complex-valued distributions}, Phys.
  Rev. D 94 (2016) 074503.

\bibitem{Seiler_2017}
E.~Seiler, J.~Wosiek, \emph{{Positive Representations of a Class of Complex
  Measures}}, J. Phys. A 50 (2017) 495403, \eprint{1702.06012}.

\bibitem{Salcedo_2018}
L.~L. Salcedo, \emph{{Positive representations of complex distributions on
  groups}}, J. Phys. A 51 (2018) 505401, \eprint{1805.01698}.

\bibitem{Wosiek:2018jht}
J.~Wosiek, B.~Ruba, \emph{{Beyond Complex Langevin Equations: a Progress
  Report}}, in: \emph{{36th International Symposium on Lattice Field Theory
  (Lattice 2018) East Lansing, MI, United States, July 22-28, 2018}} (2018),
  \eprint{1810.11519}.

\bibitem{jona-lasinio1985}
G.~Jona-Lasinio, P.~K. Mitter, \emph{On the stochastic quantization of field
  theory}, Comm. Math. Phys. 101 (1985) 409.

\bibitem{PhysRevD.32.2736}
G.~G. Batrouni, G.~R. Katz, A.~S. Kronfeld, G.~P. Lepage, B.~Svetitsky, K.~G.
  Wilson, \emph{Langevin simulations of lattice field theories}, Phys. Rev. D
  32 (1985) 2736.

\bibitem{PhysRevB.99.035114}
G.~G. Batrouni, R.~T. Scalettar, \emph{Langevin simulations of a long-range
  electron-phonon model}, Phys. Rev. B 99 (2019) 035114.

\bibitem{Duane:1987de}
S.~Duane, A.~D. Kennedy, B.~J. Pendleton, D.~Roweth, \emph{{Hybrid Monte
  Carlo}}, Phys. Lett. B195 (1987) 216.

\bibitem{PhysRevD.35.2531}
S.~Gottlieb, W.~Liu, D.~Toussaint, R.~L. Renken, R.~L. Sugar,
  \emph{Hybrid-molecular-dynamics algorithms for the numerical simulation of
  quantum chromodynamics}, Phys. Rev. D 35 (1987) 2531.

\bibitem{PTPS1993CLSimulation}
K.~Okano, L.~Schulke, B.~Zheng, \emph{{Complex Langevin simulation}}, Prog.
  Theor. Phys. Suppl. 111 (1993) 313.

\bibitem{Troyer2012}
V.~Ambegaokar, M.~Troyer, \emph{{Estimating errors reliably in Monte Carlo
  simulations of the Ehrenfest model}}, Am. J. Phys. 78 (2010) 150,
  \eprint{0906.0943}.

\bibitem{DRUMMOND1983119}
I.~Drummond, S.~Duane, R.~Horgan, \emph{The stochastic method for numerical
  simulations:: Higher order corrections}, Nucl. Phys. B 220 (1983) 119 .

\bibitem{HOROWITZ1987510}
A.~M. Horowitz, \emph{The second order Langevin equation and numerical
  simulations}, Nucl. Phys. B 280 (1987) 510.

\bibitem{CATTERALL1991177}
S.~Catterall, I.~Drummond, R.~Horgan, \emph{Langevin algorithms for spin
  models}, Phys. Lett. B 254 (1991) 177.

\bibitem{Aarts2012SU3}
G.~Aarts, F.~A. James, \emph{Complex Langevin dynamics in the SU(3) spin model
  at nonzero chemical potential revisited}, JHEP 2012 (2012) 118.

\bibitem{AartsPRD81054508}
G.~Aarts, E.~Seiler, I.-O. Stamatescu, \emph{Complex Langevin method: When can
  it be trusted?}, Phys. Rev. D 81 (2010) 054508.

\bibitem{CLZeroesFermionDet}
G.~Aarts, E.~Seiler, D.~Sexty, I.-O. Stamatescu, \emph{Complex Langevin
  dynamics and zeroes of the fermion determinant}, JHEP 2017 (2017) 44.

\bibitem{Seiler2017StatusOfCL}
E.~Seiler, \emph{{Status of Complex Langevin}}, EPJ Web Conf. 175 (2018) 01019,
  \eprint{1708.08254}.

\bibitem{2011EurPhysJC711756}
G.~Aarts, F.~A. James, E.~Seiler, I.-O. Stamatescu, \emph{{Complex Langevin:
  Etiology and Diagnostics of its Main Problem}}, Eur. Phys. J. C 71 (2011)
  1756, \eprint{1101.3270}.

\bibitem{CLJustificationPRD94114515}
K.~Nagata, J.~Nishimura, S.~Shimasaki, \emph{Argument for justification of the
  complex Langevin method and the condition for correct convergence}, Phys.
  Rev. D 94 (2016) 114515.

\bibitem{PhysRevD.99.014512}
M.~Scherzer, E.~Seiler, D.~Sexty, I.-O. Stamatescu, \emph{Complex Langevin and
  boundary terms}, Phys. Rev. D 99 (2019) 014512.

\bibitem{aarts2013}
G.~Aarts, P.~Giudice, E.~Seiler, \emph{Localised distributions and criteria for
  correctness in complex Langevin dynamics}, Annals of Physics 337 (2013) 238 .

\bibitem{scherzer2018}
M.~Scherzer, E.~Seiler, D.~Sexty, I.-O. Stamatescu, \emph{Complex Langevin:
  Boundary terms and application to QCD} (2018), \eprint{1810.09713}.

\bibitem{Scherzer2020}
M.~Scherzer, D.~Sexty, I.~O. Stamatescu, \emph{Deconfinement transition line
  with the Complex Langevin equation up to $\mu/T \sim 5$} (2020),
  \eprint{2004.05372}.

\bibitem{Scherzer2020b}
M.~Scherzer, E.~Seiler, D.~Sexty, I.-O. Stamatescu, \emph{Controlling complex
  Langevin simulations of lattice models by boundary term analysis}, Phys. Rev.
  D 101 (2020) 014501.

\bibitem{cai2020validity}
Z.~Cai, X.~Dong, Y.~Kuang, \emph{On the validity of complex Langevin method for
  path integral computations} (2020), \eprint{2007.10198}.

\bibitem{PRDSalcedoCL2016}
L.~L. Salcedo, \emph{Does the complex Langevin method give unbiased results?},
  Phys. Rev. D 94 (2016) 114505.

\bibitem{aarts2016b}
G.~Aarts, E.~Seiler, D.~Sexty, I.~O. Stamatescu, \emph{On complex Langevin
  dynamics and zeroes of the measure II: Fermionic determinant} (2016),
  \eprint{1611.02931}.

\bibitem{aarts2016c}
G.~Aarts, E.~Seiler, D.~Sexty, I.-O. Stamatescu, \emph{On complex Langevin
  dynamics and zeroes of the measure I: Formal proof and simple models} (2016),
  \eprint{1611.02930}.

\bibitem{shimasaki2016}
S.~Shimasaki, K.~Nagata, J.~Nishimura, \emph{On the condition for correct
  convergence in the complex Langevin method} (2016), \eprint{1611.10170}.

\bibitem{PhysRevD.92.011501}
J.~Nishimura, S.~Shimasaki, \emph{New insights into the problem with a singular
  drift term in the complex Langevin method}, Phys. Rev. D 92 (2015) 011501.

\bibitem{Nagata2018b}
K.~Nagata, J.~Nishimura, S.~Shimasaki, \emph{Testing the criterion for correct
  convergence in the complex Langevin method}, Journal of High Energy Physics
  2018 (2018) 4.

\bibitem{SeilerGaugeCooling}
E.~Seiler, D.~Sexty, I.-O. Stamatescu, \emph{{Gauge cooling in complex Langevin
  for QCD with heavy quarks}}, Phys. Lett. B 723 (2013) 213,
  \eprint{1211.3709}.

\bibitem{Bongiovanni:2013nxa}
L.~Bongiovanni, G.~Aarts, E.~Seiler, D.~Sexty, I.-O. Stamatescu,
  \emph{{Adaptive gauge cooling for complex Langevin dynamics}}, PoS
  LATTICE2013 (2014) 449, \eprint{1311.1056}.

\bibitem{Aarts2013b}
G.~Aarts, L.~Bongiovanni, E.~Seiler, D.~Sexty, I.-O. Stamatescu,
  \emph{Controlling complex Langevin dynamics at finite density}, The European
  Physical Journal A 49 (2013) 89.

\bibitem{PhysRevD.92.085020}
H.~Makino, H.~Suzuki, D.~Takeda, \emph{Complex Langevin method applied to the
  2D $SU(2)$ Yang-Mills theory}, Phys. Rev. D 92 (2015) 085020.

\bibitem{Nagata:2015uga}
K.~Nagata, J.~Nishimura, S.~Shimasaki, \emph{{Justification of the complex
  Langevin method with the gauge cooling procedure}}, PTEP 2016 (2016) 013B01,
  \eprint{1508.02377}.

\bibitem{Nagata:2016alq}
K.~Nagata, J.~Nishimura, S.~Shimasaki, \emph{{Gauge cooling for the
  singular-drift problem in the complex Langevin method - a test in Random
  Matrix Theory for finite density QCD}}, JHEP 07 (2016) 073,
  \eprint{1604.07717}.

\bibitem{Zhenning2020}
C.~Zhenning, Y.~Di, X.~Dong, \emph{How does Gauge Cooling Stabilize Complex
  Langevin?}, Communications in Computational Physics 27 (2020) 1344.

\bibitem{dong2020alternating}
X.~Dong, Z.~Cai, Y.~Di, \emph{Alternating Descent Method for Gauge Cooling of
  Complex Langevin Simulations} (2020), \eprint{2008.06654}.

\bibitem{Aarts:2016qhx}
G.~Aarts, F.~Attanasio, B.~J{\"a}ger, D.~Sexty, \emph{{Complex Langevin in
  Lattice QCD: dynamic stabilisation and the phase diagram}}, Acta Phys. Polon.
  Supp. 9 (2016) 621, \eprint{1607.05642}.

\bibitem{Lattice2016AttanasioJager}
F.~Attanasio, B.~J{\"a}ger, \emph{{Testing dynamic stabilisation in complex
  Langevin simulations}}, PoS LATTICE2016 (2016) 053, \eprint{1610.09298}.

\bibitem{Lattice2017AttanasioJager}
{Attanasio, Felipe}, {J{\"a}ger, Benjamin}, \emph{Improved convergence of
  Complex Langevin simulations}, EPJ Web Conf. 175 (2018) 07039.

\bibitem{attanasio2018}
F.~Attanasio, B.~Jäger, \emph{Stabilising complex Langevin simulations}
  (2018), \eprint{1810.12973}.

\bibitem{Attanasio2019}
F.~Attanasio, B.~J{\"a}ger, \emph{Dynamical stabilisation of complex Langevin
  simulations of QCD}, Eur. Phys. J. C 79 (2019) 16.

\bibitem{PRD95094502}
A.~C. Loheac, J.~E. Drut, \emph{Third-order perturbative lattice and complex
  Langevin analyses of the finite-temperature equation of state of
  nonrelativistic fermions in one dimension}, Phys. Rev. D 95 (2017) 094502.

\bibitem{Ito2016}
Y.~Ito, J.~Nishimura, \emph{The complex Langevin analysis of spontaneous
  symmetry breaking induced by complex fermion determinant}, Journal of High
  Energy Physics 2016 (2016) 9.

\bibitem{Ito2018b}
Y.~Ito, J.~Nishimura, \emph{Comparative studies of the deformation techniques
  for the singular-drift problem in the complex Langevin method}, EPJ Web Conf.
  175 (2018) 07019.

\bibitem{2018EPJWC17507017N}
K.~Nagata, J.~Nishimura, S.~Shimasaki, \emph{{Complex Langevin simulation of
  QCD at finite density and low temperature using the deformation technique}},
  EPJ Web Conf. 175 (2018) 07017, \eprint{1710.07416}.

\bibitem{Splittorff2015}
K.~Splittorff, \emph{Dirac spectrum in complex Langevin simulations of QCD},
  Phys. Rev. D 91 (2015) 034507.

\bibitem{Ichihara2016}
T.~Ichihara, K.~Nagata, K.~Kashiwa, \emph{Test for a universal behavior of
  Dirac eigenvalues in the complex Langevin method}, Phys. Rev. D 93 (2016)
  094511.

\bibitem{xue1986}
S.-S. Xue, \emph{The Fokker-Planck equations in lattice gauge theories}, Phys.
  Lett. B 180 (1986) 275 .

\bibitem{GuralnikNPB200905503213}
G.~Guralnik, C.~Pehlevan, \emph{Complex Langevin equations and Schwinger-Dyson
  equations}, Nucl. Phys. B 811 (2009) 519 .

\bibitem{Salcedo1993SpuriousCLSolutions}
L.~L. {Salcedo}, \emph{{Spurious solutions of the complex Langevin equation}},
  Phys. Lett. B 305 (1993) 125.

\bibitem{Salcedo:2018fvt}
L.~L. Salcedo, E.~Seiler, \emph{{Schwinger-Dyson equations and line
  integrals}}, J. Phys. A 52 (2019) 035201, \eprint{1809.06888}.

\bibitem{PhysRevLett.55.2242}
F.~Karsch, H.~W. Wyld, \emph{Complex Langevin Simulation of the SU(3) Spin
  Model with Nonzero Chemical Potential}, Phys. Rev. Lett. 55 (1985) 2242.

\bibitem{Flower1986}
J.~Flower, S.~W. Otto, S.~Callahan, \emph{Complex Langevin equations and
  lattice gauge theory}, Phys. Rev. D 34 (1986) 598.

\bibitem{Lattice2012Aarts}
G.~Aarts, \emph{{Complex Langevin dynamics and other approaches at finite
  chemical potential}}, PoS LATTICE2012 (2012) 017, \eprint{1302.3028}.

\bibitem{aarts2014}
G.~Aarts, F.~Attanasio, B.~Jäger, E.~Seiler, D.~Sexty, I.-O. Stamatescu,
  \emph{QCD at nonzero chemical potential: recent progress on the lattice}
  (2014), \eprint{1412.0847}.

\bibitem{attanasio2020b}
F.~Attanasio, B.~Jäger, F.~P.~G. Ziegler, \emph{Complex Langevin and the QCD
  phase diagram: Recent developments} (2020), \eprint{2006.00476}.

\bibitem{Muroya2003}
S.~Muroya, A.~Nakamura, C.~Nonaka, T.~Takaishi, \emph{{Lattice QCD at Finite
  Density: An Introductory Review}}, Prog. Theor. Phys. 110 (2003) 615,
  \eprint{https://academic.oup.com/ptp/article-pdf/110/4/615/5301553/110-4-615.pdf}.

\bibitem{forcrand2010}
P.~de~Forcrand, \emph{Simulating QCD at finite density} (2010),
  \eprint{1005.0539}.

\bibitem{Aarts2015}
G.~Aarts, \emph{Developments in lattice quantum chromodynamics for matter at
  high temperature and density}, Pramana 84 (2015) 787.

\bibitem{Aarts2016}
G.~Aarts, \emph{Introductory lectures on lattice {QCD} at nonzero baryon
  number}, J, Phys. Conf. Ser. 706 (2016) 022004.

\bibitem{deAguiar:2010ue}
T.~C. de~Aguiar, N.~F. Svaiter, G.~Menezes, \emph{{Stochastic Quantization of
  Real-Time Thermal Field Theory}}, J. Math. Phys. 51 (2010) 102304,
  \eprint{0908.3009}.

\bibitem{aarts2010}
G.~Aarts, F.~A. James, \emph{The XY model at finite chemical potential using
  complex Langevin dynamics} (2010), \eprint{1009.5838}.

\bibitem{Katz2017}
S.~D. Katz, F.~Niedermayer, D.~N\'ogr\'adi, C.~T\"or\"ok, \emph{Comparison of
  algorithms for solving the sign problem in the O(3) model in $1+1$ dimensions
  at finite chemical potential}, Phys. Rev. D 95 (2017) 054506.

\bibitem{Bilic1988}
N.~Bili\ifmmode~\acute{c}\else \'{c}\fi{}, H.~Gausterer, S.~Sanielevici,
  \emph{Complex Langevin solution to an effective theory of lattice QCD}, Phys.
  Rev. D 37 (1988) 3684.

\bibitem{Aarts:2012ft}
G.~Aarts, F.~A. James, J.~M. Pawlowski, E.~Seiler, D.~Sexty, I.-O. Stamatescu,
  \emph{{Stability of complex Langevin dynamics in effective models}}, JHEP 03
  (2013) 073, \eprint{1212.5231}.

\bibitem{Greensite2014}
J.~Greensite, \emph{Comparison of complex Langevin and mean field methods
  applied to effective Polyakov line models}, Phys. Rev. D 90 (2014) 114507.

\bibitem{greensite2014b}
J.~Greensite, \emph{Effective Polyakov line actions, and their solutions at
  finite chemical potential} (2014), \eprint{1411.0607}.

\bibitem{Mollgaard2014}
A.~Mollgaard, K.~Splittorff, \emph{Full simulation of chiral random matrix
  theory at nonzero chemical potential by complex Langevin}, Phys. Rev. D 91
  (2015) 036007.

\bibitem{Verbaarschot:2000dy}
J.~Verbaarschot, T.~Wettig, \emph{{Random matrix theory and chiral symmetry in
  QCD}}, Ann. Rev. Nucl. Part. Sci. 50 (2000) 343, \eprint{hep-ph/0003017}.

\bibitem{PRD201311116007}
A.~Mollgaard, K.~Splittorff, \emph{{Complex Langevin Dynamics for chiral Random
  Matrix Theory}}, Phys. Rev. D 88 (2013) 116007, \eprint{1309.4335}.

\bibitem{Nagata2016b}
K.~Nagata, J.~Nishimura, S.~Shimasaki, \emph{Testing a generalized cooling
  procedure in the complex Langevin simulation of chiral Random Matrix Theory},
  PoS LATTICE2015  (2016).

\bibitem{2017EPJWC13707030B}
J.~Bloch, J.~Glesaaen, O.~Philipsen, J.~Verbaarschot, S.~Zafeiropoulos,
  \emph{{Complex Langevin simulations of a finite density matrix model for
  QCD}}, EPJ Web Conf. 137 (2017) 07030, \eprint{1612.04621}.

\bibitem{Pawlowski:2013pje}
J.~M. Pawlowski, C.~Zielinski, \emph{{Thirring model at finite density in 0+1
  dimensions with stochastic quantization: Crosscheck with an exact solution}},
  Phys. Rev. D 87 (2013) 094503, \eprint{1302.1622}.

\bibitem{Pawlowski:2013gag}
J.~M. Pawlowski, C.~Zielinski, \emph{{Thirring model at finite density in 2+1
  dimensions with stochastic quantization}}, Phys. Rev. D 87 (2013) 094509,
  \eprint{1302.2249}.

\bibitem{li2016}
D.~Li, \emph{Comparison between Fermion Bag Approach and Complex Langevin
  Dynamics for Massive Thirring Model at Finite Density in 0 + 1 Dimensions}
  (2016), \eprint{1605.04623}.

\bibitem{li2016b}
D.~Li, \emph{Fermion bag approach for the massive Thirring model at finite
  density}, Phys. Rev. D 94 (2016) 114501.

\bibitem{fujii2017}
H.~Fujii, S.~Kamata, Y.~Kikukawa, \emph{Performance of Complex Langevin
  Simulation in 0+1 dimensional massive Thirring model at finite density}
  (2017), \eprint{1710.08524}.

\bibitem{aarts2014d}
G.~Aarts, F.~Attanasio, B.~Jäger, E.~Seiler, D.~Sexty, I.-O. Stamatescu,
  \emph{Exploring the phase diagram of QCD with complex Langevin simulations}
  (2014), \eprint{1411.2632}.

\bibitem{Aarts:2016qrv}
G.~Aarts, F.~Attanasio, B.~J{\"a}ger, D.~Sexty, \emph{{The QCD phase diagram in
  the limit of heavy quarks using complex Langevin dynamics}}, JHEP 09 (2016)
  087, \eprint{1606.05561}.

\bibitem{Attanasio2016}
F.~Attanasio, G.~Aarts, B.~Jaeger, E.~Seiler, D.~Sexty, I.-O. Stamatescu,
  \emph{Towards the heavy dense QCD phase diagram using Complex Langevin
  simulations}, PoS LATTICE2015  (2016).

\bibitem{aarts2016e}
G.~Aarts, F.~Attanasio, B.~Jäger, D.~Sexty, \emph{Results on the heavy-dense
  QCD phase diagram using complex Langevin} (2016), \eprint{1610.04401}.

\bibitem{Aarts:2013nja}
G.~Aarts, L.~Bongiovanni, E.~Seiler, D.~Sexty, I.-O. Stamatescu, \emph{{Complex
  Langevin simulation for QCD-like models}}, PoS LATTICE2013 (2014) 451,
  \eprint{1310.7412}.

\bibitem{POSInsightsintoHDQCDusingCL}
G.~Aarts, F.~Attanasio, B.~J{\"a}ger, E.~Seiler, D.~Sexty, I.-O. Stamatescu,
  \emph{{Insights into the heavy dense QCD phase diagram using Complex Langevin
  simulations}}, PoS LATTICE2015 (2016) 155, \eprint{1510.09100}.

\bibitem{Aarts2014b}
G.~Aarts, E.~Seiler, D.~Sexty, I.-O. Stamatescu, \emph{Simulating QCD at
  nonzero baryon density to all orders in the hopping parameter expansion},
  Phys. Rev. D 90 (2014) 114505.

\bibitem{sexty2014b}
D.~Sexty, \emph{New algorithms for finite density QCD} (2014),
  \eprint{1410.8813}.

\bibitem{Fromm2013}
M.~Fromm, J.~Langelage, S.~Lottini, M.~Neuman, O.~Philipsen, \emph{Onset
  Transition to Cold Nuclear Matter from Lattice QCD with Heavy Quarks}, Phys.
  Rev. Lett. 110 (2013).

\bibitem{Langelage2014}
J.~Langelage, M.~Neuman, O.~Philipsen, \emph{Heavy dense QCD and nuclear matter
  from an effective lattice theory}, Journal of High Energy Physics 2014 (2014)
  131.

\bibitem{aarts2014e}
G.~Aarts, B.~Jäger, E.~Seiler, D.~Sexty, I.-O. Stamatescu, \emph{Systematic
  approximation for QCD at non-zero density} (2014), \eprint{1412.5775}.

\bibitem{aarts2015b}
G.~Aarts, E.~Seiler, D.~Sexty, I.~O. Stamatescu, \emph{Hopping parameter
  expansion to all orders using the complex Langevin equation} (2015),
  \eprint{1503.08813}.

\bibitem{Bloch:2015coa}
J.~Bloch, J.~Mahr, S.~Schmalzbauer, \emph{{Complex Langevin in low-dimensional
  QCD: the good and the not-so-good}}, PoS LATTICE2015 (2016) 158,
  \eprint{1508.05252}.

\bibitem{JHEP0820101017}
G.~Aarts, K.~Splittorff, \emph{Degenerate distributions in complex Langevin
  dynamics: one-dimensional QCD at finite chemical potential}, JHEP 2010 (2010)
  17.

\bibitem{Bloch:2017sfg}
J.~Bloch, J.~Meisinger, S.~Schmalzbauer, \emph{{Reweighted complex Langevin and
  its application to two-dimensional QCD}}, PoS LATTICE2016 (2017) 046,
  \eprint{1701.01298}.

\bibitem{PhysRevD95054509}
J.~Bloch, \emph{Reweighting complex Langevin trajectories}, Phys. Rev. D 95
  (2017) 054509.

\bibitem{Schmalzbauer2017}
S.~Schmalzbauer, J.~Bloch, \emph{Complex Langevin Dynamics In 1+1d QCD At
  Non-Zero Densities}, PoS LATTICE2016  (2017).

\bibitem{Sexty:2013ica}
D.~Sexty, \emph{{Simulating full QCD at nonzero density using the complex
  Langevin equation}}, Phys. Lett. B 729 (2014) 108, \eprint{1307.7748}.

\bibitem{sexty2013}
D.~Sexty, \emph{Extending complex Langevin simulations to full QCD at nonzero
  density} (2013), \eprint{1310.6186}.

\bibitem{Sexty2014}
D.~Sexty, \emph{Progress in complex Langevin simulations of full QCD at
  non-zero density}, Nuclear Physics A 931 (2014) 856 , {Quark Matter 2014}.

\bibitem{Fodor2015}
Z.~Fodor, S.~D. Katz, D.~Sexty, C.~T\"or\"ok, \emph{Complex Langevin dynamics
  for dynamical QCD at nonzero chemical potential: A comparison with
  multiparameter reweighting}, Phys. Rev. D 92 (2015) 094516.

\bibitem{Lattice2015KogutSinclair}
D.~K. Sinclair, J.~B. Kogut, \emph{{Exploring Complex-Langevin Methods for
  Finite-Density QCD}}, PoS LATTICE2015 (2016) 153, \eprint{1510.06367}.

\bibitem{LATTICE2016PROCSinclairKogut}
D.~K. Sinclair, J.~B. Kogut, \emph{{Complex Langevin for Lattice QCD at $T=0$
  and $\mu \ge 0$}}, PoS LATTICE2016 (2016) 026, \eprint{1611.02312}.

\bibitem{Lattice2017KogutSinclair}
D.~K. Sinclair, J.~B. Kogut, \emph{{Complex Langevin Simulations of QCD at
  Finite Density. Progress Report}}, EPJ Web Conf. 175 (2018) 07031,
  \eprint{1710.08465}.

\bibitem{Sinclair:2018rbk}
D.~K. Sinclair, J.~B. Kogut, \emph{{Complex Langevin for Lattice QCD}}, PoS
  LATTICE2018 (2018) 143, \eprint{1810.11880}.

\bibitem{Kogut2019}
J.~B. Kogut, D.~K. Sinclair, \emph{Applying complex Langevin simulations to
  lattice QCD at finite density}, Phys. Rev. D 100 (2019) 054512.

\bibitem{Sinclair2019}
D.~K. Sinclair, J.~B. Kogut, \emph{Applying Complex Langevin to Lattice QCD at
  finite $\mu$} (2019), \eprint{1910.11412}.

\bibitem{Nagata2016}
K.~Nagata, H.~Matsufuru, J.~Nishimura, S.~Shimasaki, \emph{Gauge cooling for
  the singular-drift problem in the complex Langevin method --- an application
  to finite density QCD} (2016), \eprint{1611.08077}.

\bibitem{Nagata2018}
K.~Nagata, J.~Nishimura, S.~Shimasaki, \emph{Complex Langevin calculations in
  finite density QCD at large $\ensuremath{\mu}/T$ with the deformation
  technique}, Phys. Rev. D 98 (2018) 114513.

\bibitem{ito2018}
Y.~Ito, H.~Matsufuru, J.~Nishimura, S.~Shimasaki, A.~Tsuchiya, S.~Tsutsui,
  \emph{Exploring the phase diagram of finite density QCD at low temperature by
  the complex Langevin method} (2018), \eprint{1811.12688}.

\bibitem{tsutsui2019}
S.~Tsutsui, Y.~Ito, H.~Matsufuru, J.~Nishimura, S.~Shimasaki, A.~Tsuchiya,
  \emph{Exploring the QCD phase diagram at finite density by the complex
  Langevin method on a $16^3\times 32$ lattice} (2019), \eprint{1912.00361}.

\bibitem{ito2020}
Y.~Ito, H.~Matsufuru, Y.~Namekawa, J.~Nishimura, S.~Shimasaki, A.~Tsuchiya,
  S.~Tsutsui, \emph{Complex Langevin calculations in QCD at finite density}
  (2020), \eprint{2007.08778}.

\bibitem{tsutsui2018}
S.~Tsutsui, Y.~Ito, H.~Matsufuru, J.~Nishimura, S.~Shimasaki, A.~Tsuchiya,
  \emph{Can the complex Langevin method see the deconfinement phase transition
  in QCD at finite density?} (2018), \eprint{1811.07647}.

\bibitem{Bloch2018}
J.~Bloch, O.~Schenk, \emph{Selected inversion as key to a stable Langevin
  evolution across the QCD phase boundary}, EPJ Web of Conferences 175 (2018)
  07003.

\bibitem{Sexty2019}
D.~Sexty, \emph{Calculating the equation of state of dense quark-gluon plasma
  using the complex Langevin equation}, Phys. Rev. D 100 (2019) 074503.

\bibitem{Hueffel1984}
H.~Hüffel, H.~Rumpf, \emph{Stochastic quantization in Minkowski space},
  Physics Letters B 148 (1984) 104 .

\bibitem{Gozzi1985}
E.~Gozzi, \emph{Langevin simulation in Minkowski space?}, Physics Letters B 150
  (1985) 119 .

\bibitem{Hiromichi1986}
H.~Nakazato, Y.~Yamanaka, \emph{Minkowski stochastic quantization}, Phys. Rev.
  D 34 (1986) 492.

\bibitem{Fukuda1988}
R.~Fukuda, H.~Higurashi, \emph{Equilibrium limit of the stochastic quantization
  in minkowski space}, Physics Letters B 202 (1988) 541 .

\bibitem{Kolley1988}
E.~Kolley, W.~Kolley, \emph{Stochastic quantization of thermal fermions on the
  real-time Keldysh contour}, Physics Letters A 129 (1988) 208 .

\bibitem{Rumpf1986}
H.~Rumpf, \emph{Stochastic quantization of Einstein gravity}, Phys. Rev. D 33
  (1986) 942.

\bibitem{Anzaki2015}
R.~Anzaki, K.~Fukushima, Y.~Hidaka, T.~Oka, \emph{Restricted phase-space
  approximation in real-time stochastic quantization}, Annals of Physics 353
  (2015) 107 .

\bibitem{Vicari2009}
E.~Vicari, H.~Panagopoulos, \emph{Theta dependence of SU(N) gauge theories in
  the presence of a topological term}, Physics Reports 470 (2009) 93.

\bibitem{Bongiovanni:2014rna}
L.~Bongiovanni, G.~Aarts, E.~Seiler, D.~Sexty, \emph{{Complex Langevin dynamics
  for SU(3) gauge theory in the presence of a theta term}}, PoS LATTICE2014
  (2014) 199, \eprint{1411.0949}.

\bibitem{hirasawa2020}
M.~Hirasawa, A.~Matsumoto, J.~Nishimura, A.~Yosprakob, \emph{Complex Langevin
  analysis of 2D U(1) gauge theory on a torus with a $\theta$ term} (2020),
  \eprint{2004.13982}.

\bibitem{Banks1997}
T.~Banks, W.~Fischler, S.~H. Shenker, L.~Susskind, \emph{$M$ theory as a matrix
  model: A conjecture}, Phys. Rev. D 55 (1997) 5112.

\bibitem{Pallab2018}
P.~Basu, K.~Jaswin, A.~Joseph, \emph{Complex Langevin dynamics in large $N$
  unitary matrix Models}, Phys. Rev. D 98 (2018) 034501.

\bibitem{Anosh2019}
A.~Joseph, A.~Kumar, \emph{Complex Langevin simulations of zero-dimensional
  supersymmetric quantum field theories}, Phys. Rev. D 100 (2019) 074507.

\bibitem{Nishimura2019}
J.~Nishimura, A.~Tsuchiya, \emph{Complex Langevin analysis of the space-time
  structure in the Lorentzian type IIB matrix model}, Journal of High Energy
  Physics 2019 (2019) 77.

\bibitem{ito2016b}
Y.~Ito, J.~Nishimura, \emph{Spontaneous symmetry breaking induced by complex
  fermion determinant --- yet another success of the complex Langevin method}
  (2016), \eprint{1612.00598}.

\bibitem{anagnostopoulos2019}
K.~N. Anagnostopoulos, T.~Azuma, Y.~Ito, J.~Nishimura, S.~K. Papadoudis,
  \emph{Dynamical compactification of extra dimensions in the Euclidean type
  IIB matrix model: A numerical study using the complex Langevin method}
  (2019), \eprint{1906.01841}.

\bibitem{Anagnostopoulos2018}
K.~N. Anagnostopoulos, T.~Azuma, Y.~Ito, J.~Nishimura, S.~K. Papadoudis,
  \emph{Complex Langevin analysis of the spontaneous symmetry breaking in
  dimensionally reduced super Yang-Mills models}, Journal of High Energy
  Physics 2018 (2018).

\bibitem{anagnostopoulos2020}
K.~N. Anagnostopoulos, T.~Azuma, Y.~Ito, J.~Nishimura, T.~Okubo, S.~K.
  Papadoudis, \emph{Complex Langevin analysis of the spontaneous breaking of
  10D rotational symmetry in the Euclidean IKKT matrix model} (2020),
  \eprint{2002.07410}.

\bibitem{anagnostopoulos2020b}
K.~N. Anagnostopoulos, T.~Azuma, Y.~Ito, J.~Nishimura, T.~Okubo, S.~K.
  Papadoudis, \emph{Dynamical Compactification of Extra Dimensions in the
  Euclidean IKKT Matrix Model via Spontaneous Symmetry breaking} (2020),
  \eprint{2005.12567}.

\bibitem{Lin1986}
H.~Q. Lin, J.~E. Hirsch, \emph{Monte Carlo versus Langevin methods for
  nonpositive definite weights}, Phys. Rev. B 34 (1986) 1964.

\bibitem{Drut2018}
J.~E. Drut, \emph{Advances in non-relativistic matter via complex Langevin
  approaches}, J. Phys. Conf. Ser. 1041 (2018) 012005.

\bibitem{2018UFGviaCL}
L.~Rammelm\"uller, A.~C. Loheac, J.~E. Drut, J.~Braun, \emph{Finite-Temperature
  Equation of State of Polarized Fermions at Unitarity}, Phys. Rev. Lett. 121
  (2018) 173001.

\bibitem{yamamoto2015}
A.~Yamamoto, T.~Hayata, \emph{Complex Langevin simulation in condensed matter
  physics} (2015), \eprint{1508.00415}.

\bibitem{PhysRevA92043628}
T.~Hayata, A.~Yamamoto, \emph{Complex Langevin simulation of quantum vortices
  in a Bose-Einstein condensate}, Phys. Rev. A 92 (2015) 043628.

\bibitem{Berger2019}
C.~Berger, J.~Drut, \emph{{Interacting Bosons at Finite Angular Momentum Via
  Complex Langevin}}, PoS LATTICE2018 (2019) 244.

\bibitem{Attanasio2020}
F.~Attanasio, J.~E. Drut, \emph{Thermodynamics of spin-orbit-coupled bosons in
  two dimensions from the complex Langevin method}, Phys. Rev. A 101 (2020)
  033617.

\bibitem{0954-3899-44-1-015101}
S.~Kemler, M.~Pospiech, J.~Braun, \emph{Formation of self-bound states in a
  one-dimensional nuclear model: A renormalization group based density
  functional study}, J. Phys. G 44 (2017) 015101.

\bibitem{RevModPhys.85.1633}
X.-W. Guan, M.~T. Batchelor, C.~Lee, \emph{Fermi gases in one dimension: From
  {Bethe} ansatz to experiments}, Rev. Mod. Phys. 85 (2013) 1633.

\bibitem{PhysRevD.99.074511}
H.~Singh, S.~Chandrasekharan, \emph{Few-body physics on a spacetime lattice in
  the worldline approach}, Phys. Rev. D 99 (2019) 074511.

\bibitem{Rammelmueller2018}
L.~Rammelmüller, J.~E. Drut, J.~Braun, \emph{A complex Langevin approach to
  ultracold fermions}, J. Phys. Conf. Ser. 1041 (2018) 012006.

\bibitem{1742-5468-2007-06-P06011}
T.~Iida, M.~Wadati, \emph{Exact analysis of a delta-function spin-1/2
  attractive Fermi gas with arbitrary polarization}, J. Stat. Mech. Theor. Exp.
  2007 (2007) P06011.

\bibitem{doi:10.1063/1.4964252}
C.~A. Tracy, H.~Widom, \emph{On the ground state energy of the delta-function
  Fermi gas}, J. Math. Phys. 57 (2016) 103301.

\bibitem{PhysRevA.96.033635}
L.~Rammelm\"uller, W.~J. Porter, J.~Braun, J.~E. Drut, \emph{Evolution from
  few- to many-body physics in one-dimensional Fermi systems: One- and two-body
  density matrices and particle-partition entanglement}, Phys. Rev. A 96 (2017)
  033635.

\bibitem{Singh:2018pci}
H.~Singh, \emph{{Worldline approach to few-body physics on the lattice}}, PoS
  LATTICE2018 (2018) 158, \eprint{1812.02364}.

\bibitem{rammelmueller2020}
L.~Rammelm\"uller, J.~E. Drut, J.~Braun, \emph{{Pairing patterns in
  one-dimensional spin- and mass-imbalanced Fermi gases}}, SciPost Phys. 9
  (2020) 14.

\bibitem{Luescher2008}
A.~L\"uscher, R.~M. Noack, A.~M. L\"auchli,
  \emph{{Fulde-Ferrell-Larkin-Ovchinnikov state in the one-dimensional
  attractive Hubbard model and its fingerprint in spatial noise correlations}},
  Phys. Rev. A 78 (2008) 013637.

\bibitem{PhysRevA.91.033618}
M.~D. Hoffman, P.~D. Javernick, A.~C. Loheac, W.~J. Porter, E.~R. Anderson,
  J.~E. Drut, \emph{Universality in one-dimensional fermions at finite
  temperature: Density, pressure, compressibility, and contact}, Phys. Rev. A
  91 (2015) 033618.

\bibitem{Loheac2018}
A.~C. Loheac, J.~Braun, J.~E. Drut, \emph{Equation of state of non-relativistic
  matter from automated perturbation theory and complex Langevin}, EPJ Web of
  Conferences 175 (2018) 03007.

\bibitem{Shill2018}
C.~R. Shill, J.~E. Drut, \emph{Particle Projection Using a Complex Langevin
  Method}, EPJ Web of Conferences 175 (2018) 03003.

\bibitem{PhysRevD.98.054507}
A.~C. Loheac, J.~Braun, J.~E. Drut, \emph{Polarized fermions in one dimension:
  Density and polarization from complex Langevin calculations, perturbation
  theory, and the virial expansion}, Phys. Rev. D 98 (2018) 054507.

\bibitem{Alexandru2018}
A.~Alexandru, P.~F. Bedaque, N.~C. Warrington, \emph{Spin polarized
  nonrelativistic fermions in $1+1$ dimensions}, Phys. Rev. D 98 (2018) 054514.

\bibitem{Zwerger2012}
W.~Zwerger~(Ed.), \emph{The BCS-BEC Crossover and the Unitary Fermi Gas}
  (Springer-Verlag, Berlin Heidelberg).

\bibitem{PhysRevA.85.051601}
J.~E. Drut, T.~A. L\"ahde, G.~Wlaz\l{}owski, P.~Magierski, \emph{Equation of
  state of the unitary Fermi gas: An update on lattice calculations}, Phys.
  Rev. A 85 (2012) 051601.

\bibitem{PhysRevLett.110.090401}
G.~Wlaz\l{}owski, P.~Magierski, J.~E. Drut, A.~Bulgac, K.~J. Roche,
  \emph{Cooper Pairing Above the Critical Temperature in a Unitary Fermi Gas},
  Phys. Rev. Lett. 110 (2013) 090401.

\bibitem{PhysRevA.93.053604}
O.~Goulko, M.~Wingate, \emph{{Numerical study of the unitary Fermi gas across
  the superfluid transition}}, Phys. Rev. A 93 (2016) 053604,
  \eprint{1507.08230}.

\bibitem{Jensen2019}
S.~Jensen, C.~N. Gilbreth, Y.~Alhassid, \emph{{The pseudogap regime in the
  unitary Fermi gas}}, Eur. Phys. J. ST 227 (2019) 2241, \eprint{1807.03913}.

\bibitem{PhysRevA.84.061602}
J.~Carlson, S.~Gandolfi, K.~E. Schmidt, S.~Zhang, \emph{Auxiliary-field quantum
  Monte Carlo method for strongly paired fermions}, Phys. Rev. A 84 (2011)
  061602.

\bibitem{PhysRevB.73.115112}
D.~Lee, \emph{Ground-state energy of spin-$\frac{1}{2}$ fermions in the unitary
  limit}, Phys. Rev. B 73 (2006) 115112.

\bibitem{PhysRevLett.121.130405}
R.~Rossi, T.~Ohgoe, K.~Van~Houcke, F.~Werner, \emph{Resummation of Diagrammatic
  Series with Zero Convergence Radius for Strongly Correlated Fermions}, Phys.
  Rev. Lett. 121 (2018) 130405.

\bibitem{VanHoucke2012}
K.~Van~Houcke, F.~Werner, E.~Kozik, N.~Prokof'ev, B.~Svistunov, M.~J.~H. Ku,
  A.~T. Sommer, L.~W. Cheuk, A.~Schirotzek, M.~W. Zwierlein, \emph{Feynman
  diagrams versus Fermi-gas Feynman emulator}, Nature Physics 8 (2012) 366.

\bibitem{He2020}
R.~He, N.~Li, B.-N. Lu, D.~Lee, \emph{Superfluid condensate fraction and
  pairing wave function of the unitary Fermi gas}, Phys. Rev. A 101 (2020)
  063615.

\bibitem{Ku2012}
M.~J.~H. Ku, A.~T. Sommer, L.~W. Cheuk, M.~W. Zwierlein, \emph{Revealing the
  Superfluid Lambda Transition in the Universal Thermodynamics of a Unitary
  Fermi Gas}, Science 335 (2012) 563.

\bibitem{Shin2006}
Y.~Shin, M.~W. Zwierlein, C.~H. Schunck, A.~Schirotzek, W.~Ketterle,
  \emph{Observation of Phase Separation in a Strongly Interacting Imbalanced
  Fermi Gas}, Phys. Rev. Lett. 97 (2006) 030401.

\bibitem{PhysRevLett.101.070404}
Y.-i. Shin, A.~Schirotzek, C.~H. Schunck, W.~Ketterle, \emph{Realization of a
  Strongly Interacting Bose-Fermi Mixture from a Two-Component Fermi Gas},
  Phys. Rev. Lett. 101 (2008) 070404.

\bibitem{Shin2008}
Y.-i. Shin, C.~H. Schunck, A.~Schirotzek, W.~Ketterle, \emph{Phase diagram of a
  two-component Fermi gas with resonant interactions}, Nature 451 (2008) 689.

\bibitem{Nascimbene2010}
S.~Nascimbène, N.~Navon, K.~J. Jiang, F.~Chevy, C.~Salomon, \emph{Exploring
  the thermodynamics of a universal Fermi gas}, Nature 463 (2010) 1057.

\bibitem{Navon2010}
N.~Navon, S.~Nascimb{\`e}ne, F.~Chevy, C.~Salomon, \emph{The Equation of State
  of a Low-Temperature Fermi Gas with Tunable Interactions}, Science 328 (2010)
  729.

\bibitem{Lobo2006}
C.~Lobo, A.~Recati, S.~Giorgini, S.~Stringari, \emph{Normal State of a
  Polarized Fermi Gas at Unitarity}, Phys. Rev. Lett. 97 (2006) 200403.

\bibitem{Chevy2010}
F.~Chevy, C.~Mora, \emph{Ultra-cold polarized Fermi gases}, Rep. Prog. Phys. 73
  (2010) 112401.

\bibitem{Radzihovsky2010}
L.~Radzihovsky, D.~E. Sheehy, \emph{Imbalanced Feshbach-resonant Fermi gases},
  Rep. Prog. Phys. 73 (2010) 076501.

\bibitem{Strinati2018}
G.~C. Strinati, P.~Pieri, G.~Röpke, P.~Schuck, M.~Urban, \emph{The BCS–BEC
  crossover: From ultra-cold Fermi gases to nuclear systems}, Phys. Rep. 738
  (2018) 1 .

\bibitem{Gubbels:2007xc}
K.~Gubbels, H.~Stoof, \emph{{Renormalization Group Theory for the Imbalanced
  Fermi Gas}}, Phys. Rev. Lett. 100 (2008) 140407, \eprint{0711.2963}.

\bibitem{Krippa:2014kra}
B.~Krippa, \emph{{Pairing in Asymmetric Many-Fermion Systems: Functional
  Renormalisation Group Approach}}, Phys. Lett. B 744 (2015) 288,
  \eprint{1407.5438}.

\bibitem{Boettcher2014}
I.~Boettcher, J.~Braun, T.~K. Herbst, J.~M. Pawlowski, D.~Roscher,
  C.~Wetterich, \emph{Phase structure of spin-imbalanced unitary Fermi gases},
  Phys. Rev. A 91 (2015) 013610.

\bibitem{Roscher:2015xha}
D.~Roscher, J.~Braun, J.~E. Drut, \emph{{Phase structure of mass- and
  spin-imbalanced unitary Fermi gases}}, Phys. Rev. A 91 (2015) 053611,
  \eprint{1501.05544}.

\bibitem{Frank2018}
B.~Frank, J.~Lang, W.~Zwerger, \emph{Universal Phase Diagram and Scaling
  Functions of Imbalanced Fermi Gases}, JETP 127 (2018) 812.

\bibitem{LIU201337}
X.-J. Liu, \emph{Virial expansion for a strongly correlated Fermi system and
  its application to ultracold atomic Fermi gases}, Phys. Rept. 524 (2013) 37.

\bibitem{Anglani_2014}
R.~Anglani, R.~Casalbuoni, M.~Ciminale, N.~Ippolito, R.~Gatto, M.~Mannarelli,
  M.~Ruggieri, \emph{Crystalline color superconductors}, Reviews of Modern
  Physics 86 (2014) 509–561.

\bibitem{Buballa_2015}
M.~Buballa, S.~Carignano, \emph{Inhomogeneous chiral condensates}, Progress in
  Particle and Nuclear Physics 81 (2015) 39–96.

\end{thebibliography}

\end{document}